Thesis leading to
the degree of Ph.D.

# Interface Growth Processes

Oleg Kupervasser

**Under The Supervision of: Itamar Procaccia**

*Department of Chemical Physics*
*The Weizmann Institute of Science*
*76100 Rehovot, Israel*

*November 1999*

# Contents

# Chapter 1

# Introduction

Problems of interface growth have received much attention recently [1–3]. Such are, for example, the duffusion limited aggregation (DLA) [4], random sequential adsorption (RSA) [5], Laplacian growth [6–8] or flame front propagation [9]. We will mainly pay attention in this Thesis to the numerical and analytical investigation of the last two problems. In addition to the fact that flame front propagation is an interesting physical problem we feel that we can also explain experimental results on the basis of theoretical investigations. There exists possibility to use methods found for the flame front propagation, in different fields where similar problems appear such as the important model of Laplacian growth .

The premixed flame - the self-sustaining wave of an exothermic chemical reaction - is one of the basic manifestations of gaseous combustion. It is well established, however, that the simplest imaginable flame configuration - unbounded planar flame freely propagating through initially motionless homogeneous combustible mixture - is intrinsically unstable and spontaneously assumes a characteristic two- or three-dimensional structure.

In the recent paper of Gostintsev, Istratov and Shulenin [10] an interesting survey of experimental studies on outward propagating spherical and cylindrical flames in the regime of well developed hydrodynamic (Darrieus-Landau) instability is presented. The available data clearly indicate that freely expanding wrinkled flames possess two intrinsic features:

1. Multi-quasi-cusps structure of the flame front. (The flame front consists of a large number of quasi-cusps, i.e., cusps with rounded tips.)

2. Noticeable acceleration of the flame front

Moreover, the temporal dependence of the flame radius is nearly identical for all premixtures discussed and correlates well with the simple relation:

$$R_0(t) = At^{3/2} + B \qquad (1.1)$$

Here $R_0(t)$ is the effective (average) radius of the wrinkled flame and A, B are empirical constants.

In this Thesis we study the spatial and temporal behavior of a nonlinear continuum model (i.e., a model which possesses an infinite number of degrees of freedom) which embodies all the characteristics deemed essential to premixed flame systems; namely, dispersiveness, nonlinearity and linear instability. Sivashinsky, Filyand and Frankel [11] recently obtained an equation, denoted by SFF in what follows, to describe how two-dimensional wrinkles of the cylindrical premixed flame grow as a consequence of the well-known Landau-Darrieus hydrodynamic instability. The SFF equation reads as follows:

$$\frac{\partial R}{\partial t} = \frac{U_b}{2R_0^2(t)}\left(\frac{\partial R}{\partial \theta}\right)^2 + \frac{D_M}{R_0^2(t)}\frac{\partial^2 R}{\partial \theta^2} + \frac{\gamma U_b}{2R_0(t)}I\{R\} + U_b \ . \quad (1.2)$$

where $0 < \theta < 2\pi$ is an angle, R($\theta$,t) is the modulus of the radius-vector on the flame interface, $U_b, D_M, \gamma$ are constants.

$$I(R) = \frac{1}{\pi}\sum_{n=1}^{\infty} n \int_0^{2\pi} \cos[n(\theta - \theta^*)]R(\theta^*, t)d\theta^* =$$
$$= -\frac{1}{\pi}P\int_{-\infty}^{+\infty} \frac{\frac{\partial R(\theta^*,t)}{\partial \theta^*}}{\theta^* - \theta}d\theta^* \quad (1.3)$$

$$R_0(t) = \frac{1}{2\pi}\int_0^{2\pi} R(\theta, t)d\theta \ . \quad (1.4)$$

Sivashinsky, Filyand and Frankel [11] made a direct numerical simulation of this nonlinear evolution equation for the cylindrical flame interface dynamics. The result obtained shows that the two mentioned experimental effects take place. Moreover, the evaluated acceleration rate is not incompatible with the power law given by eq.(1.1). For comparison, numerical simulations of freely expanding diffusively unstable flames were presented as well. In this case no tendency towards acceleration has been observed.

In the absence of surface tension, whose effect is to stabilize the short-wavelength perturbations of the interface, the problem of 2D Laplacian growth is described as follows

$$(\partial_x^2 + \partial_y^2)u = 0 \ . \quad (1.5)$$

$$u\mid_{\Gamma(t)} = 0 \ , \partial_n u\mid_\Sigma = 1 \ . \quad (1.6)$$

$$v_n = \partial_n u\mid_{\Gamma(t)} \ . \quad (1.7)$$

Here $u(x, y; t)$ is the scalar field mentioned, $\Gamma(t)$ is the moving interface, $\Sigma$ is a fixed external boundary, $\partial_n$ is a component of the gradient normal to the boundary (i.e. the normal derivative), and $v_n$ is a normal component of the velocity of the front.

To obtain results for radial flame growth it is necessary to investigate the channel case first. The channel version of equation for flame front propagation is the so-called Michelson-Sivashinsky equation [12,13] and looks like

$$\frac{\partial H}{\partial t} = \frac{1}{2}\left(\frac{\partial H}{\partial x}\right)^2 + \nu\frac{\partial^2 H}{\partial x^2} + I\{H\} \ . \quad (1.8)$$

$$I(H) = -\frac{1}{\pi}P\int_{-\infty}^{+\infty} \frac{\frac{\partial H(x^*,t)}{\partial x^*}}{x^* - x}dx^* \ . \quad (1.9)$$

with periodic boundary condition on the interval x [0,L], where L is size of the system. $\nu$ is constant, $\nu > 0$. H is the height of the flame front point, $P\int$ is the usual principal value integral.

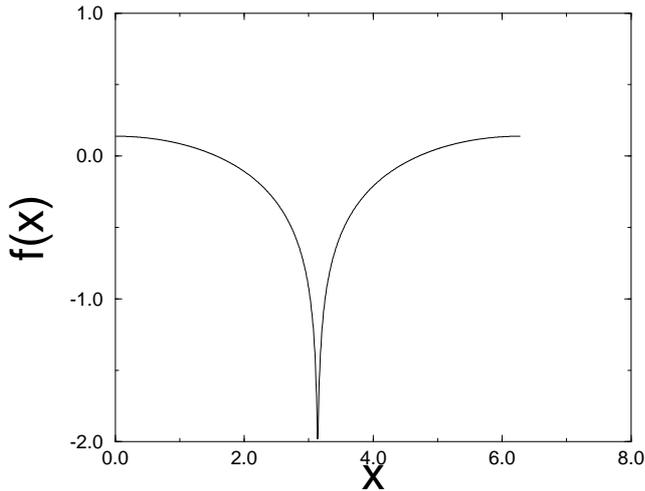

Figure 1.1: Giant cusp solution

Equations for flame front propagation and Laplacian growth with zero surface tension have remarkable property : these equations can be solved in terms of poles in the complex plane [6,12,14–16]. So we obtain a set of ordinary differential equations for the coordinates of these poles. The number of the poles is constant value in the system, but to explain such effect as growth of the velocity flame front we need to consider some noise that is a source of new poles. So we need to solve the problem of interaction of the random fluctuations and the pole motion.

The simplest case is the channel geometry. Main results for this case is existence of the giant cusp solution [12](Fig.1.1), which is represented in configuration space by poles which are organized on a line parallel to the imaginary axis. This pole solution is an attractor for pole dynamics.

A complete analysis of this steady-state solution was first presented in Ref. [12] and the main results are summarized as follows:

1. There is only one stable stationary solution which is geometrically represented by a giant cusp (or equivalently one finger) and analytically by $N(L)$ poles which are aligned on one line parallel to the imaginary axis. The existence of this solution is made clearer with the following remarks.

2. There exists an attraction between the poles along the real line. The resulting dynamics merges all the $x$ positions of poles whose $y$-position remains finite.

3. The $y$ positions are distinct, and the poles are aligned above each other in positions $y_{j-1} < y_j < y_{j+1}$ with the maximal being $y_{N(L)}$. This can be understood from equations for the poles motion in which the interaction is seen to be repulsive at short ranges, but changes sign at longer ranges.

4. If one adds an additional pole to such a solution, this pole (or another) will be pushed to infinity along the imaginary axis. If the system has less than $N(L)$ poles it is unstable to the addition of poles, and any noise will drive the system towards this unique state. The number $N(L)$ is

$$N(L) = \left[\frac{1}{2}\left(\frac{L}{\nu} + 1\right)\right], \tag{1.10}$$

where $[\ldots]$ is the integer part and $2\pi L$ is a system size. To see this consider a system with $N$ poles and such that all the values of $y_j$ satisfy the condition $0 < y_j < y_{max}$. Add now one additional pole whose coordinates are $z_a \equiv (x_a, y_a)$ with $y_a \gg y_{max}$. From the equation of motion for $y_a$, we see that the terms in the sum are all of the order of unity as is also the $\cot(y_a)$ term. Thus the equation of motion of $y_a$ is approximately

$$\frac{dy_a}{dt} \approx \nu \frac{2N+1}{L^2} - \frac{1}{L} \ . \tag{1.11}$$

The fate of this pole depends on the number of other poles. If $N$ is too large the pole will run to infinity, whereas if $N$ is small the pole will be attracted towards the real axis. The condition for moving away to infinity is that $N > N(L)$ where $N(L)$ is given by (1.10). On the other hand the $y$ coordinate of the poles cannot hit zero. Zero is a repulsive line, and poles are pushed away from zero with infinite velocity. To see this consider a pole whose $y_j$ approaches zero. For any finite $L$ the term $\coth(y_j)$ grows unboundedly whereas all the other terms in the equation for the poles motion remain bounded.

5. The height of the cusp is proportional to $L$. The distribution of positions of the poles along the line of constant $x$ was worked out in [12].

We will refer to the solution with all these properties as the Thual-Frisch-Henon (TFH)-cusp solution.

The main results of our own work are as follow. Traditional linear analysis was made for this giant cusp solution. This analysis demonstrates the existence of negative eigenvalues that go to zero when the system size goes to infinity.

1. There exists an obvious Goldstone or translational mode with eigenvalue $\lambda_0 = 0$. This eigenmode stems from the Galilean invariance of the equation of motion.

2. The rescaled eigenvalues ($L^2 \lambda_i$) oscillate periodically between values that are $L$-independent in this presentation. In other words, up to the oscillatory behavior the eigenvalues depend on $L$ like $L^{-2}$.

3. The eigenvalues $\lambda_1$ and $\lambda_2$ hit zero periodically. The functional dependence in this presentation appears almost piecewise linear.

4. The higher eigenvalues also exhibit similar qualitative behavior, but without reaching zero. We note that the solution becomes marginally stable for every value of $L$ for which the eigenvalues $\lambda_1$ and $\lambda_2$ hit zero. The $L^{-2}$ dependence of the spectrum indicates that the solution becomes more and more sensitive to noise as $L$ increases.

It was proved that arbitrary initial conditions can be written in the term of poles in the complex plane. Inverse cascade process of giant cusp formation was investigated numerically and analytically. Dependencies of the flame front width and mean velocity were found. The next step in investigation of the channel case was the influence of random noise on the pole dynamics. The main effect of the external noise is the appearance of new poles in the minima of the flame front and the merging these poles with the giant cusp. The dependence of the mean flame front velocity on the noise and the system size was found. The velocity is almost independent on the noise until the noise achieves some critical value. In the dependence of the velocity on the system size we see growth of the velocity with some exponent until the velocity achieves some saturation value.

Denoting $v$ as the velocity of the flame front and L the system size:

1. We can see two different regimes of behavior the average velocity $v$ as a function of noise $f$ for fixed system size L. For the noise $f$ smaller then same fixed value $f_{cr}$
$$v \sim f^{\xi} . \tag{1.12}$$
For these values of $f$ this dependence is very weak, and $\xi \approx 0.02$. For large values of $f$ the dependence is much stronger

2. We can see growth of the average velocity $v$ as a function of the system size L. After some values of L we can see saturation of the velocity. For regime $f < f_{cr}$ the growth of the velocity can be written as
$$v \sim L^{\mu}, \quad \mu \approx 0.35 \pm 0.03 . \tag{1.13}$$

The dependence of the number of poles in the system and the number of the poles that appear in the system in unit time was investigated numerically as a function of the noise and the system parameters. The life time of a pole was found numerically. Theoretical discussion of the effect of noise on the pole dynamics and mean velocity was made [17].

Pole dynamics can be used also to analyze small perturbation of the flame front and make the full stability analysis of the giant cusp. Two kinds of modes were found. The first one is eigenoscillations of the poles in the giant cusp. The second one is modes connected to the appearance of the new poles in the system. The eigenvalues of these modes were found. The results are in good agreement wit h the traditional stability analysis [18].

The results found for the cannel case can be used to analyze flame front propagation in the radial case [19, 20]. Main feature of this case is a competition between attraction of the poles and expanding of the flame front. So in this case we obtain not only one giant cusp but a set of cusps. New poles that appear in the system because of the noise form these cusps. On the basis of the equation of poles motion we can find connection between acceleration of the flame front and the width of the interface. On the basis of the result for mean velocity in the channel case the acceleration of the flame front can be found. So we obtain full picture of the flame front propagation in the radial case.

The next step in the investigation of the problem is considering Laplacian growth with zero surface tension that also has pole solutions. In the case of Laplacian growth we obtain result that is analogous to the merging of the poles in the channel case of the flame front propagation: all poles coalesce into one pole in the case of periodic boundary condition or two poles on the boundaries in the case of no-flux boundary conditions. This result can be proved theoretically [21].

In papers [22–26] self-acceleration without involvement of the external forcing is considered. No self-acceleration exist for the finite number of poles. So we can explain the self-acceleration and the appearance of new poles or by the noise or by the "rain" of poles from the "cloud" in infinity. Indeed, any given initial condition can be written as a sum of infinite number of poles( Sec. 2.4.1). Let us consider one pole that appears from the "cloud" in infinity. We neglect by the repulse force from the rest of poles in the system and consider only attraction force in Eqs.4.13 $-\frac{\gamma}{2r_0}$ For $r_0$ we can write in the case of self-acceleration $r_0(\tau) = (a+\tau)^{\beta}, r_0(0) = a^{\beta}, \beta > 1$. So from $\tau = 0$ to $\tau = \infty$ pole comes down a distance $\Delta y = \int_0^\infty \frac{\gamma}{2r_0(\tau)} d\tau = \frac{\gamma}{2} \frac{1}{\beta-1} \frac{1}{r_0(0)^{\frac{\beta-1}{\beta}}}$.
So the "rain" come down the finite distance after the infinite time and this distance converges to zero if $r_0(0) \mapsto \infty$! So we think that the appearance of new poles from the infinity can be explained only by the external noise. The characteristic size of cusp in the system $\mathcal{L} \sim r_0^{\frac{1}{\beta}}$. So from Fig 2.23 the noise $f \sim \frac{1}{\mathcal{L}^5} \sim \frac{1}{r_0^{\frac{5}{\beta}}}$ is necessary for the appearance of new cusps in the system. If the noise is larger than this value the dependence on the noise is very slow( $f^0.2$ for regime II and $f^0.02$ for regime III). This result explains the weak dependence of numerical simulations on the noise reduction ( [22], Fig.2).

Joulin et al. [27–30] use a vary similar approach for the channel and radial flame growth. But the main attention in our work is made to the velocity of flame front (self-acceleration for the radial case) and the flame front width. Main attention in the channel case in Joulin's work was made to the investigation of mean-spacing between cusps (crests). For the radial case only the linear dependence of radius on time (no self-acceleration) is considered in Joulin's work. Our works greatly complement each other but don't compete with each other.For example, for the distance between cusps (the mean value of cusp) without any proof we use Eq.(2.70). Fig.9( [28]) give us a excellent proof of this equation.

The structure of this Thesis is as follow. Chapter 1 is this Introduction.

In Chapter 2 we obtain main results for the channel case of the flame front propagation. We give results about steady state solutions, present traditional linear analysis of the problem and investigate analytically and numerically the influence of noise on the mean velocity of the front and pole dynamics.

In Chapter 3 we obtain results of the linear stability analysis by the help of pole solutions.

In Chapter 4 we use the result obtained for the channel case for analysis of the flame front propagation in the radial case

In Chapter 5 we investigate asymptotic behavior of the poles in the complex plane for the Laplacian growth with the zero surface-tension in the case of periodic and no-flux boundary condition.

Chapter 6 is a summary.

# Chapter 2

# Pole-Dynamics in Unstable Front Propagation: the Case of the Channel Geometry

## 2.1 Introduction

The aim of this chapter is to examine the role of random fluctuations on the dynamics of growing wrinkled interfaces which are governed by non-linear equations of motion. We are interested in those examples for which the growth of a flat or smooth interface are inherently unstable. A famous example of such growth phenomena is provided by Laplacian growth patterns [1–3]. The experimental realization of such patterns is seen for example in Hele-Shaw cells [1] in which air or another low viscosity fluid is displacing oil or some other high viscosity fluid. Under normal conditions the advancing fronts do not remain flat; in channel geometries they form in time a stable finger whose width is determined by delicate effects that arise from the existence of surface tension. In radial geometry, the growth the interface forms a contorted and ramified fractal shape. A related phenomenon has been studied in a model equation for flame propagation which has the same linear stability properties as the Laplacian growth problem [9]. The physical problem in this case is that of pre-mixed flames which exist as self-sustaining fronts of exothermic chemical reactions in gaseous combustion. Experiments [10] on flame propagation in radial geometry show that the flame front accelerates as time goes on, and roughens with characteristic exponents. Both observations did not receive proper theoretical explanations. It is notable that the channel and radial growth are markedly different; the former leads to a single giant cusp in the moving front, whereas the latter exhibits infinitely many cusps that appear in a complex hierarchy as the flame front develops ( [11, 19] and chapter 4).

Analytic techniques to study such processes are available [38]. In the context of flame propagation [12, 15, 19, 39], and in Laplacian growth in the zero surface-tension limit [6, 35, 36] one can examine solutions that are described in terms of poles in the complex plane. This description is very useful in providing a set of ordinary differential equations for the positions of the poles, from which one can deduce the geometry of the developing front in an extremely economical and efficient way. Unfortunately this description is not available in the case of Laplacian growth with surface tension, and this makes the flame propagation problem very attractive. However, it suffers from one fundamental drawback. For the noiseless equation the pole-dynamics always conserves the number of poles that existed in the initial conditions. As a result there is a final degree of ramification that is afforded by every set of initial conditions even in the radial geometry, and it is not obvious how to describe the continuing self-similar growth that is seen in experimental conditions or numerical simulations. Furthermore, as mentioned before, at least in the case of flame propagation one observes [10] an *acceleration* of the flame front with time. Such a phenomenon is impossible when the number of poles is conserved. It is therefore tempting to conjecture that noise may have an important role in affecting the actual growth phenomena that are observed in such systems. In fact, the effect of noise on unstable front dynamics has not been adequately addressed in the literature. From the point of view of analytic techniques noise can certainly generate new poles even if the initial conditions had a finite number of poles. The subject of pole dynamics with the existence of random noise, and the interaction between random fluctuations and deterministic front propagation are the main issues of this chapter.

We opt to study the example of flame propagation rather than Laplacian growth, simply because the former has an analytic description in terms of poles also in the experimentally relevant case of finite viscosity. We choose to begin the study with channel geometry. The reason is that in radial geometry it is more difficult to disentangle the effects of external noise from those of initial conditions. After all, initially the system can contain infinitely many poles,very far away near infinity in the complex plane (and therefore having an infinitely small contribution to the interface). Since the growth of the radius changes the stability of the system, more and more of these poles might fall down to the real axis and become observable. In channel geometry the analysis of the effect of initial conditions is relatively straightforward, and one can understand it before focusing on the (more interesting) effects of external noise [12]. The basic reason for this is that in this geometry the noiseless steady state solution for the developed front is known analytically. As described in Section II, in a channel of width $L$ the steady-state solution is given in terms of $N(L)$ poles that are organized on a line parallel to the imaginary axis. It can be shown that for any number of poles in the initial conditions this is the only attractor of the pole dynamics. After the establishment of this steady state we can begin to systematically examine the effects of external noise on this solution. As stated before, in radial conditions there is no stable steady state with a finite number of poles, and the disentanglement of initial vs. external perturbations is less straightforward ( [19] and chapter 4). We show later that the insights provided in this chapter have relevance for radial growth as well as will be discussed in the sequel.

We have a number of goals in this chapter. Firstly, after introducing the pole decomposition, the pole dynamics, and the basic steady state, we will present stability analysis of the solutions of the flame propagation problem in a channel geometry. It will be shown that the giant cusp solution is linearly stable, but non-linearly unstable. These results, which are described in Section III, can be obtained either by linearizing the dynamics around the giant cusp solutions in order to study the stability eigenvalues, or by examining perturbations in the form of poles in the complex plane. The main result of Section III is that there exists one Goldstone mode and two modes whose eigenvalues hit the real axis periodically when the system size $L$ increases. Thus the system is marginally stable at particular values of $L$, and it is always nonlinearly unstable, allowing finite size perturbations to introduce new poles into the system. This insight allows us to understand the relation between the system size and the effects of noise. In Section IV we discuss the relaxation dynamics that ensues after starting the system with "small" initial data. We study the coarsening process that leads in time to the final solution of the giant cusp, and understand from this what are the typical time scales that exist in our dynamics. We offer in this Section some results of numerical simulations that are interpreted in the later sections. In Section V we focus on the phenomenon of acceleration of the flame front and its relation to the existence of noise. In noiseless conditions the velocity of the flame front in a finite channel is bounded [12]. This can be shown either by using the pole dynamics or directly from the equation of motion. We will present the results of numerical simulations where the noise is controlled, and show how the velocity of the flame front is affected by the level of the noise and the system size. The main results are: (i) Noise is responsible for introducing new poles to the system; (ii) For low levels of noise the velocity of the flame front scales with the system size with a characteristic exponent; (iii) There is a phase transition at a sharp (but system-size dependent) value of the noise-level, after which the behavior of the system changes qualitatively; (iv) After the phase transition the velocity of the flame front changes very rapidly with the noise level. In the last Section we remark on the implications of these observations for the scaling behavior of the radial growth problem, and present a summary and conclusions.

## 2.2 Equations of Motion and Pole-decomposition in the Channel Geometry

It is known that planar flames freely propagating through initially motionless homogeneous combustible mixtures are intrinsically unstable. It was reported that such flames develop characteristic structures which include cusps, and that under usual experimental conditions the flame front accelerates as time goes on. A model in $1+1$ dimensions that pertains to the propagation of flame fronts in channels of width $\tilde{L}$ was proposed in [9]. It is written in terms of position $h(x,t)$ of the flame front above the $x$-axis. After appropriate rescalings it takes the form:

$$\frac{\partial h(x,t)}{\partial t} = \frac{1}{2}\left[\frac{\partial h(x,t)}{\partial x}\right]^2 + \nu \frac{\partial^2 h(x,t)}{\partial x^2} + I\{h(x,t)\} + 1 \ . \tag{2.1}$$

The domain is $0 < x < \tilde{L}$, $\nu$ is a parameter and we use periodic boundary conditions. The functional $I[h(x,t)]$ is the Hilbert transform which is conveniently defined in terms of the spatial Fourier transform

$$h(x,t) = \int_{-\infty}^{\infty} e^{ikx}\hat{h}(k,t)dk \tag{2.2}$$

$$I[h(k,t)] = |k|\hat{h}(k,t) \tag{2.3}$$

For the purpose of introducing the pole-decomposition it is convenient to rescale the domain to $0 < \theta < 2\pi$. Performing this rescaling and denoting the resulting quantities with the same notation we have

$$\frac{\partial h(\theta,t)}{\partial t} = \frac{1}{2L^2}\left[\frac{\partial h(\theta,t)}{\partial \theta}\right]^2 + \frac{\nu}{L^2}\frac{\partial^2 h(\theta,t)}{\partial \theta^2}$$
$$+\frac{1}{L}I\{h(\theta,t)\} + 1 \ . \tag{2.4}$$

In this equation $L = \tilde{L}/2\pi$. Next we change variables to $u(\theta,t) \equiv \partial h(\theta,t)/\partial \theta$. We find

$$\frac{\partial u(\theta,t)}{\partial t} = \frac{u(\theta,t)}{L^2}\frac{\partial u(\theta,t)}{\partial \theta} + \frac{\nu}{L^2}\frac{\partial^2 u(\theta,t)}{\partial \theta^2} + \frac{1}{L}I\{u(\theta,t)\} \ . \tag{2.5}$$

It is well known that the flat front solution of this equation is linearly unstable. The linear spectrum in $k$-representation is

$$\omega_k = |k|/L - \nu k^2/L^2 \ . \tag{2.6}$$

There exists a typical scale $k_{max}$ which is the last unstable mode

$$k_{max} = \frac{L}{\nu} \ . \tag{2.7}$$

Nonlinear effects stabilize a new steady-state which is discussed next.

The outstanding feature of the solutions of this equation is the appearance of cusp-like structures in the developing fronts. Therefore a representation in terms of Fourier modes is very inefficient. Rather, it appears very worthwhile to represent such solutions in terms of sums of functions of poles in the complex plane. It will be shown below that the position of the cusp along the front is determined by the real coordinate of the pole, whereas the height of the cusp is in correspondence with the imaginary coordinate. Moreover, it will be seen that the dynamics of the developing front can be usefully described in terms of the dynamics of the poles. Following [12, 19, 38, 39] we expand the solutions $u(\theta, t)$ in functions that depend on $N$ poles whose position $z_j(t) \equiv x_j(t) + i y_j(t)$ in the complex plane is time dependent:

$$\begin{aligned} u(\theta, t) &= \nu \sum_{j=1}^{N} \cot\left[\frac{\theta - z_j(t)}{2}\right] + c.c. \\ &= \nu \sum_{j=1}^{N} \frac{2 \sin[\theta - x_j(t)]}{\cosh[y_j(t)] - \cos[\theta - x_j(t)]} \ , \end{aligned} \tag{2.8}$$

$$h(\theta, t) = 2\nu \sum_{j=1}^{N} \ln\left[\cosh(y_j(t)) - \cos(\theta - x_j(t))\right] + C(t) \ . \tag{2.9}$$

In (2.9) $C(t)$ is a function of time. The function (2.9) is a superposition of quasi-cusps (i.e. cusps that are rounded at the tip). The real part of the pole position (i.e. $x_j$) is the coordinate (in the domain $[0, 2\pi]$) of the maximum of the quasi-cusp, and the imaginary part of the pole position (i.e $y_j$) is related to the depth of the quasi-cusp. As $y_j$ decreases the depth of the cusp increases. As $y_j \to 0$ the depth diverges to infinity. Conversely, when $y_j \to \infty$ the depth decreases to zero.

The main advantage of this representation is that the propagation and wrinkling of the front can be described via the dynamics of the poles. Substituting (2.8) in (2.5) we derive the following ordinary differential equations for the positions of the poles:

$$-L^2 \frac{dz_j}{dt} = \left[\nu \sum_{k=1, k \neq j}^{2N} \cot\left(\frac{z_j - z_k}{2}\right) + i\frac{L}{2} sign[Im(z_j)]\right]. \tag{2.10}$$

We note that in (2.8), due to the complex conjugation, we have $2N$ poles which are arranged in pairs such that for $j < N$ $z_{j+N} = \bar{z}_j$. In the second sum in (2.8) each pair of poles contributed one term. In Eq.(2.10) we again employ $2N$ poles since all of them interact. We can write the pole dynamics in terms of the real and imaginary parts $x_j$ and $y_j$. Because of the arrangement in pairs it is sufficient to write the equation for either $y_j > 0$ or for $y_j < 0$. We opt for the first. The equations for the positions of the poles read

$$-L^2 \frac{dx_j}{dt} = \nu \sum_{k=1,k\neq j}^{N} \sin(x_j - x_k) \Big[ [\cosh(y_j - y_k) \quad (2.11)$$

$$- \cos(x_j - x_k)]^{-1} + [\cosh(y_j + y_k) - \cos(x_j - x_k)]^{-1} \Big]$$

$$L^2 \frac{dy_j}{dt} = \nu \sum_{k=1,k\neq j}^{N} \Big( \frac{\sinh(y_j - y_k)}{\cosh(y_j - y_k) - \cos(x_j - x_k)}$$

$$+ \frac{\sinh(y_j + y_k)}{\cosh(y_j + y_k) - \cos(x_j - x_k)} \Big) + \nu \coth(y_j) - L. \quad (2.12)$$

We note that if the initial conditions of the differential equation (2.5) are expandable in a finite number of poles, these equations of motion preserve this number as a function of time. On the other hand, this may be an unstable situation for the partial differential equation, and noise can change the number of poles. This issue will be examined at length in Section 2.5.

## 2.3 Linear Stability Analysis in Channel Geometry

In this section we discuss the linear stability of the TFH-cusp solution. To this aim we first use Eq.(2.8) to write the steady solution $u_s(\theta)$ in the form:

$$u_s(\theta) = \nu \sum_{j=1}^{N} \frac{2\sin[\theta - x_s]}{\cosh[y_j] - \cos[\theta - x_s]}, \quad (2.13)$$

where $x_s$ is the real (common) position of the stationary poles and $y_j$ their stationary imaginary position. To study the stability of this solution we need to determine the actual positions $y_j$. This is done numerically by integrating the equations of motion for the poles starting from $N$ poles in initial positions and waiting for relaxation. Next one perturbs this solution with a small perturbation $\phi(\theta,t)$: $u(\theta,t) = u_s(\theta) + \phi(\theta,t)$. Linearizing the dynamics for small $\phi$ results in the equation of motion

$$\frac{\partial \phi(\theta,t)}{\partial t} = \frac{1}{L^2} \Big[ \partial_\theta [u_s(\theta)\phi(\theta,t)] + \nu \partial_\theta^2 \phi(\theta,t) \Big]$$

$$+ \frac{1}{L} I(\phi(\theta,t)) . \quad (2.14)$$

### 2.3.1 Fourier decomposition and eigenvalues

The linear equation can be decomposed in Fourier modes according to

$$\phi(\theta,t) = \sum_{k=-\infty}^{\infty} \hat{\phi}_k(t) e^{ik\theta} \quad (2.15)$$

$$u_s(\theta) = -2\nu i \sum_{k=-\infty}^{\infty} \sum_{j=1}^{N} \text{sign}(k) e^{-|k|y_j} e^{ik\theta} \quad (2.16)$$

In these sums the discrete $k$ values run over all the integers. Substituting in (2.14) we get the equations:

$$\frac{d\hat{\phi}_k(t))}{dt} = \sum_n a_{kn} \hat{\phi}_n(t) , \quad (2.17)$$

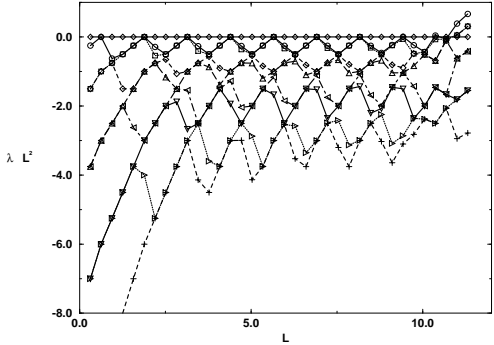

Figure 2.1: The first 10 highest eigenvalues of the stability matrix with $\nu = \pi/5$, multiplied by the square of the system size $L^2$ vs. the system size $L$. Note that all the eigennvalues oscillate around fixed values in this presentation, and that the highest two eigenvalues hit zero periodically.

where $a_{kn}$ is a infinite matrix whose entries are given by

$$a_{kk} = \frac{|k|}{L} - \frac{\nu}{L^2}k^2 \tag{2.18}$$

$$a_{kn} = \frac{k}{L^2}\text{sign}(k-n)(2\nu \sum_{j=1}^{N} e^{-|k-n|y_j}) \quad k \neq n . \tag{2.19}$$

To solve for the eigenvalues of this matrix we need to truncate it at some cutoff $k$-vector $k^*$. The choice of $k^*$ can be based on the linear stability analysis of the flat front. The scale $k_{max}$, cf. (2.7), is the largest $k$ which is still linearly unstable. We must choose $k^* > k_{max}$ and test the choice by the converegence of the eigenvalues. The chosen value of $k^*$ in our numerics was $4k_{max}$. The results for the low order eigenvalues of the matrix $a_{kn}$ that were obtained from a converged numerical calculation are presented in Fig.2.1.

The eigenvalues are multiplied by $L^2$ and are plotted as a function of $L$. We order the eigenvalues in decreasing order and denote them as $|\lambda_0| \leq |\lambda_1| \leq |\lambda_2| \ldots$.

Fig 2.1 contains a strange result on the positive eigenvalues at large L. One of methods to check some numerical result is to do analytic investigation. For example, in Chapter 3 we make detailed analytic investigation for the numerical result on Fig. 2.1 and obtain that all eigenvalues are not positive. Indeed, two types of modes exists. The first one is connected to the displacement of poles in the giant cusp. Because of the pole attraction the giant cusp is stable with respect to the longitudinal displacement of poles and so the correspondent eigenvalues are not positive. For the transversal displacement the Lyapunov function exists and so the giant cusp is stable with respect to the transversal displacement and the correspondent eigenvalues are not positive. The second type of modes is connected to additional poles. These poles go to infinity because of the repulsion from the giant cusp poles $N(L)$. So the correspondent eigenvalues are also not positive. So the positive eigenvalues at large L are a numerical artifact.

The figure offers a number of qualitative observations:

1. There exists an obvious Goldstone or translational mode $u'_s(\theta)$ with eigenvalue $\lambda_0 = 0$, which is shown with rhombes in Fig.2.1. This eigenmode stems from the Galilean invariance of the equation of motion.

2. The eigenvalues oscillate periodically between values that are $L$-independent in this presentation (in which we multiply by $L^2$). In other words, up to the oscillatory behavior the eigenvalues depend on $L$ like $L^{-2}$.

3. The eigenvalues $\lambda_1$ and $\lambda_2$, which are represented by squares and circles in Fig.2.1, hit zero periodically. The functional dependence in this presentation appears almost piecewise linear.

4. The higher eigenvalues also exhibit similar qualitative behaviour, but without reaching zero. We note that the solution becomes marginally stable for every value of $L$ for which the eigenvalues $\lambda_1$ and $\lambda_2$ hit zero. The $L^{-2}$ dependence of the spectrum indicates that the solution becomes more and more sensitive to noise as $L$ increases.

### 2.3.2 Qualitative understanding using pole-analysis

The most interesting qualitative aspects are those enumerated above as item 2 and 3. To understand them it is useful to return to the pole description, and to focus on Eq.(1.11). This equation describes the dynamics of a single far-away pole. We remarked before that this equation shows that for *fixed $L$* the stable number of poles is the integer part (1.10). Define now the number $\alpha$, $0 \leq \alpha \leq 1$, according to

$$\alpha = \left[\frac{1}{2}\left(\frac{L}{\nu}+1\right)\right] - \frac{1}{2}\left(\frac{L}{\nu}-1\right) \ . \qquad (2.20)$$

Using this number we rewrite Eq.(1.11) as

$$\frac{dy_a}{dt} \approx \frac{2\nu}{L^2}\alpha \ . \qquad (2.21)$$

As $L$ increases, $\alpha$ oscillates piecewise linearly and periodically between zero and unity. This shows that a distant pole which is added to the giant cusp solution is usually repelled to infinity except when $\alpha$ hits zero and the system becomes marginally unstable to the addition of a new pole.

To connect this to the linear stability analysis we note from Eq.(2.8) that a single far-away pole solution (i.e with $y$ very large) can be written as

$$u(\theta, t) = 4\nu e^{-y(t)} \sin(\theta - x(t)) \ . \qquad (2.22)$$

Suppose that we add to our giant cusp solution a perturbation of this functional form. From Eq.(2.21) we know that $y$ grows linearly in time, and therefore this solution decays exponentially in time. The rate of decay is a linear eigenvalue of the stability problem, and from Eq.(2.21) we understand both the $1/L^2$ dependence and the periodic marginality. We should note that this way of thinking gives us a significant part of the $L$ dependence of the eigenvalues, but not all. The variable $\alpha$ is rising from zero to unity periodically, but after reaching unity it hits zero instantly. Accordingly, if the highest non zero eigenvalue were fully determined by the pole analysis, we would expect this eigenvalue to behave as the solid line shown in Fig.2.2.

The actual highest eigenvalue computed from the stability matrix is shown in rhombuses connected by dotted line. It is clear that the pole analysis gives us a great deal of qualitative and quantitative understanding, but not all the features agree.

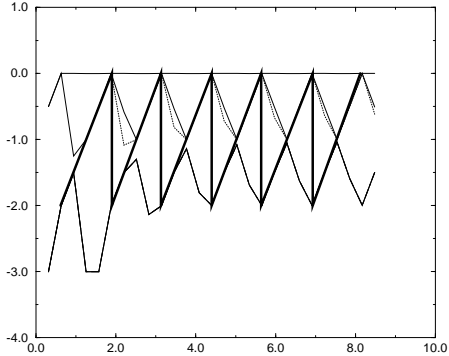

Figure 2.2: Comparison of the numerically determined highest 4 eigenvalues of the stability matrix with the prediction of the pole analysis. The eigenvalues of the stability matrix are : $\lambda_0$, $\lambda_1$, $\lambda_2$ and $\lambda_3$. The pole analysis (solid line) provides a qualitative understanding of the stability, and appears to overlap with the high est eigenvector over half of the range, and with the fourth eigenvalue over the other half.

### 2.3.3 Dynamics near marginality

The discovery of marginality at isolated values of $L$ poses questions regarding the fate of poles that are added at very large $y$'s at certain $x$-positions. We will argue now that when the system becomes marginally stable, a new pole can be added to those existing in the giant cusp. We remember that these poles have a common $\theta$ position that we denote as $\theta = \theta_c$. The fate of a new pole added at infinity depends on its $\theta$ position. If the position of the new pole is again denoted as $y_a$, and $\infty \gg y_a \gg y_{max}$, we can see from Eq.(2.12) that $dy_a/dt$ is maximal when $\theta_a = \theta_c$, whereas it is minimal when $\theta_a - \theta_c = \pi$. This follows from the fact that the cosine term has a value $+1$ when $\theta_a = \theta_c$ and a value $-1$ when $\theta_a - \theta_c = \pi$. For large $y$ differences the terms in the sum take on their minimal value when the cos term is $-1$ and their maximal values at $+1$. For infinitely large $y_a$ the equation of motion is (1.11) which is independent of $\theta_a$. Since the RHS of this equation becomes zero at marginality, we conclude that for very large but finite $y_a$ $dy_a/dt$ changes sign from positive to negative when $\theta_a - \theta_c$ changes from zero to $\pi$. The meaning of this observation is that the most unstable points in the system are those points which are furthest away from the giant cusp. It is interesting to discuss the fate of a pole that is added to the system at such a position. From the point of view of the pole dynamics $\theta = \theta_c + \pi$ is an unstable fixed point for the motion along the $\theta$ axis. The attraction to the giant cusp exactly vanishes at this point. If we start with a pole at a very large $y_a$ close to this value of $\theta$ the down-fall along the $y$ coordinate will be faster than the lateral motion towards the giant cusp. We expect to see therefore the creation of a small cusp at $\theta$ values close to $\pi$ that precedes a later stage of motion in which the small cusp moves to merge with the giant cusp. Upon the approach of the new pole to the giant cusp all the existing poles will move up and the furthest pole at $y_{max}$ will be kicked off to infinity. We will later explain that this type of dynamics occurs in stable systems that are driven by noise. The noise generates far away poles (in the imaginary direction) that get attracted around $\theta = \theta_c + \pi$ to create small cusps that run continuously towards the giant cusp.

### 2.3.4 Excitable System.

The intuition gained so far can be used to discuss the issue of stability of a stable system to *larger* perturbations. In other words, we may want to add to the system poles at finite values of $y$ and ask about their fate. We first show in this subsection that poles whose initial $y$ value is below $y_{max} \sim \log(L^2/\nu^2)$ will be attracted towards the real axis. The scenario is similar to the one described in the last paragraph.

Suppose that we generate a stable system with a giant cusp at $\theta_c = 0$ with poles distributed along the $y$ axis up to $y_{max}$. We know that the sum of all the forces that act on the upper pole is zero. Consider then an additional pole inserted in the position $(\pi, y_{max})$. It is obvious from Eq.(2.12) that the forces acting on this pole will pull it downward. On the other hand if its initial position is much above $y_{max}$ the force on it will be repulsive towards infinity. We see that this simple argument identifies $y_{max}$ as the typical scale for nonlinear instability.

Next we estimate $y_{max}$ and interpret our result in terms of the *amplitude* of a perturbation of the flame front. We explained that uppermost pole's position fluctuates between a minimal value and infinity as $L$ is changing. We want to estimate the characteristic scale of the minimal value of $y_{max}(L)$. To this aim we employ the result of ref. [12] regarding the stable distribution of pole positions in a stable large system. The parametrization of [12] differs from ours; to go from our parametrization in Eq.(2.5) to theirs we need to rescale $u$ by $L^{-1}$ and $t$ by $L$. The parameter $\nu$ in their parameterizations is $\nu/L$ in ours. According to [12] the number of poles between $y$ and $y + dy$ is given by the $\rho(y)dy$ where the density $\rho(y)$ is

$$\rho(y) = \frac{L}{\pi^2 \nu} \ln[\coth(|y|/4)] \ . \tag{2.23}$$

To estimate the minimal value of $y_{max}$ we require that the tail of the distribution $\rho(y)$ integrated between this value and infinity will allow one single pole. In other words,

$$\int_{y_{max}}^{\infty} dy \rho(y) \approx 1 \ . \tag{2.24}$$

Expanding (2.23) for large $y$ and integrating explicitly the result in (2.24) we end up with the estimate

$$y_{max} \approx 2 \ln\left[\frac{4L}{\pi^2 \nu}\right] \tag{2.25}$$

For large $L$ this result is $y_{max} \approx \ln(\frac{L^2}{\nu^2})$. If we now add an additional pole in the position $(\theta, y_{max})$ this is equivalent to perturbing the solution $u(\theta, t)$ with a function $\nu e^{-y_{max}} \sin(\theta)$, as can be seen directly from (2.8). We thus conclude that the system is unstable to a perturbation *larger* than

$$u(\theta) \sim \nu^3 \sin(\theta)/L^2 \ . \tag{2.26}$$

This indicates a very strong size dependence of the sensitivity of the giant cusp solution to external perturbations. This will be an important ingredient in our discussion of noisy systems.

## 2.4 Initial Conditions, Pole Decomposition and Coarsening

In this section we show first that any initial conditions can be approximated by pole decomposition. Later, we show that the dynamics of sufficiently smooth initial data can be well understood from the pole decomposition. Finally we employ this picture to describe the *inverse cascade* of cusps into the giant cusp which is the final steady state. By inverse cascade we mean a nonlinear coarsening process in which the small scales coalesce in favor of larger scales and finally the system saturates at the largest available scale [40].

### 2.4.1 Pole Expansion: General Comments

The fundamental question is how many poles are needed to describe any given initial condition. The answer, of course, depends on how smooth are the initial conditions. Suppose also that we have an initial function $u(\theta, t = 0)$ that is $2\pi$-periodic and which at time $t = 0$ admits a Fourier representation

$$u(\theta) = \sum_{k=1}^{\infty} A_k \sin(k\theta + \phi_k) , \qquad (2.27)$$

with $A_k > 0$ for all $k$. Suppose that we want to find a pole-decomposition representation $u_p(\theta)$ such that

$$|u_p(\theta) - u(\theta)| \leq \epsilon \quad \text{for every } \theta , \qquad (2.28)$$

where $\epsilon$ is a given wanted accuracy. If $u(\theta)$ is differentiable we can cut the Fourier expansion at some finite $k = K$ knowing that the remainder is smaller than, say, $\epsilon/2$. Choose now a large number $M$ and a small number $\Delta \ll 1/M$ and write the pole representation for $u_p(\theta)$ as

$$u_p(\theta) = \sum_{k=1}^{K} \sum_{p=0}^{M-1} \frac{2k \sin(k\theta + \phi_k)}{\cosh[k(y_k + p\Delta)] - \cos(k\theta + \phi_k)} . \qquad (2.29)$$

To see that this representation is a particular form of the general formula (2.8) We use the following two identities

$$\sum_{k=0}^{\infty} e^{-kt} \sin xk = \frac{1}{2} \frac{\sin x}{\cosh t - \cos x} , \qquad (2.30)$$

$$\sum_{k=0}^{K-1} \sin(x + ky) = \sin\left(x + \frac{K-1}{2}y\right) \sin \frac{Ky}{2} \operatorname{cosec} \frac{y}{2} . \qquad (2.31)$$

From these follows a third identity

$$\sum_{j=0}^{K-1} \frac{2 \sin\left(x - \frac{2\pi j}{K} + \phi\right)}{\cosh y - \cos\left(x - \frac{2\pi j}{K} + \phi\right)}$$
$$= \frac{2K \sin(Kx + \phi)}{\cosh Ky - \cos(Kx + \phi)} . \qquad (2.32)$$

Note that the LHS of (2.32) is of the form (2.8) with $K$ poles whose positions are all on the line $y_j = y$ and whose $x_j$ are on the lattice points $2\pi j/K - \phi$. On the other hand every term in (2.29) is of this form.

Next we use (2.30) to rewrite (2.29) in the form

$$u_p(\theta) = \sum_{k=1}^{K} \sum_{p=0}^{M-1} \sum_{n=1}^{\infty} 4k e^{-nk(y_k + p\Delta)} \sin(nk\theta + n\phi_k) . \qquad (2.33)$$

Exchanging order of summation between $n$ and $p$ we can perform the geometric sum on $p$. Denoting

$$b_{n,k} \equiv \sum_{p=0}^{M-1} e^{-nkp\Delta} = \frac{1 - e^{-Mkn\Delta}}{1 - e^{-kn\Delta}} \ , \tag{2.34}$$

we find

$$\begin{aligned} u_p(\theta) &= \sum_{k=1}^{K} \sum_{n=1}^{\infty} 4k b_{n,k} e^{-nky_k} \sin(nk\theta + n\phi_k) \\ &= \sum_{k=1}^{K} \sum_{n=2}^{\infty} 4k b_{n,k} e^{-nky_k} \sin(nk\theta + n\phi_k) \\ &+ \sum_{k=1}^{K} 4k b_{1,k} e^{-ky_k} \sin(k\theta + \phi_k) \ . \end{aligned} \tag{2.35}$$

Compare now the second term on the RHS of (2.35) with (2.27). We can identify

$$e^{-ky_k} = \frac{A_k}{4k b_{1,k}} \tag{2.36}$$

The first term can be then bound from above as

$$\left| \sum_{k=1}^{K} \sum_{n=2}^{\infty} 4k b_{n,k} e^{-nky_k} \sin(nk\theta + n\phi_k) \right| \tag{2.37}$$

$$\leq \sum_{k=1}^{K} \sum_{n=2}^{\infty} \left| 4k b_{n,k} \left[ \frac{A_k}{4k b_{1,k}} \right]^n \sin(nk\theta + n\phi_k) \right|.$$

The sine function and the factor $(4K)^{1-n}$ can be replaced by unity and we can bound the RHS of (2.37) by

$$\sum_{k=1}^{K} \sum_{n=2}^{\infty} \left[ \frac{A_k}{b_{1,k}} \right]^n b_{n,k} \leq \sum_{k=1}^{K} A_k \sum_{n=1}^{\infty} \left[ \frac{A_k}{b_{1,k}} \right]^n , \tag{2.38}$$

where we have used the fact that $b_{n,k} \leq b_{1,k}$ which follows directly from (2.34). Using now the facts that $b_{1,K} \leq b_{1,k}$ for every $k \leq K$ and that $A_k$ is bounded by some finite $C$ since it is a Fourier coefficient, we can bound (2.38) by $C^2 K/(b_{1,K} - C)$. Since we can select the free parameters $\Delta$ and $M$ to make $b_{1,K}$ as large as we want, we can make the remainder series smaller in absolute value than $\epsilon/2$.

The conclusion of this demonstration is that any initial condition that can be represented in Fourier series can be approximated to a desired accuracy by pole-decomposition. The number of needed poles is of the order $K^2 \times M$. Of course, the number of poles thus generated by the initial conditions may exceed the number $N(L)$ found in Eq.(1.10). In such a case the excess poles will move to infinity and will become irrelevant for the short time dynamics. Thus a smaller number of poles may be needed to describe the state at larger times than at $t = 0$. We need to stress at this point that the pole decomposition is over complete; for example, if there is exactly one pole at $t = 0$ and we use the above technique to reach a pole decomposition we would get a large number of poles in our representation.

### 2.4.2 The initial stages of the front evolution: the exponential stage and the inverse cascade

In this section we employ the connection between Fourier expansion and pole decomposition to understand the initial exponential stage of the evolution of the flame front with small initial data $u(\theta, t = 0)$. Next we employ our knowledge of the pole interactions to explain the slow dynamics of coarsening into the steady state solution.

Suppose that initially the expansion (2.27) is available with all the coefficients $A_k \ll 1$. We know from the linear instability of the flat flame front that each Fourier component changes exponentially in time according to the linear spectrum (2.6). The components with wave vector larger than (2.7) decrease, whereas those with lower wave vectors increase. The fastest growing mode is $k_c = L/2\nu$. In the linear stage of growth this mode will dominate the shape of the flame front, i.e.

$$u(\theta, t) \approx A_{k_c} e^{\omega_{k_c} t} \sin(k_c \theta) . \qquad (2.39)$$

Using Eq.(2.32) for a large value of $y$ (which is equivalent to small $A_{k_c}$) we see that to the order of $O(A_{k_c}^2)$ (2.39) can be represented as a sum over $L/2\nu$ poles arranged periodically along the $\theta$ axis. Other unstable modes will contribute similar arrays of poles but at much higher values of $y$, since their amplitude is exponentially smaller. In addition we have nonlinear corrections to the identification of the modes in terms of poles. These corrections can be again expanded in terms of Fourier modes, and again identified with poles, which will be further away along the $y$ axis, and with higher frequencies. To see this one can use Eq.(2.35), subtract from $u_p(\theta)$ the leading pole representations, and reexpand in Fourier series. Then we identify the leading order with double the number of poles that are situated twice further away along the $y$ axis.

We note that even when all the unstable modes are present, the number of poles in the first order identification is finite for finite $L$, since there are only $L/\nu$ unstable modes. Counting the number of poles that each mode introduces we get a total number of $\left(L/\nu\right)^2$ poles. The number $L/2\nu$ of poles which are associated with the most unstable mode is precisely the number allowed in the stable stationary solution, cf.(1.10). When the poles approach the real axis and cusps begin to develop, the linear analysis no longer holds, but the pole description does.

We now describe the qualitative scenario for the establishment of the steady state. Firstly, we understand that all the poles that belong to less unstable modes will be pushed towards infinity. To see this think of the system at this stage as an array of uncoupled systems with a scale of the order of unity. Each such system will have a characteristic value of $y$. As we discussed before poles that are further away along the $y$ axis will be pushed to infinity. Therefore the system will remain with the $L/2\nu$ poles of the most unstable mode. The net effect of the poles belonging to the (nonlinearly) stable modes is to destroy the otherwise perfect periodicity of the poles of the unstable mode. The see the effect of the higher order correction to the pole identification we again recall that they can be represented as further away poles with higher frequencies, whose dynamics is similar to the less unstable modes that were just discussed. They do not become more relevant when time goes on.

Once the poles of the stable modes get sufficiently far from the real axis, the dynamics of the remaining poles will begin to develop according to the interactions that are directed along the real axis. These interactions are much weaker and the resulting dynamics occur on much longer time scales. The qualitative picture is of an inverse cascade of merging the $\theta$ positions of the poles. We note that the system has a set of unstable fixed points which are 'cellular solutions' described by a periodic arrangement of poles along the real axis with a frequency $k$. These fixed points are not stable and they collapse, under perturbations, with a characteristic time scale (that depends on $k$) to the next unstable fixed point at $k' = k/2$. This process then goes on indefinitely until $k \sim 1/L$ i.e. we reach the giant cusp, the steady-state stable solution [40].

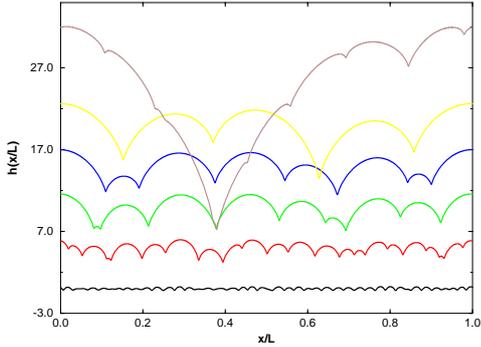

Figure 2.3: The inverse cascade process of coarsening that occurs after preparing the system with random, small initial conditions. One sees that at successive times the typical scale increases until the giant cusp forms, and attracts all the other side-poles. The effect of the existing numerical additive noise is to introduce poles that appear as side cusps that are continuously attracted to the giant cusp. This effect is obvious to the eye only after the typical scale is sufficiently large, as is seen in the last time (see text for further details).

This scenario is seen very clearly in the numerical simulations. In Fig.2.3 we show the time evolution of the flame front starting from small white-noise initial conditions. The bottom curve pertains to the earliest time in this picture, just after the fast exponential growth, and one sees clearly the periodic array of cusps that form. The successive images show the progress of the flame front in time, and one observes the development of larger scales with deeper cusps that represent the partial coalescence of poles onto the same $\theta$ positions. In Fig.2.4 we show the width and the velocity of this front as a function of time. One recognizes the exponential stage of growth in which the $L/2\nu$ poles approach the $\theta$ axis, and then a clear cross-over to much slower dynamics in which the effective scale in the system grows with a slower rate.

The slow dynamics stage can be understood qualitatively using the previous interpretation of the cascade as follows: if the initial number of poles belonging to the unstable mode is $L/2\nu$, the initial effective linear scale is $2\nu$. Thus the first step of the inverse cascade will be completed in a time scale of the order of $2\nu$. At this point the effective linear scale doubles to $4\nu$, and the second step will be completed after such a time scale. We want to know what is the typical length scale $l_t$ seen in the system at time $t$. The definition of front width is $l_t = \sqrt{\frac{1}{\tilde{L}} \int_0^{\tilde{L}} [h(x,t) - \bar{h}]^2 dx}$, $\bar{h} = \frac{1}{\tilde{L}} \int_0^{\tilde{L}} h(x,t) dx$. The typical width of the system at this stage will be proportional to this scale.

Denote the number of cascade steps that took place until this scale is achieved

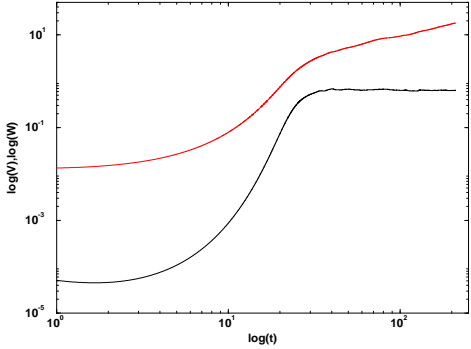

Figure 2.4: log-log plots of the front velocity (lower curve) and width (upper curve) as a function of time in the inverse cascade process seen in Fig.2.3 in a system o f size 2000 and $\nu = 1$. Both quantities exhibit an initial exponential growth that turns to a power law growth (after $t \approx 30$). The velocity is constant after this time, and the width increases like $t^\zeta$. Note that at the earliest time there is a slight decrease in the velocity; this is due to the decay of linearly stable modes that exist in random initial conditions.

by $s_l$. The total time elapsed, $t(l_t)$ is the sum

$$t(l_t) \sim \sum_{i=1}^{s_l} 2^i \ . \tag{2.40}$$

The geometric sum is dominated by the largest term and we therefore estimate $t(l_t) \sim l_t$. We conclude that the scale and the width are linear in the time elapsed from the initial conditions ($l_t \sim t^\zeta$, $\zeta = 1$). In noiseless simulations we find (see Fig.2.4) a value of $\zeta$ which is $\zeta \approx 0.95 \pm 0.1$.

### 2.4.3 Inverse cascade in the presence of noise

An interesting consequence of the discussion in the last section is that the inverse cascade process is an effective "clock" that measures the typical time scales in this system. For future purposes we need to know the typical time scales when the dynamics is perturbed by random noise. To this aim we ran simulations following the inverse cascade in the *presence* of external noise. The main result that will be used in later arguments is that now the appearance of a typical scale $l_t$ occurs not after time $t$, but rather according to

$$l_t \sim t^\zeta \ , \quad \zeta \approx 1.2 \pm 0.1 \ . \tag{2.41}$$

The numerical confirmation of this law is exhibited in Fig.2.5 .

We also find that the front velocity in this case increases with time according to

$$v \sim t^\gamma \ , \quad \gamma \approx 0.48 \pm 0.05 \ . \tag{2.42}$$

This result will be related to the acceleration of the flame front in noisy simulations, as will be seen in the next Sections.

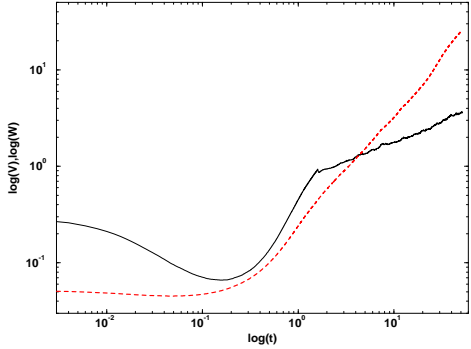

Figure 2.5: The same as Fig.2.4 but with additive random noise for a system of size 1000, $\nu = 0.1$ and $f = 10^{-13}$. The velocity does not saturate now, and the exponent $\zeta$ characterizing the increase of the width with time changes to $\zeta = 1.2 \pm 0.1$. The velocity increases in time like $t^\gamma$ with $\gamma \approx 0.48 \pm 0.04$.

## 2.5 Acceleration of the Flame Front, Pole Dynamics and Noise

A major motivation of this Section is the observation that in radial geometry the same equation of motion shows an acceleration of the flame front. The aim of this section is to argue that this phenomenon is caused by the noisy generation of new poles. Moreover, it is our contention that a great deal can be learned about the acceleration in radial geometry by considering the effect of noise in channel growth. In Ref. [12] it was shown that any initial condition which is represented in poles goes to a unique stationary state which is the giant cusp which propagates with a constant velocity $v = 1/2$ up to small $1/L$ corrections. In light of our discussion of the last section we expect that any smooth enough initial condition will go to the same stationary state. Thus if there is no noise in the dynamics of a finite channel, no acceleration of the flame front is possible. What happens if we add noise to the system?

For concreteness we introduce an additive white-noise term $\eta(\theta, t)$ to the equation of motion (2.5) where

$$\eta(\theta, t) = \sum_k \eta_k(t) \exp(ik\theta) , \qquad (2.43)$$

and the Fourier amplitudes $\eta_k$ are correlated according to

$$<\eta_k(t)\eta_{k'}^*(t')> = \frac{f}{L}\delta_{k,k'}\delta(t-t') . \qquad (2.44)$$

We will first examine the result of numerical simulations of noise-driven dynamics, and later return to the theoretical analysis.

### 2.5.1 Noisy Simulations

Previous numerical investigations [11, 13] did not introduce noise in a controlled fashion. We will argue later that some of the phenomena encountered in these simulations can be ascribed to the (uncontrolled) numerical noise. We performed numerical simulations of Eq.(2.5 using a pseudo-spectral method. The time-stepping scheme was chosen as Adams-Bashforth with 2nd order precision in time. The additive white noise was generated in Fourier-space by choosing $\eta_k$ for every $k$ from a flat distribution in the interval $[-\sqrt{2\frac{f}{L}}, \sqrt{2\frac{f}{L}}]$. We examined the average steady state velocity of the front as a function of $L$ for fixed $f$ and as a function of $f$ for fixed $L$. We found the interesting phenomena that are summarized here:

1. In Fig.2.7 we can see two different regimes of the behavior of the average velocity $v$ as a function of the noise $f^{0.5}$ for the fixed system size L. For the noise $f$ smaller then same fixed value $f_{cr}$

$$v \sim f^\xi . \qquad (2.45)$$

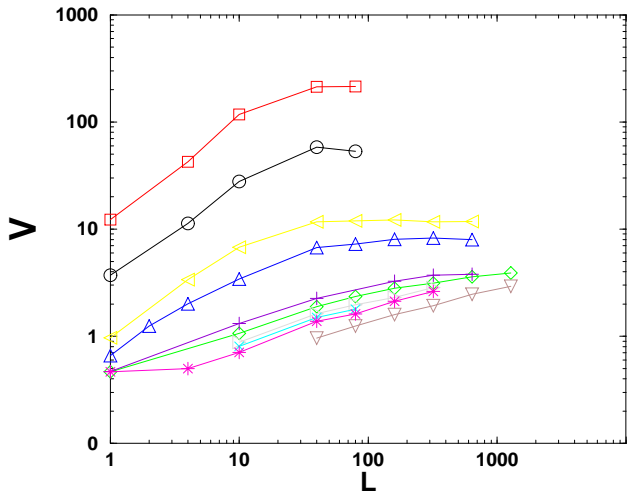

Figure 2.6: The dependence of the average velocity $v$ on the system size $L$ for $f^{0.5} = 0, 2.7\times 10^{-6}, 2.7\times 10^{-5}, 2.7\times 10^{-4}, 2.7\times 10^{-3}, 2.7\times 10^{-2}, 2.7\times 10^{-1}, 0.5, 1.3, 2.7$.

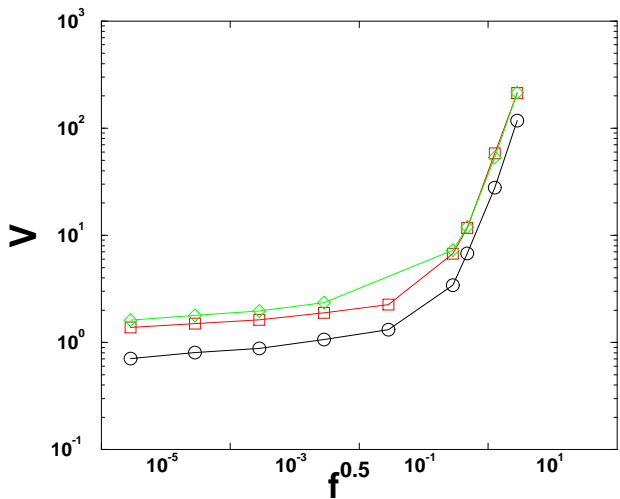

Figure 2.7: The dependence of the average velocity $v$ on the noise $f^{0.5}$ for L=10, 40, 80.

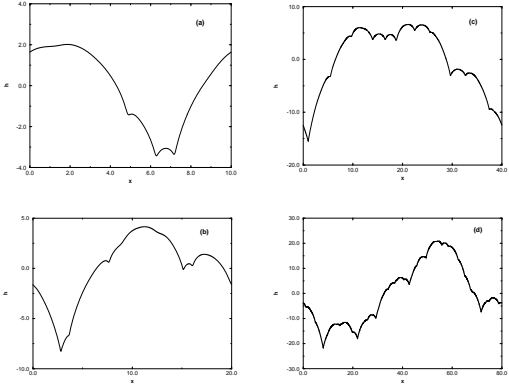

Figure 2.8: Typical flame fronts for $f < f_{cr}$ where the system is sufficiently small not to be terribly affected by the noise. The effect of noise in this regime is to add additional small cusps to the giant cusp. In figures a-d we present fronts for growing system sizes $\tilde{L} = 10, 20, 40$ and $80$ respectively, $\nu = 0.1$. One can observe that when the system size grows there are more cusps with a more complex structure.

   For these values of $f$ this dependence is very weak, and $\xi \approx 0.02$. For the large values of $f$ the dependence is much stronger

2. In Fig.2.6 we can see the growth of the average velocity $v$ as a function of the system size L. After some values of L we can see saturation of the velocity. For regime $f < f_{cr}$ the growth of the velocity can be written as

$$v \sim L^\mu, \quad \mu \approx 0.35 \pm 0.03 . \qquad (2.46)$$

3. In Fig.2.8 and Fig.2.9 we can see flame fronts for $f < f_{cr}$ and $f > f_{cr}$.

### 2.5.2 Calculation of the Number of Poles in the System

The interesting problem that we would like to solve here to better understand the dynamics of poles, is to determine those that exist in our system outside the giant cusp. This can be done by calculating the number of cusps (points of minimum or inflectional points) and their position on the interval $\theta : [0, 2\pi]$ in every moment of time and drawing the positions of the cusps like functions of time, see Fig. 2.10. In this picture we can see the x-positions of all cusps in the system as a function of time.

We have assumed that our system is in a "quasi-stable" state most of the time, i.e. every new cusp that appears in the system includes only one pole. Using pictures obtained in this way we can find:

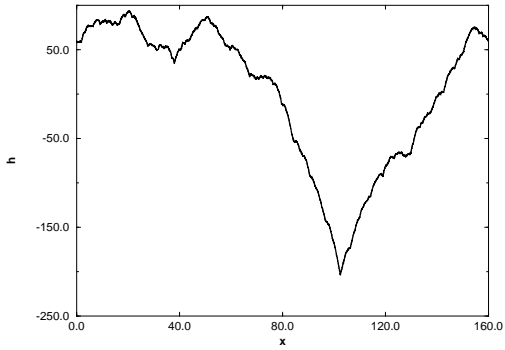

Figure 2.9: A typical flame front for $f > f_{cr}$. The system size is 160. This is sufficient to cause a qualitative change in the appearance of the flame front: the noise introduces significant levels of small scales structure in addition to the cusps.

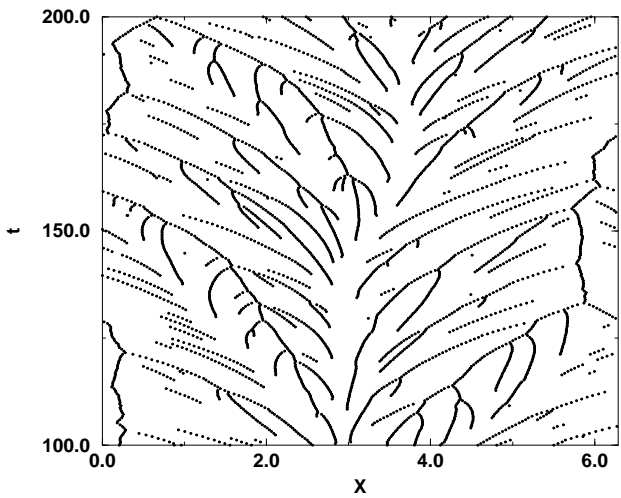

Figure 2.10: The dependence of the cusps positions on time. $L = 80$ $\nu = 0.1$ $f = 9 \times 10^{-6}$

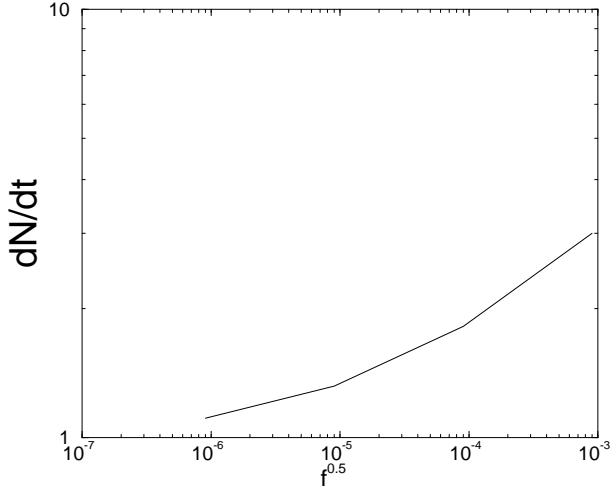

Figure 2.11: The dependence of the pole number in the unit time $dN/dt$ on the noise $f^{0.5}$. $\nu = 0.1$ $L = 80$

1. The mean number of poles in the system. By calculating the number of cusps in some moment of time and by investigating the history of every cusp (except the giant cusp), i.e. how many initial cusps take part in formatting this cusp, and after averaging the number of poles found with respect to different moments of time, we can find the mean number of poles that exist in our system outside the giant cusp. Let us denote this number by $\delta N$. There are four regimes that can be defined with respect to the dependence of this number on the noise $f$:

   (i) Regime I: Such little noise that no new cusps exist in our system outside the giant cusp;

   (ii) Regime II: Strong dependence of the pole number $\delta N$ on the noise $f$;

   (iii) Regime III: Saturation of the pole number $\delta N$ on the noise $f$, so that this number depends very little on the noise (Fig. 2.12);

$$\delta N \sim f^{0.03} \qquad (2.47)$$

The saturated value of $\delta N$ is defined by next formula (Fig. 2.14, Fig. 2.16)

$$\delta N \approx N(L)/2 \approx \frac{1}{4}\frac{L}{\nu} \qquad (2.48)$$

where $N(L) \approx \frac{1}{2}\frac{L}{\nu}$ is the number of poles in the giant cusp.

(iv) Regime IV: We again see a strong dependence of the pole number $\delta N$ on the noise $f$ (Fig. 2.12);

$$\delta N \sim f^{0.1} \qquad (2.49)$$

Because of the numerical noise we can see in most of the simulations only regime III and IV. In the future if no new evidence is seen we will discuss regime III.

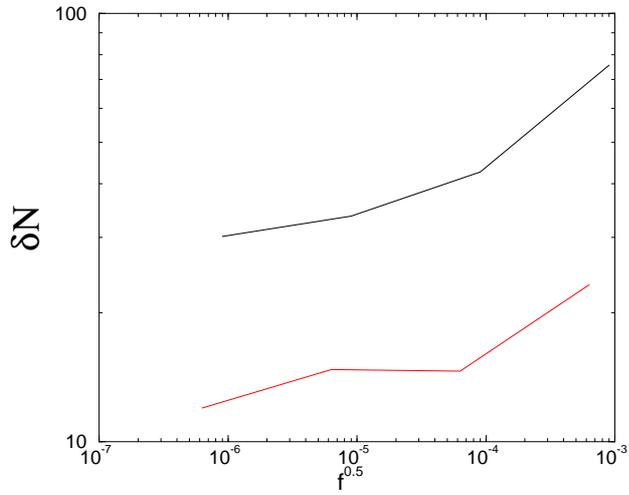

Figure 2.12: The dependence of the excess pole number $\delta N$ on the noise $f^{0.5}$. $\nu = 0.1$ $L = 40, 80$.

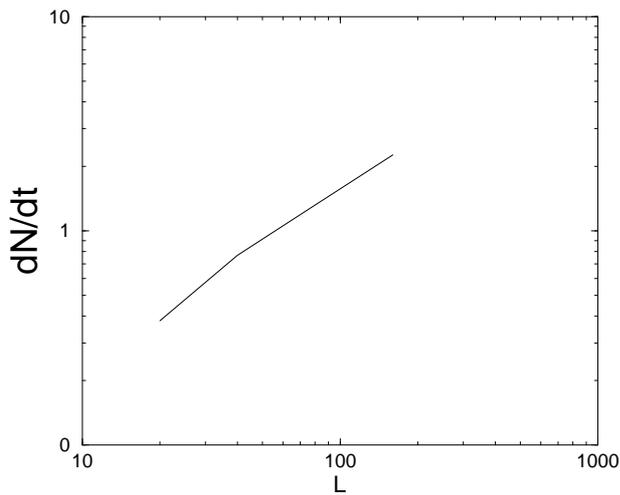

Figure 2.13: The dependence of the pole number in the unit time $dN/dt$ on the system size L. $\nu = 0.1$ $f^{0.5} = 9 \times 10^{-6}$.

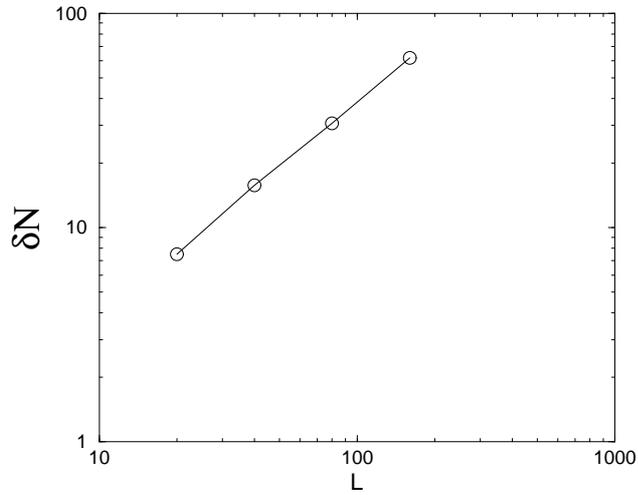

Figure 2.14: The dependence of the excess pole number $\delta N$ on the system size L.
$\nu = 0.1$  $f^{0.5} = 9 \times 10^{-6}$

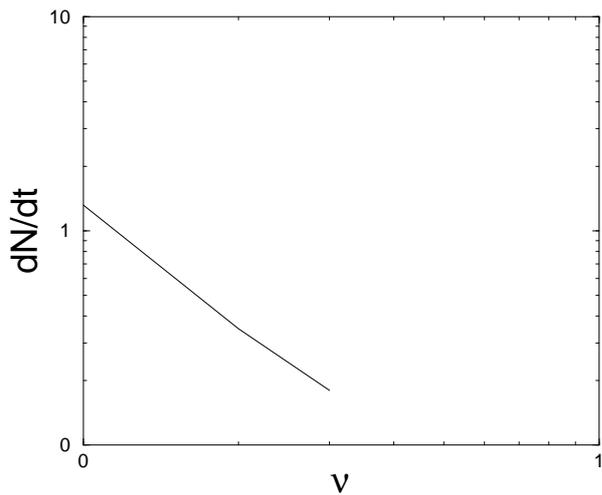

Figure 2.15: The dependence of the pole number in the unit time $dN/dt$ on the parameter $\nu$. $L = 80$  $\nu = 0.1$.

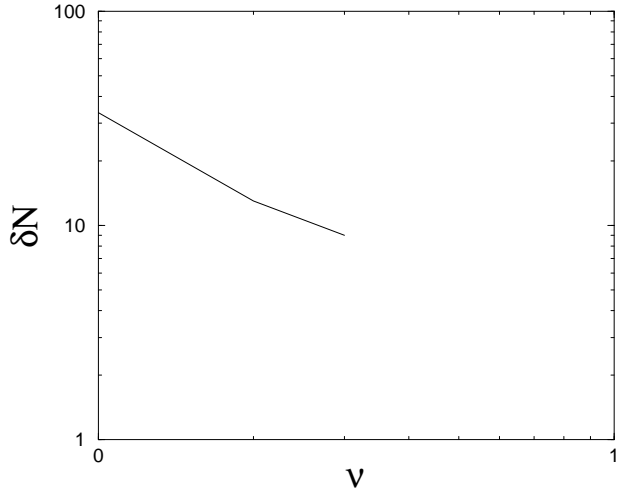

Figure 2.16: The dependence of the excess pole number $\delta N$ on the the parameter $\nu$. $L = 80$ $\nu = 0.1$.

2. By calculating the new cusp number that appears in the system in the unit time we can find the number of poles that appear in the system in the unit time $\frac{dN}{dt}$. In regime III (Fig. 2.11)

$$\frac{dN}{dt} \sim f^{0.03} \qquad (2.50)$$

The dependence on $L$ and $\nu$ is defined by (Fig. 2.13 and Fig. 2.15)

$$\frac{dN}{dt} \sim L^{0.8} \qquad (2.51)$$

$$\frac{dN}{dt} \sim \frac{1}{\nu^2} \qquad (2.52)$$

In regime IV, the dependence on the noise is defined by the following: (Fig. 2.11)

$$\frac{dN}{dt} \sim f^{0.1} \qquad (2.53)$$

### 2.5.3 Theoretical Discussion of the Effect of Noise

**The Threshold of Instability to Added Noise. Transition from regime I to regime II**

First we present the theoretical arguments that explain the sensitivity of the giant cusp solution to the effect of added noise. This sensitivity increases dramatically with increasing the system size $L$. To see this we use again the relationship between the linear stability analysis and the pole dynamics.

Our additive noise introduces perturbations with all $k$-vectors. We showed previously that the most unstable mode is the $k = 1$ component $A_1 \sin(\theta)$. Thus the most effective noisy perturbation is $\eta_1 \sin(\theta)$ which can potentially lead to a growth of the most unstable mode. Whether or not this mode will grow depends on the amplitude of the noise. To see this clearly we return to the pole description. For small values of the amplitude $A_1$ we represent $A_1 \sin(\theta)$ as a single pole solution of the functional form $\nu e^{-y} \sin\theta$. The $y$ position is determined from $y = -\log|A_1|/\nu$, and the $\theta$-position is $\theta = \pi$ for positive $A_1$ and $\theta = 0$ for negative $A_1$. From the analysis of Section III we know that for very small $A_1$ the fate of the pole is to be pushed to infinity, independently of its $\theta$ position; the dynamics is symmetric in $A_1 \to -A_1$ when $y$ is large enough. On the other hand when the value of $A_1$ increases the symmetry is broken and the $\theta$ position and the sign of $A_1$ become very important. If $A_1 > 0$ there is a threshold value of $y$ below which the pole is attracted down. On the other hand if $A_1 < 0$, and $\theta = 0$ the repulsion from the poles of the giant cusp grows with decreasing $y$. We thus understand that qualitatively speaking the dynamics of $A_1$ is characterized by an asymmetric "potential" according to

$$\dot{A}_1 = -\frac{\partial V(A_1)}{\partial A_1}, \tag{2.54}$$

$$V(A_1) = \lambda A_1^2 - a A_1^3 + \ldots. \tag{2.55}$$

>From the linear stability analysis we know that $\lambda \approx \nu/L^2$, cf. Eq.(1.11). We know further that the threshold for nonlinear instability is at $A_1 \approx \nu^3/L^2$, cf. Eq(2.26). This determines that value of the coefficient $a \approx 2/3\nu^2$. The magnitude of the "potential" at the maximum is

$$V(A_{max}) \approx \nu^7/L^6. \tag{2.56}$$

The effect of the noise on the development of the mode $A_1 \sin\theta$ can be understood from the following stochastic equation

$$\dot{A}_1 = -\frac{\partial V(A_1)}{\partial A_1} + \eta_1(t). \tag{2.57}$$

It is well known [41] that for such dynamics the rate of escape $R$ over the "potential" barrier for small noise is proportional to

$$R \sim \frac{\nu}{L^2} \exp^{-\nu^7/fL^5}. \tag{2.58}$$

The conclusion is that any arbitrarily tiny noise becomes effective when the system size increase and when $\nu$ decreases. If we drive the system with noise of amplitude $\frac{f}{L}$ the system can always be sensitive to this noise when its size exceeds a critical value $L_c$ that is determined by $f/L_c \sim \nu^7/L_c^6$. This formula defines transition from regime I (no new cusps) to regime II. For $L > L_c$ the noise will introduce new poles into the system. Even numerical noise in simulations involving large size systems may have a macroscopic influence.

The appearance of new poles must increase the velocity of the front. The velocity is proportional to the mean of $(u/L)^2$. New poles distort the giant cusp by additional smaller cusps on the wings of the giant cusp, increasing $u^2$. Upon increasing the noise amplitude more and more smaller cusps appear in the front, and inevitably the velocity increases. This phenomenon is discussed quantitatively in Section 2.5.

**Numerical verifying of the asymmetric "potential" form and dependence of the noise on $L_c$**

From the equations of the motion for poles we can find the distribution of poles in the giant cusp [12]. If we know the distribution of poles in the giant cusp we can then find the form of the "potential" and verify numerically expressions for values $\lambda$, $A_{max}$ and $\frac{\partial V(A_1)}{\partial A_1}$ discussed previously. The connection between amplitude $A_1$ and the position of the pole $y$ is defined by $A_1 = 4\nu e^{-y}$ and the connection between the potential function $\frac{\partial V(A_1)}{\partial A_1}$ and the position of the pole $y$ is defined by formula $\frac{\partial V(A_1)}{\partial A_1} = 4\nu \frac{dy}{dt} e^{-y}$, where $\frac{dy}{dt}$ can be determined from the equation of the motion of the poles. We can find $A_{max}$ as the zero-point of $\frac{\partial V(A_1)}{\partial A_1}$ and $\lambda$ can be found as $\frac{1}{2}\frac{\partial^2 V(A_1)}{\partial A_1^2}$ for $A_1 = 0$. Numerical measurements were made for the set of values $L = 2n\nu$, where $n$ is a integer and $n > 2$. For our numerical measurements we use the constant $\nu = 0.005$ and the variable $L$, where $L$ changes in the interval $[1,150]$, or variable $\nu$ that changes in the interval $[0.005,0.05]$ and the constant $L = 1$. The results obtained follow:

1. $\frac{A_{max}L^2}{\nu^3}$ as a function of $L$ is almost a constant. (Fig. 2.17)

2. $\frac{A_{max}L^2}{\nu^3}$ as a function of $\nu$ is almost a constant. (Fig. 2.18)

3. $\frac{A_{max}}{A_{N(L)}}$ as a function of $L$ is almost a constant. ($A_{N(L)}$ is defined by the position of the upper pole.) (Fig. 2.19)

4. $\frac{A_{max}}{A_{N(L)}}$ as a function of $\nu$ is almost a constant. (Fig. 2.20)

5. The value of $\frac{\lambda L^2}{\nu}$ as a function of $L$ is a constant (Fig. 2.21).

6. The value of $\frac{\lambda L^2}{\nu}$ as a function of $\nu$ is a constant ( Fig. 2.22 ).

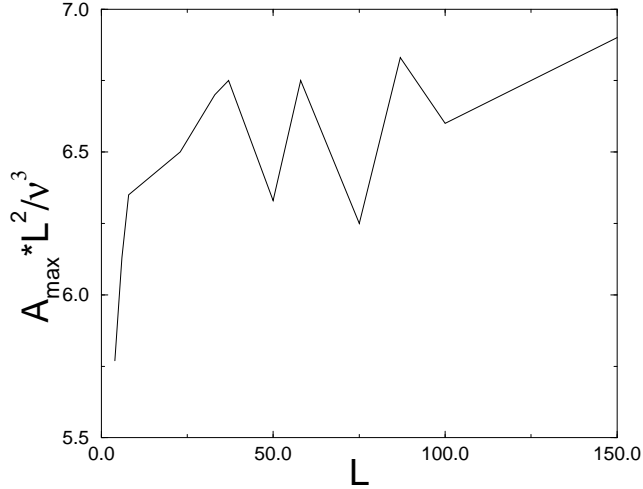

Figure 2.17: The dependence of the normalized amplitude $A_{max}L^2/\nu^3$ on the system size $L$.

We also verify the boundary between regime I (no new cusps) and regime II (new cusps appear). Fig. 2.23 shows the dependence of $\frac{f}{L_c}$ on $L_c$. We can see that $f/L_c \sim 1/L_c^6$. These results are in good agreement with the theory.

### The Noisy Steady State and its Collapse with Large Noise and System Size

In this subsection we discuss the response of the giant cusp solution to noise levels that are able to introduce a large number of excess poles in addition to those existing in the giant cusp. We will denote the excess number of poles by $\delta N$. The first question that we address is how difficult is it to insert yet an additional pole when there is already a given excess $\delta N$. To this aim we estimate the effective potential $V_{\delta N}(A_1)$ which is similar to (2.55) but is taking into account the existence of an excess number of poles. A basic approximation that we employ is that the fundamental form of the giant cusp solution is not seriously modified by the existence of an excess number of poles. Of course this approximation breaks down quantitatively already with one excess pole. Qualitatively however it holds well until the excess number of poles is of the order of the original number $N(L)$ of the giant cusp solution. Another approximation is that the rest of the linear modes play no role in this case. At this point we limit the discussion therefore to the situation $\delta N \ll N(L)$ (regime II).

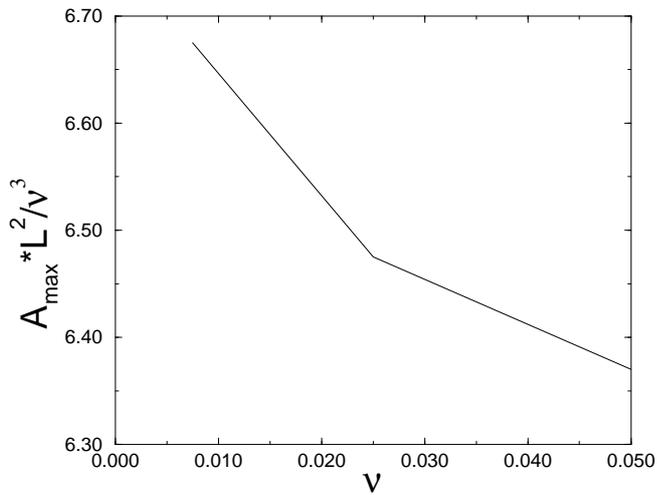

Figure 2.18: The dependence of the normalized amplitude $A_{max}L^2/\nu^3$ on the parameter $\nu$.

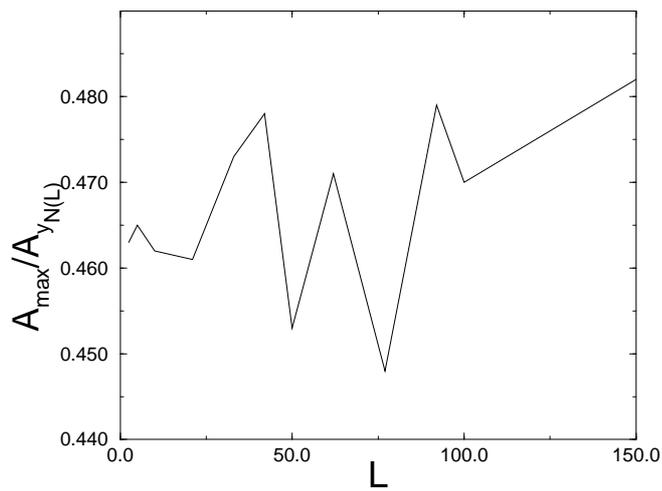

Figure 2.19: The relationship between the amplitude defined by the minimum of the potential $A_{max}$ and the amplitude defined by the position of the upper pole $A_{N(L)}$ as a function of the system size $L$.

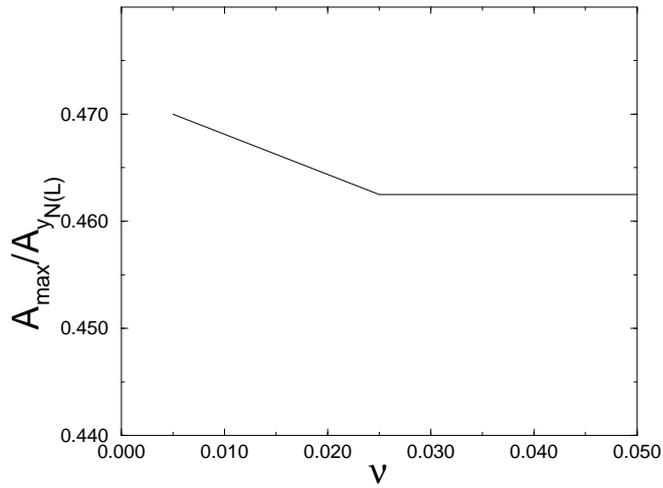

Figure 2.20: The relationship between the amplitude defined by the minimum of the potential $A_{max}$ and the amplitude defined by the position of the upper pole $A_{N(L)}$ as a function of the parameter $\nu$.

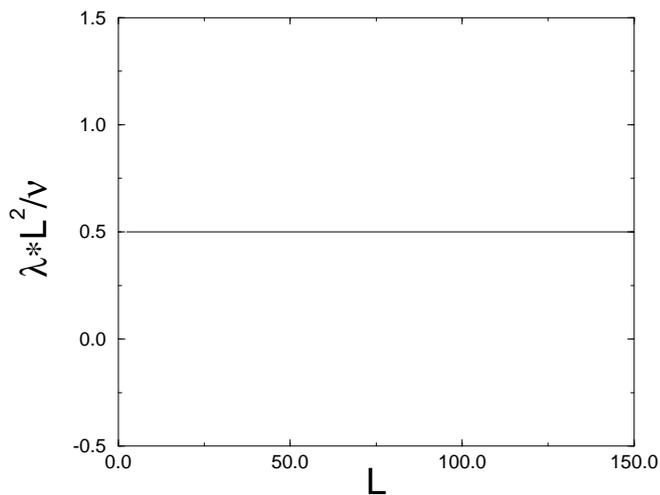

Figure 2.21: The dependence of the normalized parameter $\lambda L^2/\nu$ on the system size $L$.

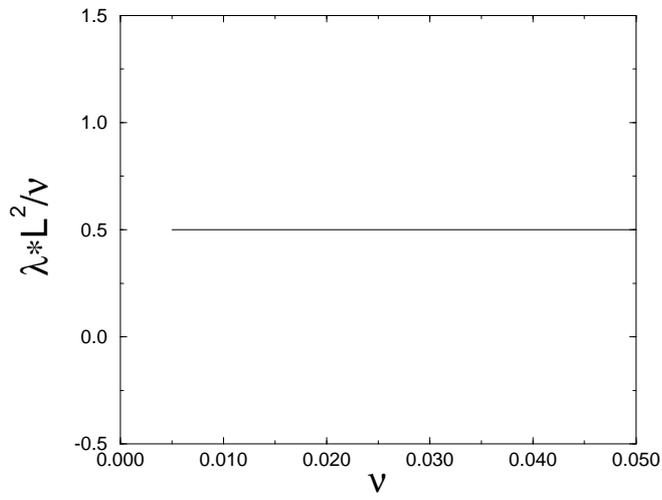

Figure 2.22: The dependence of the normalized parameter $\lambda L^2/\nu$ on the parameter $\nu$.

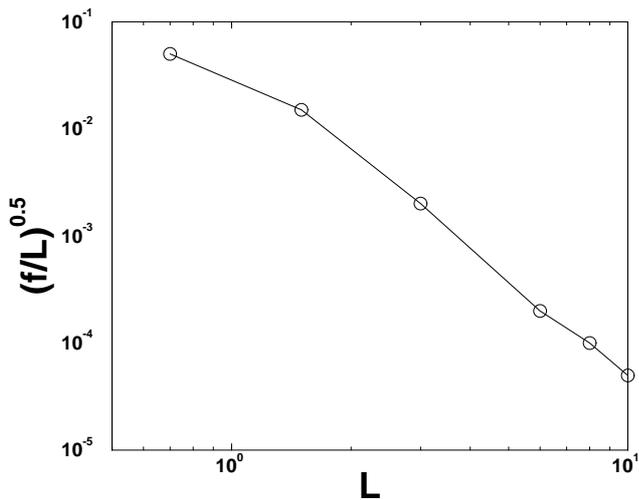

Figure 2.23: The dependence of the critical noise on the system size.

To estimate the parameter $\lambda$ in the effective potential we consider the dynamics of one pole whose $y$ position $y_a$ is far above $y_{max}$. According to Eq.(1.11) the dynamics reads

$$\frac{dy_a}{dt} \approx \frac{2\nu(N(L)+\delta N)}{L^2} - \frac{1}{L} \tag{2.59}$$

Since the $N(L)$ term cancels against the $L^{-1}$ term (cf. Sec. II A), we remain with a repulsive term that in the effective potential translates to

$$\lambda = \frac{\nu \delta N}{L^2} \ . \tag{2.60}$$

Next we estimate the value of the potential at the break-even point between attraction and repulsion. In the last subsection we saw that a foreign pole has to be inserted below $y_{max}$ in order to be attracted towards the real axis. Now we need to push the new pole below the position of the existing pole whose index is $N(L)-\delta N$. This position is estimated as in Sec III C by employing the TFH distribution function (2.23). We find

$$y_{\delta N} \approx 2 \ln \left[ \frac{4L}{\pi^2 \nu \delta N} \right] \ . \tag{2.61}$$

As before, this implies a threshold value of the amplitude of single pole solution $A_{max} \sin \theta$ which is obtained from equating $A_{max} = \nu e^{-y_{\delta N}}$. We thus find in the present case $A_{max} \sim \nu^3 (\delta N)^2 / L^2$. Using again a cubic representation for the effective potential we find $a = 2/(3\nu^2 \delta N)$ and

$$V(A_{max}) = \frac{1}{3} \frac{\nu^7 (\delta N)^5}{L^6} \ . \tag{2.62}$$

Repeating the calculation of the escape rate over the potential barrier we find in the present case

$$R \sim \frac{\nu \delta N}{L^2} \exp^{-\nu^7(\delta N)^5/fL^5} \ . \tag{2.63}$$

For a given noise amplitude $f$ there is always a value of $L$ and $\nu$ for which the escape rate is of $O(1)$ as long as $\delta N$ is not too large. When $\delta N$ increases the escape rate decreases, and eventually no additional poles can creep into the system. The typical number $\delta N$ for fixed values of the parameters is estimated from equating the argument in the exponent to unity

$$\delta N \approx \left( fL^5/\nu^7 \right)^{1/5} \ . \tag{2.64}$$

We can see that $\delta N$ is strongly dependent on noise $f$, in contrast to regime III. Let us find the conditions of transition from regime II to III, where we see the saturation of $\delta N$ with respect to noise $f$.

(i) We use the expression $A_{max} = 4\nu e^{-y_{\delta N}}$ for the amplitude of the pole solution that equals to $\frac{2\nu \sin\theta}{\cosh(y_{\delta N}) - \cos\theta}$; however, this is correct only for the large number $y_{\delta N}$. When $y_{\delta N} < 1$, a better approximation is $A_{max} = \frac{4\nu}{y_{\delta N}^2}$. From the equation (2.61) we find that the boundary value $y_{\delta N} = 1$ corresponds to $\delta N \approx N(L)/2$.

(ii) We use the expression $y_{\delta N} \approx 2\ln\left[\frac{4L}{\pi^2 \nu \delta N}\right]$, but for a large value of $\delta N$ a better approximation that can be found the same way is $y_{\delta N} \approx \frac{\pi^2 \nu}{2L}(N(L) - \delta N)\ln\left[\frac{8eL}{\pi^2 \nu (N(L) - \delta N)}\right]$ [12]. These expressions give us nearly lthe same result for $\delta N \approx N(L)/2$.

From (i) and (ii) we can make the following conclusions:

(a) The transition from regime II to regime III generally occurs for $\delta N \approx N(L)/2$;

(b) Using the new expressions in (i) and (ii) for the amplitude $A_{max}$ and $y_{\delta N}$, we can determine the noise $\frac{f}{L}$ in regime III by

$$\frac{f}{L} \sim V(A_{max}) \sim \lambda A_{max}^2 \sim \frac{\nu \delta N}{L^2}\left(\frac{4\nu}{y_{\delta N}^2}\right)^2 \sim \frac{L^2}{\nu} \frac{\delta N}{(N(L) - \delta N)^4} \quad (2.65)$$

This expression defines a very slight dependence of $\delta N$ on the noise $f$ for $\delta N > N(L)/2$, which explains the noise saturation of $\delta N$ for regime III.

(c) The form of the giant cusp solution is governed by the poles that are close to zero with respect to $y$. For the regime III, $N(L)/2$ poles that have positions $y < y_{\delta N = N(L)/2} = 1$ remain at this position. This result explains why the giant cusp solution cannot be seriously modified for regime III.

From eq. (2.64) by using the condition

$$\delta N \approx N(L)/2 \quad (2.66)$$

the boundary noise $f_b$ between regimes II and III can be found as

$$f_b \sim \nu^2 . \quad (2.67)$$

The basic equation describing pole dynamics follows

$$\frac{dN}{dt} = \frac{\delta N}{T} , \quad (2.68)$$

where $\frac{dN}{dt}$ is the number of poles that appear in the unit time in our system, $\delta N$ is the excess number of poles, and T is the mean lifetime of a pole (between appearing and merging with the giant cusp). Using the result of numerical simulations for $\frac{dN}{dt}$ and (2.66) we can find for $T$

$$T = \frac{\delta N}{\frac{dN}{dt}} \sim \nu L^{0.2} . \quad (2.69)$$

Thus the lifetime is proportional to $\nu$ and depends on the system size $L$ very slightly.

Moreover, the lifetime of a pole is defined by the lifetime of the poles that are in a cusp. From the maximum point of the linear part of Eq.(2.1 ), we can find the mean character size (Fig.9( [28]))

$$\lambda_m \sim \nu \quad (2.70)$$

that defines the size of our cusps. The mean number of poles in a cusp

$$n_{big} \approx \frac{\lambda_m}{2\nu} \sim const \quad (2.71)$$

does not depend on $L$ and $\nu$. The mean number of cusps is

$$N_{big} \sim \frac{\delta N}{n_{big}} \sim \frac{L}{\nu} . \tag{2.72}$$

Let us assume that some cusp exists in the main minimum of the system. The lifetime of a pole in such a cusp is defined by three parts.

(I) Time of the cusp formation. This time is proportional to the cusp size (with ln-corrections) and the pole number in the cusp (from pole motion equations)

$$T_1 \sim \lambda_m n_{big} \sim \nu \tag{2.73}$$

(II) Time that the cusp is in the minimum neighborhood. This time is defined by

$$T_2 \sim \frac{a}{v} \tag{2.74}$$

where $a$ is a neighborhood of minimum, such that the force from the giant cusp is smaller than the force from the fluctuations of the excess pole number $\delta N$, and $v$ is the velocity of a pole in this neighborhood. Fluctuations of excess pole number $\delta N$ are expressed as

$$N_{fl} = \sqrt{\delta N} . \tag{2.75}$$

From this result and the pole motion equations we find that

$$v \sim \frac{\nu}{L} N_{fl} \sim \frac{\nu}{L}\sqrt{\frac{L}{\nu}} \sim \sqrt{\frac{\nu}{L}} . \tag{2.76}$$

The velocity from the giant cusp is defined by

$$v \sim \frac{\nu}{L} N(L) \frac{a}{L} \sim \frac{a}{L} . \tag{2.77}$$

So from equating these two equations we obtain

$$a \sim \sqrt{\nu L} . \tag{2.78}$$

Thus for $T_2$ we obtain

$$T_2 \sim \frac{a}{v} \sim L . \tag{2.79}$$

(III) Time of attraction to the giant cusp. From the equations of motion for the poles we get

$$T_3 \sim L \ln(\frac{L}{a}) \sim L \ln \sqrt{L} \sim L . \tag{2.80}$$

The investigated domain of the system size was found to be

$$T_1 \gg T_2, T_3 \tag{2.81}$$

Therefore full lifetime is

$$T = T_1 + T_2 + T_3 \sim \nu + sL , \tag{2.82}$$

where $s$ is a constant and

$$0 < s \ll 1 . \tag{2.83}$$

This result qualitatively and partly quantatively explains dependence (2.69). From (2.69), (2.68), (2.66) we can see that in regime III $\frac{dN}{dt}$ is saturated with the system size $L$.

### 2.5.4 The acceleration of the flame front because of noise

In this section we estimate the scaling exponents that characterize the velocity of the flame front as a function of the system size. To estimate the velocity of the flame front we need to create an equation for the mean of $<dh/dt>$ given an arbitrary number $N$ of poles in the system. This equation follows directly from (2.4)

$$\left\langle \frac{dh}{dt} \right\rangle = \frac{1}{L^2} \frac{1}{2\pi} \int_0^{2\pi} u^2 d\theta \ . \tag{2.84}$$

After substitution of (2.8) in (2.84) we get, using (2.11) and (2.12)

$$\left\langle \frac{dh}{dt} \right\rangle = 2\nu \sum_{k=1}^{N} \frac{dy_k}{dt} + 2\left(\frac{\nu N}{L} - \frac{\nu^2 N^2}{L^2}\right) \ . \tag{2.85}$$

Estimating the second and third terms in this equation are straightforward. Writing $N = N(L) + \delta N(L)$ and remembering that $N(L) \sim L/\nu$ and $\delta N(L) \sim N(L)/2$, we find that these terms contribute $O(1)$. The first term contributes only when the current of the poles is asymmetric. Noise introduces poles at a finite value of $y_{min}$, whereas the rejected poles stream towards infinity and disappear at the boundary of nonlinearity defined by the position of the highest pole as

$$y_{max} \approx 2 \ln \left[\frac{4L}{\pi^2 \nu}\right] \ . \tag{2.86}$$

Thus we have an asymmetry that contributes to the velocity of the front. To estimate the first term let us define

$$d(\sum \frac{dy_k}{dt}) = \sum_{l}^{l+dl} \frac{dy_k}{dt} \ , \tag{2.87}$$

where $\sum_{l}^{l+dl} \frac{dy_k}{dt}$ is the sum over the poles that are on the interval $y : [l, l+dl]$. We can write

$$d(\sum \frac{dy_k}{dt}) = d(\sum \frac{dy_k}{dt})_{up} + d(\sum \frac{dy_k}{dt})_{down} \ , \tag{2.88}$$

where $d(\sum \frac{dy_k}{dt})_{up}$ is the flux of poles moving up and $d(\sum \frac{dy_k}{dt})_{down}$ is the flux of poles moving down.

For these fluxes we can write

$$d(\sum \frac{dy_k}{dt})_{up}, -d(\sum \frac{dy_k}{dt})_{down} \leq \frac{dN}{dt} dl \ . \tag{2.89}$$

So for the first term

$$0 \leq \sum_{k=1}^{N} \frac{dy_k}{dt} = \int_{y_{min}}^{y_{max}} \frac{d(\sum \frac{dy_k}{dt})}{dl} dl \qquad (2.90)$$
$$= \int_{y_{min}}^{y_{max}} \frac{d(\sum \frac{dy_k}{dt})_{up} + d(\sum \frac{dy_k}{dt})_{down}}{dl} dl$$
$$\leq \frac{dN}{dt}(y_{max} - y_{min})$$
$$\leq \frac{dN}{dt} y_{max}$$

Because of slight (ln) dependence of $y_{max}$ on $L$ and $\nu$, $\frac{dN}{dt}$ term determines order of nonlinearity for the first term in eq (2.85). This term equals zero for the symmetric current of poles and achieves the maximum for the maximal asymmetric current of poles. A comparison of $v \sim L^{0.42} f^{0.02}$ and $\frac{dN}{dt} \sim L^{0.8} f^{0.03}$ confirms this calculation.

## 2.6 Summary and Conclusions

The main two messages of this chapter are: (i) There is an important interaction between the instability of developing fronts and random noise; (ii) This interaction and its implications can be understood qualitatively and sometimes quantitatively using the description in terms of complex poles.

The pole description is natural in this context firstly because it provides an exact (and effective) representation of the steady state without noise. Once one succeeds to describe also the *perturbations* about this steady state in terms of poles, one achieves a particularly transparent language for the study of the interplay between noise and instability. This language also allows us to describe in qualitative and semi-quantitative terms the inverse cascade process of increasing typical lengths when the system relaxes to the steady state from small, random initial conditions.

The main conceptual steps in this chapter are as follows: firstly one realizes that the steady state solution, which is characterized by $N(L)$ poles aligned along the imaginary axis is marginally stable against noise in a periodic array of $L$ values. For all values of $L$ the steady state is nonlinearly unstable against noise. The main and foremost effect of noise of a given amplitude $f$ is to introduce an excess number of poles $\delta N(L, f)$ into the system. The existence of this excess number of poles is responsible for the additional wrinkling of the flame front on top of the giant cusp, and for the observed acceleration of the flame front. By considering the noisy appearance of new poles we rationalize the observed scaling laws as a function of the noise amplitude and the system size.

Theoretically we therefore concentrate on estimating $\delta N(L, f)$. We note that some of our consideration are only qualitative. For example, we estimated $\delta N(L, f)$ by assuming that the giant cusp solution is not seriously perturbed. On the other hand we find a flux of poles going to infinity due to the introduction of poles at finite values of $y$ by the noise. The existence of poles spread between $y_{max}$ and infinity *is* a significant perturbation of the giant cusp solution. Thus also the comparison between the various scaling exponents measured and predicted must be done with caution; we cannot guarantee that those cases in which our prediction hit close to the measurement mean that the theory is quantitative. However we believe that our consideration extract the essential ingredients of a correct theory.

The "phase diagram" as a function of $L$ and $f$ in this system consists of four regimes (in contradiction with our previous results [17]). In the first one, discussed in Section 2.5.3 , the noise is too small to have any effect on the giant cusp solution. The second regime (very small excess number of poles ) can not be observed because of numerical noise and discussed only theoretically. In the third regime the noise introduces excess poles that serve to decorate the giant cusp with side cusps. In this regime we find scaling laws for the velocity as a function of $L$ and $f$ and we are reasonably successful in understanding the scaling exponents. In the fourth regime the noise is large enough to create small scale structures that are not neatly understood in terms of individual poles. It appears from our numerics that in this regime the roughening of the flame front gains a contribution from the the small scale structure in a way that is reminiscent of *stable*, noise driven growth models like the Kardar-Parisi-Zhang model.

One of our main motivations in this research was to understand the phenomena observed in radial geometry with expanding flame fronts. . We note that many of the insights offered above translate immediately to that problem. Indeed, in radial geometry the flame front accelerates and cusps multiply and form a hierarchic structure as time progresses. Since the radius (and the typical scale) increase in this system all the time, new poles will be added to the system even by a vanishingly small noise. The marginal stability found above holds also in this case, and the system will allow the introduction of excess poles as a result of noise. The results discussed in Ref. [19] can be combined with the present insights to provide a theory of radial growth ( chapter 4).

Finally, the success of this approach in the case of flame propagation raises hope that Laplacian growth patterns may be dealt with using similar ideas. A problem of immediate interest is Laplacian growth in channels, in which a finger steady-state solution is known to exist. It is documented that the stability of such a finger solution to noise decreases rapidly with increasing the channel width. In addition, it is understood that noise brings about additional geometric features on top of the finger. There are enough similarities here to indicate that a careful analysis of the analytic theory may shed as much light on that problem as on the present one.

# Chapter 3

# Using of Pole Dynamics for Stability Analysis of Flame Fronts: Dynamical Systems Approach in the Complex Plane

## 3.1 Introduction

In this chapter we discuss the stability of steady flame fronts in channel geometry. We write shortly about this topic in chapter 2 (Sec. 2.3) and we want to consider it in detail in this chapter. Traditionally [1–3] one studies stability by considering the linear operator which is obtained by linearizing the equations of motion around the steady solution. The eigenfunctions obtained are *delocalized* and in certain cases are not easy to interpret. In the case of flame fronts the steady state solution is space dependent and therefore the eigenfunctions are very different from simple Fourier modes. We show in this chapter that a good understanding of the nature of the eigenspectrum and eigenmodes can be obtained by doing almost the opposite of traditional stability analysis, i.e., studying the *localized* dynamics of singularities in the complex plane. By reducing the stability analysis to a study of a finite dimensional dynamical system one can gain considerable intuitive understanding of the nature of the stability problem.

The analysis is based on the understanding that for a given channel width $L$ the steady state solution for the flame front is given in terms of $N(L)$ poles that are organized on a line parallel to the imaginary axis [12]. Stability of this solution can then be considered in two steps. In the first step we examine the response of this set of $N(L)$ poles to perturbations in their positions. This procedure yields an important part of the stability spectrum. In the second step we examine general perturbations, which can also be described by the addition of extra poles to the system of $N(L)$ poles. The response to these perturbations gives us the rest of the stability spectrum; the combinations of these two steps rationalizes all the qualitative features found by traditional stability analysis.

In Sec.2 we present the results of traditional linear stability analysis, and show the eigenvalues and eigenfunctions that we want to interpret by using the pole decomposition. Sec. 3 presents the analysis in terms of complex singularities, in two steps as discussed above. A summary and discussion is presented in Sec.4.

## 3.2 Linear Stability Analysis in Channel Geometry

The standard technique to study the linear stability of the steady solution is to perturb it by a small perturbation $\phi(\theta,t)$: $u(\theta,t) = u_s(\theta) + \phi(\theta,t)$. Linearizing the dynamics for small $\phi$ results in the following equation of motion

$$\frac{\partial \phi(\theta,t)}{\partial t} = \frac{1}{L^2}\Big[\partial_\theta[u_s(\theta)\phi(\theta,t)] + \nu \partial_\theta^2 \phi(\theta,t)\Big] + \frac{1}{L}I(\phi(\theta,t)) \ . \tag{3.1}$$

were the linear operator contains $u_s(\theta)$ as a coefficient. Accordingly simple Fourier modes do not diagonalize it. Nevertheless, we proceed to decompose $\phi(x)$ in Fourier modes according to ,

$$\phi(\theta,t) = \sum_{k=-\infty}^{\infty} \hat{\phi}_k(t) e^{ik\theta} \tag{3.2}$$

$$u_s(\theta) = -2\nu i \sum_{k=-\infty}^{\infty} \sum_{j=1}^{N} sign(k) e^{-|k|y_j} e^{ik\theta} \tag{3.3}$$

The last equation follows from (2.13) by expanding in a series of $\sin k\theta$. In these sums the discrete $k$ values run over all the integers. Substituting in Eq.(3.1) we get:

$$\frac{d\hat{\phi}_k(t))}{dt} = \sum_n a_{kn}\hat{\phi}_n(t) \ , \tag{3.4}$$

where $a_{kn}$ are entires of an infinite matrix:

$$a_{kk} = \frac{|k|}{L} - \frac{\nu}{L^2}k^2 \ , \tag{3.5}$$

$$a_{kn} = \frac{k}{L^2}sign(k-n)(2\nu \sum_{j=1}^{N} e^{-|k-n|y_j}) \quad k \neq n \ . \tag{3.6}$$

To solve for the eigenvalues of this matrix we need to truncate it at some cutoff $k$-vector $k^*$. The scale $k^*$ can be chosen on the basis of Eq.(3.5) from which we see that the largest value of $k$ for which $a_{kk} \geq 0$ is a scale that we denote as $k_{max}$, which is the integer part of $L/\nu$. We must choose $k^* > k_{max}$ and test the choice by the convergence of the eigenvalues. The chosen value of $k^*$ in our numerics was $4k_{max}$. One should notice that this cutoff limits the number of eigenvalues, which should be infinite. However the lower eigenvalues will be well represented. The results for the low order eigenvalues of the matrix $a_{kn}$ that were obtained from the converged numerical calculation are presented in Fig.3.1 The eigenvalues are multiplied by $L^2/\nu$ and are plotted as a function of $L$. We order the eigenvalues in decreasing order and denote them as $\lambda_0 \geq \lambda_1 \geq \lambda_2 \ldots$. In addition to the eigenvalues, the truncated matrix also yields eigenvectors that we denote as $A^{(\ell)}$. Each such vector has $k^*$ entries, and we can compute the eigenfunctions $f^{(\ell)}(\theta)$ of the linear operator (3.1), using (3.2), as

$$f^{(\ell)}(\theta) \equiv \sum_{-k^*}^{k^*} e^{ik\theta} A_k^{(\ell)} \ . \tag{3.7}$$

Eq.(3.1) does not mix even with odd solutions in $\theta$, as can be checked by inspection. Consequently the available solutions have even or odd parity, expandable in either cos or sin functions. The first two nontrivial eigenfunctions $f^{(1)}(\theta)$ and $f^{(2)}(\theta)$ are shown in Figs.3.2,3.3. a It is evident that the function in Fig.3.2 is odd around zero whereas in Fig.3.3 it is even. Similarly we can numerically generate any other eigenfunction of the linear operator, but we understand neither the physical significance of these eigenfunction nor the $L$ dependence of their associated eigenvalues shown in Fig.3.1 In the next section we will demonstrate how the dynamical system approach in terms of singularities in the complex plane provides us with considerable intuition about these issues.

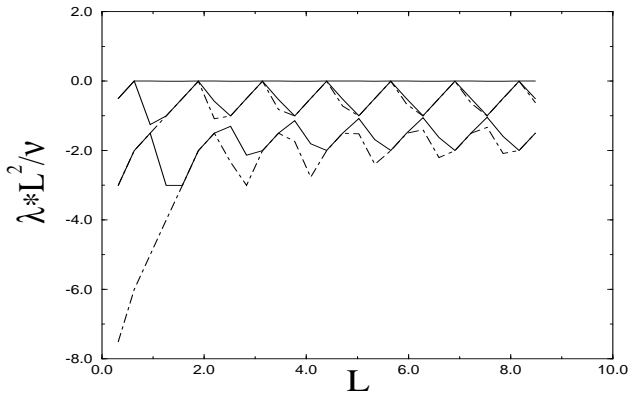

Figure 3.1: A plot of the first five eigenvalues obtained by diagonalizing the matrix obtained by traditional stability analysis, against the system size. The eigenvalues are normalized by $L^2/\nu$. The largest eigenvalue is zero, which is a Goldstone mode. All the other eigenvalues are negative except for the second and third that touch zero periodically. The second and fourth eigenvalues are represented by a solid line and the third and fifth eigenvalues are represented by a dot-dashed line.

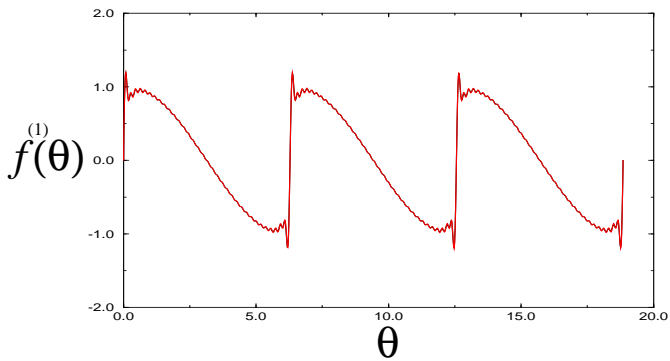

Figure 3.2: The first odd eigenfunction obtained from traditional stability analysis.

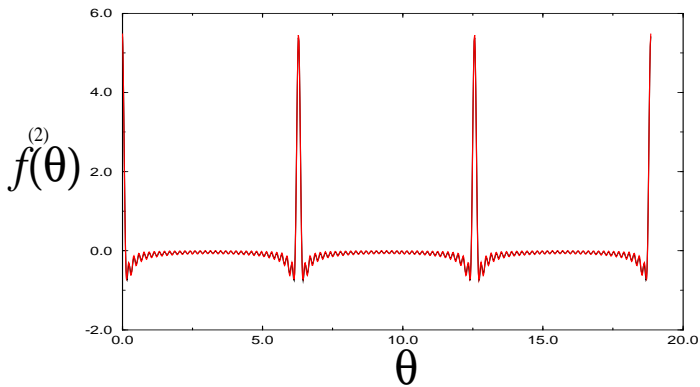

Figure 3.3: The first even eigenfunction obtained from traditional stability analysis.

## 3.3 Linear Stability in terms of complex singularities

Since the partial differential equation is continuous there is an infinite number of modes. To understand this in terms of pole dynamics we consider the problem in two steps: First, we consider the $2N(L)$ modes associated with the dynamics of the $N(L)$ poles of the giant cusp. In the second step we explain that all the additional modes result from the introduction of additional poles, including the reaction of the $N(L)$ poles of the giant cusp to the new poles. After these two steps we will be able to identify all the linear modes that were found by diagonalizing the stability matrix in the previous section.

### 3.3.1 The modes associated with the giant cusp

In the steady solution all the poles occupy stable equibirilium positions. The forces operating on any given pole cancel exactly, and we can write matrix equations for small perturbations in the pole positions $\delta y_i$ and $\delta x_i$.

Following [12] we rewrite the equations of motion (2.12) using the Lyapunov function $U$:

$$L\dot{y}_i = \frac{\partial U}{\partial y_i} \tag{3.8}$$

where $i = 1, ..., N$ and

$$U = \frac{\nu}{L}[\sum_i \ln \sinh y_i \ + \ 2\sum_{i<k}(\ln \sinh \frac{y_k - y_i}{2}$$
$$+ \ \ln \sinh \frac{y_k + y_i}{2})] - \sum_i y_i \tag{3.9}$$

The linearized equations of motion for $\delta y_i$ are:

$$L\dot{\delta y_i} = \sum_k \frac{\partial^2 U}{\partial y_i \partial y_k} \delta y_k \ . \tag{3.10}$$

The matrix $\partial^2 U/\partial y_i \partial y_k$ is real and symmetric of rank $N$. We thus expect to find $N$ real eigenvalues and N orthogonal eigenvectors.

For the deviations $\delta x_i$ in the $x$ positions we find the following linearized equations

of motion

$$L\delta\dot{x}_j = -\frac{\nu}{L}\delta x_j \sum_{k=1,k\neq j}^{N} (\frac{1}{\cosh(y_j - y_k) - 1}$$
$$+ \frac{1}{\cosh(y_j + y_k) - 1})$$
$$+ \frac{\nu}{L}\sum_{k=1,k\neq j}^{N} \delta x_k (\frac{1}{\cosh(y_j - y_k) - 1} + \frac{1}{\cosh(y_j + y_k) - 1}) \quad (3.11)$$

In shorthand:
$$L\frac{d\delta x_i}{dt} = V_{ik}\delta x_k . \quad (3.12)$$

The matrix $V$ is also real and symmetric. Thus $V$ and $\partial^2 U/\partial y_i \partial y_k$ together supply $2N(L)$ real eigenvalues and $2N(L)$ orthogonal eigenvectors. The explicit form of the matrices $V$ and $\partial^2 U/\partial y_i \partial y_k$ is as follows: For $i \neq k$:

$$\frac{\partial^2 U}{\partial y_i \partial y_k} = \frac{\nu}{L}[\frac{1/2}{\sinh^2(\frac{y_k - y_i}{2})} - \frac{1/2}{\sinh^2(\frac{y_k + y_i}{2})}] \quad (3.13)$$

$$V_{ik} = \frac{\nu}{L}(\frac{1}{\cosh(y_i - y_k) - 1} + \frac{1}{\cosh(y_i + y_k) - 1}) \quad (3.14)$$

and for $i = k$ one gets:

$$\frac{\partial^2 U}{\partial y_i^2} = -\frac{\nu}{L}[\sum_{k\neq i}^{N} \left(\frac{1}{2\sinh^2(\frac{y_k - y_i}{2})} + \frac{1}{2\sinh^2(\frac{y_k + y_i}{2})}\right)$$
$$+ \frac{1}{\sinh^2(y_i)}] \quad (3.15)$$

$$V_{ii} = \sum_{k\neq i}^{N}[-\frac{\nu}{L}(\frac{1}{\cosh(y_i - y_k) - 1} + \frac{1}{\cosh(y_i + y_k) - 1})] \quad (3.16)$$

Using the known steady state solutions $y_i$ at any given $L$ we can diagonalize the $N(L) \times N(L)$ matrices numerically. In Fig.3.4 we present the eigenvalues of the lowest order modes obtained from this procedure. The least negative eigenvalues touch zero periodically. This eigenvalue can be fully identified with the motion of the highest pole $y_{N(L)}$ in the giant cusp. At isolated values of $L$ the position of this pole tends to infinity, and then the row and the column in our matrices that contain $y_{N(L)}$ vanish identically, leading to a zero eigenvalue. The rest of the upper eigenvalues match perfectly with half of the observed eigenvalues in Fig.3.1. In other words, the eigenvalues observed here agree perfectly with the ones plotted in this Fig.3.1 until the discontinuous increase from their minimal points. The "second half" of the oscillation in the eigenvalues as a function of $L$ is not contained in this spectrum of the $N(L)$ poles of the giant cusp. To understand the rest of the spectrum we need to consider perturbation of the giant cusp by additional poles. The eigenfunctions can be found using the knowledge of the eigenvectors of these matrices. Let us denote the eigenvectors of $\partial^2 U/\partial y_i \partial y_k$ and $V$ as $a^{(\ell)}$ and $b^{(\ell)}$ respectively. The perturbed solution is explicitly given as (taken for $x_s = 0$):

$$u_s(\theta) + \delta u = 2\nu \sum_{i=1}^{N} \frac{\sin(\theta - \delta x_i)}{\cosh(y_i + \delta y_i) - \cos(\theta - \delta x_i)} \quad (3.17)$$

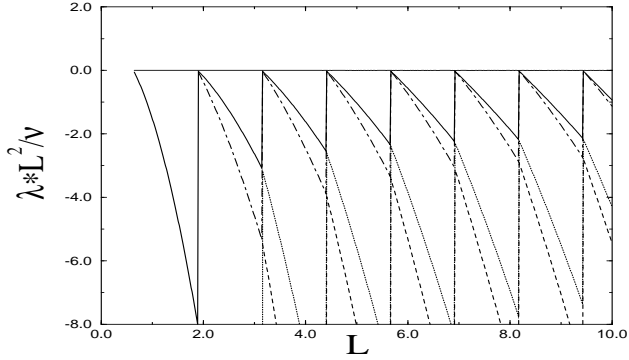

Figure 3.4: The eigenvalues associated with perturbing the positions of the poles that consist the giant cusp. The largest eigenvalue is zero. The second, third, fourth and fifth eigenvalues are represented by a solid line, dot-dashed line, dotted line and dashed line respectively.

where $\delta u$ is

$$\delta u = - 4\nu \sum_{i=1}^{N} \sum_{k=1}^{\infty} \delta y_i k e^{-ky_i} \sin k\theta$$
$$- 4\nu \sum_{i=1}^{N} \sum_{k=1}^{\infty} \delta x_i k e^{-ky_i} \cos k\theta \qquad (3.18)$$

So knowing the eigenvectors $a^{(\ell)}$ and $b^{(\ell)}$ we can estimate the eigenvectors $f^{(\ell)}(\theta)$ of (3.7):

$$f_{\sin}^{(\ell)}(\theta) = -4\nu \sum_{i=1}^{N} \sum_{k=1}^{\infty} a_i^{(j)} k e^{-ky_i} \sin k\theta \;, \quad j = 1, ..., N \qquad (3.19)$$

or

$$f_{\cos}^{(\ell)}(\theta) = -4\nu \sum_{i=1}^{N} \sum_{k=1}^{\infty} b_i^{(j)} k e^{-ky_i} \cos k\theta \;, \quad j = 1, ..., N \qquad (3.20)$$

where we display separately the sin expansion and the cos expansion. For the case $j = 1$, the eigenvalue is zero, and a uniform translation of the poles in any amount $\delta x_i$ results in a Goldstone mode. This is characterized by an eigenvector $b_i^{(1)} = 1$ for all $i$. The eigenvectors $f^{(\ell)}$ (Fig.3.5,3.6)computed this way are identical to numerical precision with those shown in Figs.3.2,3.3, and observe the agreement.

### 3.3.2 Modes related to additional poles

In this subsection we identify the rest of the modes that were not found in the previous subsection. To this aim we study the response of the TFH solution to the introduction of additional poles. We choose to add $M$ new poles all positioned at the same imaginary coordinate $y_p \ll y_{max}$, distributed at equidistant real positions $\{x_j = x_0 + (2\pi/M)j\}_{j=1}^{M}$. For $x_0 = 0$ we use (2.8) and the Fourier expansion to obtain a perturbation of the form

$$\delta u(\theta, t) \simeq 4\nu M e^{-M y_p(t)} \sin M\theta \qquad (3.21)$$

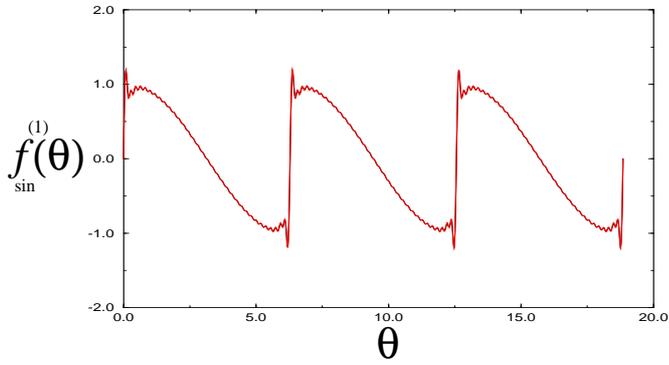

Figure 3.5: The first odd eigenfunction associated with perturbing the positions of the poles in the giant cusp.

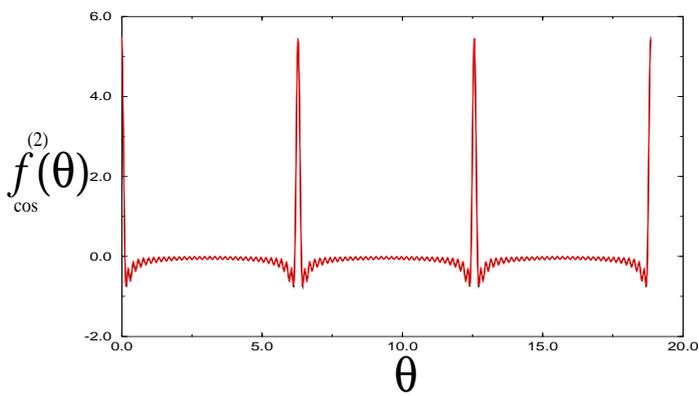

Figure 3.6: The first even eigenfunction associated with perturbing the positions of the poles in the giant cusp.

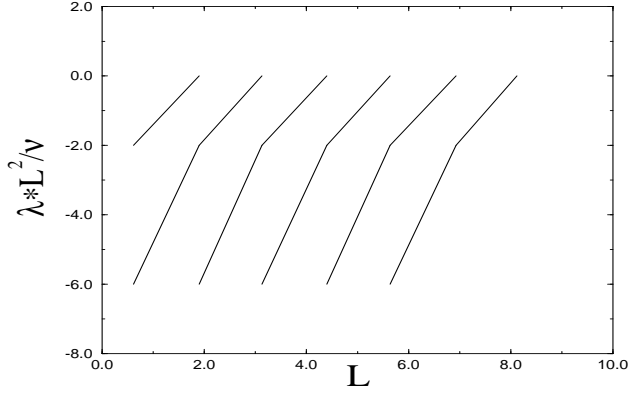

Figure 3.7: Spectrum of eigenvalues associated with the reaction of the poles in the giant cusp to the addition of new poles.

For $x_0 = -\pi/2M$ we get

$$\delta u(\theta, t) \simeq 4\nu M e^{-My_p(t)} \cos M\theta \qquad (3.22)$$

in both cases the equations for the dynamics of $y_p$ follow from Eqs.(2.11)-(2.12):

$$\frac{dy_p}{dt} \simeq 2\frac{\nu}{L^2}\alpha(M) , \qquad (3.23)$$

where $\alpha(M)$ is given as:

$$\alpha(M) = [\frac{1}{2}(\frac{L}{\nu} + 1)] - \frac{1}{2}(\frac{L}{\nu} - M) \qquad (3.24)$$

Since (3.23) is linear, we can solve it and substitute in Eqs.(3.21)-(3.22). Seeking a form $\delta u(\theta, t) \sim exp(-\lambda(M)t)$ we find that the eigenvalue $\lambda(M)$ is

$$\lambda(M) = 2M\frac{\nu}{L^2}\alpha(M) \qquad (3.25)$$

These eigenvalues are plotted in Fig.3.7 At this point we consider the dynamics of the poles in the giant cusp under the influence of the additional $M$ poles. From Eqs.(3.10), (3.12), (2.11), (2.12) we obtain, after some obvious algebra,

$$L\dot{\delta y_i} = \sum_j \frac{\partial^2 U}{\partial y_i \partial y_j}\delta y_j - 4\frac{\nu}{L}Me^{-My_p(t)} \sinh(My_i) \qquad (3.26)$$

or

$$L\dot{\delta x_i} = \sum_j V_{ij}\delta x_j - 4\frac{\nu}{L}Me^{-My_p(t)} \cosh(My_i) \qquad (3.27)$$

It is convenient now to transform from the basis $\delta y_i$ to the natural basis $w_i$ which is obtained using the linear transformation $w = A^{-1}\delta y$. Here the matrix $A$ has columns which are the eigenvectors of $\partial^2 U/\partial y_i \partial y_j$ which were computed before. Since the matrix was real symmetric, the matrix $A$ is orthogonal, and $A^{-1} = A^T$. Define $C = 4\frac{\nu}{L^2}Me^{-My_p(0)}$ and write

$$\dot{w_i} = -\lambda_i w_i - Ce^{-\lambda(M)t}\xi_i , \qquad (3.28)$$

where $-\lambda_i$ are the eigenvalues associated with the columns of $A$, and

$$\xi_i = \sum_j A_{ji} \sinh M y_j \ . \tag{3.29}$$

We are looking now for a solution that decays exponentially at the rate $\lambda(M)$:

$$w_i(t) = w_i(0) e^{-\lambda(M)t} \tag{3.30}$$

Substituting the desired solution in (3.28) we find a condition on the initial value of $w_i$:

$$w_i(0) = -\frac{C}{\lambda_i - \lambda(M)} \xi_i \tag{3.31}$$

Transforming back to $\delta y_i$ we get

$$\delta y_i(0) = \sum_k A_{ik} w_k(0) = -\sum_k A_{ik} \frac{C}{\lambda_k - \lambda(M)} \sum_l A_{lk} \sinh M y_l$$
$$= -C \sum_l \sinh M y_l \sum_k \frac{A_{ik} A_{lk}}{\lambda_k - \lambda_p^M} \tag{3.32}$$

We can get the eigenfunctions of the linear operator, as before, using Eqs.(3.18), (3.21), (3.22), (3.32). We get

$$f_{\sin}^{(M)}(\theta) = 4C\nu \sum_{i=1}^{N(L)} \sum_{k=1}^{\infty} (\sum_l \sinh M y_l \sum_m \frac{A_{im} A_{lm}}{\lambda_m - \lambda(M)})$$
$$\times k e^{-k y_i} \sin k\theta + L^2 C \sin M\theta \tag{3.33}$$

An identical calculation to the one started with Eq. (3.28) can be followed for the deviations $\delta x_i$. The final result reads

$$f_{\cos}^{(M)}(\theta) = 4C\nu \sum_{i=1}^{N(L)} \sum_{k=1}^{\infty} (\sum_l \cosh M y_l \sum_m \frac{\tilde{A}_{im} \tilde{A}_{lm}}{\tilde{\lambda}_m - \lambda(M)})$$
$$\times k e^{-k y_i} \cos k\theta + L^2 C \cos M\theta \ , \tag{3.34}$$

where $\tilde{A}$ is the matrix whose columns are the eigenvectors of $V$, and $-\tilde{\lambda}_i$ its eigenvalues.

We are now in position to explain the entire linear spectrum using the knowledge that we have gained. The spectrum consists of two separate types of contributions. The first type has $2N$ modes that belong to the dynamics of the unperturbed $N(L)$ poles in the giant cusp. The second part, which is most of the spectrum, is built from modes of the second type since $M$ can go to infinity. This structure is seen in the Fig.3.4 and Fig.3.7.

We can argue that the set of eigenfunctions obtained above is complete and exhaustive. To do this we show that any arbitrary periodic function of $\theta$ can be expanded in terms of these eigenfunctions. Start with the standard Fourier series in terms of sin and cos functions. At this point solve for $\sin k\theta$ and $\cos k\theta$ from Eqs.(3.33-3.34). Substitute the results in the Fourier sums. We now have an expansion in terms of the eigenmodes $f^{(M)}$ and in terms of the triple sums. The triple sums however can be expanded, using Eqs.(3.19-3.20), in terms of the eigenfunctions $f^{(\ell)}$. We can thus decompose any function in terms of the eigenfunctions $f^{(M)}$ and $f^{(\ell)}$.

## 3.4 Conclusions

We discussed the stability of flame fronts in channel geometry using the representation of the solutions in terms of singularities in the complex plane. In this language the stationary solution, which is a giant cusp in configuration space, is represented by $N(L)$ poles which are organized on a line parallel to the imaginary axis. We showed that the stability problem can be understood in terms of two types of perturbations. The first type is a perturbation in the positions of the poles that make up the giant cusp. The longitudinal motions of the poles give rise to odd modes, whereas the transverse motions to even modes. The eigenvalues associated with these modes are eigenvalues of a finite, real and symmetric matrices, cf. Eqs.(3.13), (3.14), (3.15), (3.16). The second type of perturbations is obtained by adding poles to the set of $N(L)$ poles representing the giant cusp. The reaction of the latter poles is again separated into odd and even functions as can be seen from Eqs.(3.21), (3.22). Together the two types of perturbations rationalize and explain all the features of the eigenvalues and eigenfunctions obtained from the standard linear stability analysis.

# Chapter 4

# Dynamics and Wrinkling of Radially Propagating Fronts Inferred from Scaling Laws in Channel Geometries

## 4.1 Introduction

The main idea of this chapter is that in order to derive scaling laws for unstable front propagation in radial geometry, it is useful to study noisy propagation in channel geometries, in which the noiseless dynamics results usually in simple shapes of the advancing fronts [1].

The understanding of radial geometries requires control of the effect of noise on the unstable dynamics of propagation. It is particularly difficult to achieve such a control in radial geometries due to the vagueness of the distinction between external noise and noisy initial conditions. Channel geometries are simpler when they exhibit a stable solution for growth in the noiseless limit. One can then study the effects of external noise in such geometries without any ambiguity. If one finds rules to translate the resulting understanding of the effects of noise in channel growth to radial geometries, one can derive the scaling laws in the later situation in a satisfactory manner. We will exemplify the details of such a translation in the context of premixed flames that exist as self sustaining fronts of exothermic chemical reactions in gaseous combustion. But our contention is that similar ideas should be fruitful also in other contexts of unstable front propagation. Needless to say, there are aspects of the front dynamics and statistics in the radial geometry that *cannot* be explained from observations of fronts in a channel geometry; examples of such aspects are discussed at the end of this chapter.

Mathematically our example is described [11] by an equation of motion for the angle- dependent modulus of the radius vector of the flame front, $R(\theta, t)$:

$$\begin{aligned}\frac{\partial R}{\partial t} &= \frac{U_b}{2R_0^2(t)} \left(\frac{\partial R}{\partial \theta}\right)^2 + \frac{D_M}{R_0^2(t)} \frac{\partial^2 R}{\partial \theta^2} \\ &+ \frac{\gamma U_b}{2R_0(t)} I(R) + U_b \ . \end{aligned} \quad (4.1)$$

Here $0 < \theta < 2\pi$ is an angle and the constants $U_b, D_M$ and $\gamma$ are the front velocity for an ideal cylindrical front, the Markstein diffusivity and the thermal expansion coefficient respectively. $R_0(t)$ is the mean radius of the propagating flame:

$$R_0(t) = \frac{1}{2\pi} \int_0^{2\pi} R(\theta, t) d\theta \ . \quad (4.2)$$

The functional $I(R)$ is best represented in terms of its Fourier decomposition. Its Fourier component is $|k|R_k$ where $R_k$ is the Fourier component of $R$. Simulations of this equation, as well as experiments in the parameter regime for which this equation is purportedly relevant, indicate that for large times $R_0$ grows as a power in time

$$R_0(t) = (const + t)^\beta , \qquad (4.3)$$

with $\beta > 1$, and that the width of the interface $W$ grows with $R_0$ as

$$W(t) \sim R_0(t)^\chi , \qquad (4.4)$$

with $\chi < 1$.

## 4.2 The Geometry of Developing Flame Fronts: Analysis with Pole Decomposition

The study of growing fronts in nonlinear physics [1] offers fascinating examples of spontaneous generation of fractal geometry [2, 3]. Advancing fronts rarely remain flat; usually they form either fractal objects with contorted and ramified appearance, like Laplacian growth patterns and diffusion limited aggregates (DLA) [31], or they remain graphs, but they "roughen" in the sense of producing self-affine fractals whose "width" diverges with the linear scale of the system with some characteristic exponent. The study of interface growth where the roughening is caused by the noisy environment, with either annealed or quenched noise, was a subject of active research in recent years [32, 33]. These studies met considerable success and there is significant analytic understanding of the nature of the universality classes that can be expected. The study of interface roughening in system in which the flat surface is inherently unstable is less developed. One interesting example that attracted attention is the Kuramoto-Sivashinsky equation [9, 34] which is known to roughen in 1+1 dimensions but is claimed not to roughen in higher dimensions [42]. Another outstanding example is Laplacian growth patterns [35]. This chapter is motivated by a new example of the dynamics of outward propagating flames whose front wrinkles and fractalizes [11]. We will see that this problem has many features that closely resemble Laplacian growth, including the existence of a single finger in channel growth versus tip splitting in cylindrical outward growth, extreme sensitivity to noise, etc. In the case of flame fronts the equation of motion is amenable to analytic solutions and as a result we can understand some of these issues.

The physical problem that motivates this analysis is that of pre-mixed flames which exist as self-sustaining fronts of exothermic chemical reactions in gaseous combustion. It had been known for some time that such flames are intrinsically unstable [43]. It was reported that such flames develop characteristic structures which includes cusps, and that under usual experimental conditions the flame front accelerates as time goes on [10]. In recent work Filyand et al. [11] proposed an equation of motion that is motivated by the physics and seems to capture a number of the essential features of the observations. The equation is written in cylindrical geometry and is for $R(\theta, t)$ which is the modulus of the radius vector on the flame front:

$$\begin{aligned}\frac{\partial R}{\partial t} &= \frac{U_b}{2R_0^2(t)}\left(\frac{\partial R}{\partial \theta}\right)^2 + \frac{D_M}{R_0^2(t)}\frac{\partial^2 R}{\partial \theta^2} \\ &+ \frac{\gamma U_b}{2R_0(t)}I(R) + U_b .\end{aligned} \qquad (4.5)$$

Here $0 < \theta < 2\pi$ is an angle and the constants $U_b, D_M$ and $\gamma$ are the front velocity for an ideal cylindrical front, the Markstein diffusivity and the thermal expansion coefficient respectively. $R_0(t)$ is the mean radius of the propagating flame:

$$R_0(t) = \frac{1}{2\pi}\int_0^{2\pi} R(\theta, t)d\theta .\quad (4.6)$$

The functional $I(R)$ is best represented in terms of its Fourier decomposition. Its Fourier component is $|k|R_k$ where $R_k$ is the Fourier component of $R$.

Numerical simulations of the type reported in ref. [11] are presented in Fig.4.1. The two most prominent features of these simulations are the wrinkled multi-cusp appearance of the fronts and its acceleration as time progresses. One observes the phenomenon of tip splitting in which new cusps are added to the growing fronts between existing cusps. Both experiments and simulations indicate that for large times $R_0$ grows as a power in time

$$R_0(t) = (const + t)^\beta ,\quad (4.7)$$

with $\beta > 1$, (of the order of 1.5) and that the width of the interface $W$ grows with $R_0$ as

$$W(t) \sim R_0(t)^\chi ,\quad (4.8)$$

with $\chi < 1$ (of the order of 2/3). The understanding of these two features and the derivation of the scaling relation between $\beta$ and $\chi$ are the main aims of this chapter.

Equation (4.6) can be written as a one-parameter equation by rescaling $R$ and $t$ according to $r \equiv RU_b/D_M$, $\tau \equiv tU_b^2/D_M$. Computing the derivative of Eq.(4.6) with respect to $\theta$ and substituting the dimensionless variables one obtains:

$$\frac{\partial u}{\partial \tau} = \frac{u}{r_0^2}\frac{\partial u}{\partial \theta} + \frac{1}{r_0^2}\frac{\partial^2 u}{\partial \theta^2} + \frac{\gamma}{2r_0}I\{u\} .\quad (4.9)$$

where $u \equiv \frac{\partial r}{\partial \theta}$. To complete this equation we need a second one for $r_0(t)$, which is obtained by averaging (4.6) over the angles and rescaling as above. The result is

$$\frac{dr_0}{d\tau} = \frac{1}{2r_0^2}\frac{1}{2\pi}\int_0^{2\pi} u^2 d\theta + 1 .\quad (4.10)$$

These two equations are the basis for further analysis

Following [12, 14–16, 38, 39] we expand now the solutions $u(\theta, \tau)$ in poles whose position $z_j(\tau) \equiv x_j(\tau) + iy_j(\tau)$ in the complex plane is time dependent:

$$\begin{aligned}u(\theta, \tau) &= \sum_{j=1}^N \cot\left[\frac{\theta - z_j(\tau)}{2}\right] + c.c.\quad (4.11)\\ &= \sum_{j=1}^N \frac{2\sin[\theta - x_j(\tau)]}{\cosh[y_j(\tau)] - \cos[\theta - x_j(\tau)]} ,\end{aligned}$$

$$r(\theta, \tau) = 2\sum_{j=1}^N \ln\left[\cosh(y_j(\tau)) - \cos(\theta - x_j(\tau))\right] + C(\tau) .\quad (4.12)$$

In (4.12) $C(\tau)$ is a function of time. The function (4.12) is a superposition of quasi-cusps (i.e. cusps that are rounded at the tip). The real part of the pole position (i.e. $x_j$) describes the angle coordinate of the maximum of the quasi-cusp, and the imaginary part of the pole position (i.e $y_j$) is related the height of the quasi-cusp. As $y_j$ decreases (increases) the height of the cusp increases (decreases). The physical motivation for this representation of the solutions should be evident from Fig.4.1.

The main advantage of this representation is that the propagation and wrinkling of the front can be described now via the dynamics of the poles and of $r_0(t)$. Substituting (4.11) in (4.9) we derive the following ordinary differential equations for the positions of the poles:

$$-r_0^2 \frac{dz_j}{d\tau} = \sum_{k=1, k \neq j}^{2N} \cot\left(\frac{z_j - z_k}{2}\right) + i\frac{\gamma r_0}{2} sign[Im(z_j)] \ . \quad (4.13)$$

After substitution of (4.11) in (4.10) we get, using (4.13) the ordinary differential equation for $r_0$,

$$\frac{dr_0}{d\tau} = 2 \sum_{k=1}^{N} \frac{dy_k}{d\tau} + 2\left(\frac{\gamma}{2}\frac{N}{r_0} - \frac{N^2}{r_0^2}\right) + 1 \ . \quad (4.14)$$

In our problem the outward growth introduces important modifications to the channel results. The number of poles in a stable configuration is proportional here to the radius $r_0$ instead of $L$, but the former grows in time. The system becomes therefore unstable to the addition of new poles. If there is noise in the system that can generate new poles, they will not be pushed toward infinite $y$. It is important to stress that any infinitesimal noise (either numerical or experimental) is sufficient to generate new poles. These new poles do not necessarily merge their $x$-positions with existing cusps. Even though there is attraction along the real axis as in the channel case, there is a stretching of the distance between the poles due to the radial growth. This may counterbalance the attraction. Our first new idea is that these two opposing tendencies define a typical scale denoted as $\mathcal{L}$. if we have a cusp that is made from the $x$-merging of $N_c$ poles on the line $x = x_c$ and we want to know whether a $x$-nearby pole with real coordinate $x_1$ will merge with this large cusp, the answer depends on the distance $D = r_0|x_c - x_1|$. There is a length $\mathcal{L}(N_c, r_0)$ such that if $D > \mathcal{L}(N_c, r_0)$ then the single cusp will never merge with the larger cusp. In the opposite limit the single cusp will move towards the large cusp until their $x$-position merges and the large cusp will have $N_c + 1$ poles.

This finding stems directly from the equations of motion of the $N_c$ $x$-merged poles and the single pole at $x_1$. First note that from Eq.4.7 (which is not explained yet) it follows that asymptotically $r_0(\tau) = (a + \tau)^\beta$ where $r_0(0) = a^\beta$. Next start from 4.13 and write equations for the angular distance $x = x_1 - x_c$. It follows that for any configuration $y_j$ along the imaginary axis

$$\frac{dx}{d\tau} \leq -\frac{2N_c \sin x [1 - \cos x]^{-1}}{(a+\tau)^{2\beta}} = -\frac{2N_c \cot(\frac{x}{2})}{(a+\tau)^{2\beta}} \ . \quad (4.15)$$

For small x we get

$$\frac{dx}{d\tau} \leq -\frac{4N_c}{x(a+\tau)^{2\beta}} \ . \quad (4.16)$$

The solution of this equation is

$$x(0)^2 - x(\tau)^2 \geq \frac{8N_c}{2\beta - 1}(a^{1-2\beta} - (a+\tau)^{1-2\beta}) \ . \quad (4.17)$$

To find $\mathcal{L}$ we set $x(\infty) > 0$ from which we find that the angular distance will remain finite as long as

$$x(0)^2 > \frac{8N_c}{2\beta - 1}a^{1-2\beta} \ . \quad (4.18)$$

Since $r_0 \sim a^\beta$ we find the threshold angle $x^*$

$$x^* \sim \sqrt{N_c} r_0^{\frac{(1-2\beta)}{2\beta}} \;, \tag{4.19}$$

above which there is no merging between the giant cusp and the isolated pole. To find the actual distance $\mathcal{L}(N_c, r_0)$ we multiply the angular distance by $r_0$ and find

$$\mathcal{L}(N_c, r_0) \equiv r_0 x^* \sim \sqrt{N_c} r_0^{\frac{1}{2\beta}} \;. \tag{4.20}$$

To understand the geometric meaning of this result we recall the features of the TFH cusp solution. Having a typical length $L$ the number of poles in the cusp is linear in $L$. Similarly, if we have in this problem two cusps a distance $2\mathcal{L}$ apart, the number $N_c$ in each of them will be of the order of $\mathcal{L}$. From (4.20) it follows that

$$\mathcal{L} \sim r_0^{\frac{1}{\beta}} \;. \tag{4.21}$$

For $\beta > 1$ the circumference grows faster than $\mathcal{L}$, and therefore at some points in time poles that appear between two large cusps would not be attracted toward either, and new cusps will appear. We will show later that the most unstable positions to the appearance of new cusps are precisely the midpoints between existing cusps. This is the mechanism for the addition of cusps in analogy with tip splitting in Laplacian growth.

We can now estimate the width of the flame front as the height of the largest cusps. Since this height is proportional to $\mathcal{L}$ (cf. property (v) of the TFH solution), Eq.(4.21) and Eq.(4.8) lead to the scaling relation

$$\chi = 1/\beta \;. \tag{4.22}$$

This scaling law is expected to hold all the way to $\beta = 1$ for which the flame front does not accelerate and the size of the cusps becomes proportional to $r_0$.

In channels there is a natural lengthscale, the width $\tilde{L}$ of the channel. The translation of channel results to radial geometry will be based on the identification in the latter context of the time dependent scale $\mathcal{L}(t)$ that plays the role of $\tilde{L}$ in the former. To do this we need first to review briefly the main pertinent results for noisy channel growth. In channel geometry the equation of motion is written in terms of the position $h(x,t)$ of the flame front above the $x$-axis. After appropriate rescalings [12] it reads:

$$\frac{\partial h(x,t)}{\partial t} = \frac{1}{2}\left[\frac{\partial h(x,t)}{\partial x}\right]^2 + \nu \frac{\partial^2 h(x,t)}{\partial x^2} + I\{h(x,t)\} + 1. \tag{4.23}$$

It is convenient to rescale the domain size further to $0 < \theta < 2\pi$, and to change variables to $u(\theta, t) \equiv \partial h(\theta, t)/\partial \theta$. In terms of this function we find

$$\frac{\partial u(\theta,t)}{\partial t} = \frac{u(\theta,t)}{L^2}\frac{\partial u(\theta,t)}{\partial \theta} + \frac{\nu}{L^2}\frac{\partial^2 u(\theta,t)}{\partial \theta^2} + \frac{1}{L}I\{u(\theta,t)\} \tag{4.24}$$

where $L = \tilde{L}/2\pi$. In noiseless conditions this equation admits exact solutions that are represented in terms of $N$ poles whose position $z_j(t) \equiv x_j(t) + iy_j(t)$ in the complex plane is time dependent:

$$u(\theta, t) = \nu \sum_{j=1}^{N} \cot\left[\frac{\theta - z_j(t)}{2}\right] + c.c. \;, \tag{4.25}$$

The steady state for channel propagation is unique and linearly stable; it consists of $N(L)$ poles which are aligned on one line parallel to the imaginary axis. The geometric appearance of the flame front is a giant cusp, analogous to the single finger in the case of Laplacian growth in a channel. The height of the cusp is proportional to $L$, and the propagation velocity is a constant of the motion. The number of poles in the giant cusp is linear in $L$ Eq. (1.10,

The introduction of additive random noise to the dynamics changes the picture qualitatively. It is convenient to add noise to the equation of motion in Fourier representation by adding a white noise $\eta_k$ for every $k$ mode. The noise correlation function satisfies the relation $<\eta_k(t)\eta_{k'}(t')> = \delta_{k,k'}\delta(t-t')f/L$. The noise in our simulations is taken from a flat distribution in the interval $[-\sqrt{2f/L}, \sqrt{2f/L}]$; this guarantees that when the system size changes, the typical noise per unit length of the flame front remains constant. It was shown in chapter 2 and [17] that for moderate but fixed noise levels the average velocity $v$ of the front increases with $L$ as a power law. In our present simulations we found

$$v \sim L^\mu, \quad \mu \approx 0.35 \pm 0.03 \ . \tag{4.26}$$

For a fixed system size $L$ the velocity has also a power law dependence on the level of the noise, but with a much smaller exponent: $v \sim f^\xi$, $\xi \approx 0.02$. These results were understood theoretically by analysing the noisy creation of new poles that interact with the poles defining the giant cusp (in chapter 2 and [17]).

Next we shed light on phenomenon of tip splitting that here is seen as the addition of new cusps roughly in between existing ones. We mentioned the instability toward the addition of new poles. We argue now that the tip between the cusps is most sensitive to pole creation. This can be shown in both channel and radial geometry. For example consider a TFH-giant cusp solution in which all the poles are aligned (without loss of generality) on the $x = 0$ line. Add a new pole in the complex position $(x_a, y_a)$ to the existing $N(L)$ poles, and study its fate. It can be shown that in the limit $y_a \to \infty$ (which is the limit of a vanishing perturbation of the solution) the equation of motion is

$$\frac{dy_a}{dt} = \frac{2\pi\nu}{L^2}(2N(L)+1) - 1 \qquad y_a \to \infty \ . \tag{4.27}$$

Since $N(L)$ satisfies (1.10) this equation can be rewritten as

$$\frac{dy_a}{dt} = \frac{4\pi\nu}{L^2}(1-\alpha) \qquad y_a \to \infty \ , \tag{4.28}$$

where $\alpha = (L/(2\pi\nu)+1)/2 - N(L)$. Obviously $\alpha \le 1$ and it is precisely 1 only when $L$ is $L = (2n+1)2\pi\nu$. Next it can be shown that for $y_a$ much larger than $y_{N(L)}$ but not infinite the following is true:

$$\frac{dy_a}{dt} > \lim_{y_a \to \infty} \frac{dy_a}{dt} \quad x_a = 0 \tag{4.29}$$

$$\frac{dy_a}{dt} < \lim_{y_a \to \infty} \frac{dy_a}{dt} \quad x_a = \pi \tag{4.30}$$

We learn from these results that there exist values of $L$ for which a pole that is added at infinity will have marginal attraction $(dy_a/dt = 0)$. Similar understanding can be obtained from a standard stability analysis without using pole decomposition. Perturbing a TFH-cusp solution we find linear equations whose eigenvalues $\lambda_i$ can be obtained by standard numerical techniques: (i) all $Re(\lambda_i)$ are non-positive. (ii) at the isolated values of $L$ for which $L = (2n+1)2\pi\nu\ Re(\lambda_1)$ and $Re(\lambda_2)$ become zero (note that due to the logarithmic scale the zero is not evident) (iii) There exists a general tendency of all $Re(\lambda_i)$ to approach zero in absolute magnitude such as $\frac{1}{L^2}$ from below as $L$ increases. This indicates a growing sensitivity to noise when the system size increases. (iv) There exists a Goldsone mode $\lambda_0 = 0$ due to translational invariance.

The upshot of this discussion is that finite perturbations (i.e. poles at finite $y_a$) will grow if the $x$ position of the pole is sufficiently near the tip. The position $x = \pi$ (the tip of the finger) is the most unstable one. In the channel geometry this means that noise results in the appearance of new cusps at the tip of the fingers, but due to the attraction to the giant cusp they move toward $x = 0$ and disappear in the giant cusp. In fact, one sees in numerical simulations a train of small cusps that move toward the giant cusp. Analysis shows that at the same time the furthest pole at $y_{N(L)}$ is pushed towards infinity. Also in cylindrical geometry the most sensitive position to the appearance of new cusps is right between two existing cusps independently if the system is marginal (the total number of poles fits the radius) or unstable (total number of poles is too small at a given radius). Whether or not the addition of a new pole results in tip splitting depends on their $x$ position. When the distance from existing cusps is larger than $\mathcal{L}$ the new poles that are generated by noise will remain near the tip between the two cusps and will cause tip splitting.

The picture used remains valid as long as the poles that are introduced by the noisy perturbation do not destroy the identity of the giant cusp. Indeed, the numerical simulations show that in the presence of moderate noise the additional poles appear as smaller cusps that are constantly running towards the giant cusp. Our point here, is not to predict the numerical values of the scaling exponents in the *channel* (this was done in ref. [17] and chapter 2, but to use them to predict the scaling exponents characterizing the acceleration and the geometry of the flame front in *radial* geometry.

Superficially it seems that in radial geometry the growth pattern is qualitatively different. In fact, close observation of the growth patterns (see Fig.4.1) shows that most of the time there exist some big cusps that attract other smaller cusps, but that every now and then "new" big cusps form and begin to act as local absorbers of small cusps that appear randomly. The understanding of this phenomenon gives the clue how to translate results from channels to radial growth.

Equations (4.9), (4.10) again admit exact solutions in terms of poles, of the form of Eq.(4.13). It is easy write down the equations of motion of the poles and check that the poles are attractive along the real direction (which means physically that they are attracted along the angular coordinate) but they are repulsive along the imaginary direction, which is associated with the radial coordinate. If it were not for the stretching that is caused by the increase of the radius (and with it the perimeter), all the poles would have coalesced into one giant cusp. Thus we have a competition between pole attraction and stretching. Since the attraction decreases with the distance between the poles in the angular directions, there is always an initial critical length scale above which poles cannot coalesce their real coordinates when time progresses.

Suppose now that noise adds new poles to the system. The poles do not necessarily merge their real positions with existing cusps. If we have a large cusp made from the merging of the real coordinates $x_c$ of $N_c$ poles, we want to know whether a nearby pole with real coordinate $x_1$ will merge with this large cusp. The answer will depend of course on the distance $D \equiv r_0|x_c - x_1|$. A direct calculation [19], using the equation of motions for the poles shows that there exists a critical length $\mathcal{L}(r_0)$ such that if $D > \mathcal{L}(r_0)$ the single pole never merges with the giant cusp. The result of the calculation is that

$$\mathcal{L} \sim r_0^{1/\beta} \ . \quad (4.31)$$

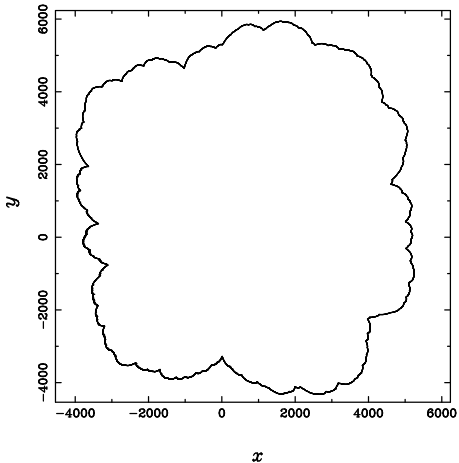

Figure 4.1: Simulations of the outward propagating flame front. Note that there is a wide distribution of cusp sizes.

Note that a failure of a single pole to be attracted to a large cusp means that tip-splitting has occurred. This is the exact analog of tip-splitting in Laplacian growth.

It is now time to relate the channel and radial geometries. We identify the typical scale in the radial geometry as $\mathcal{L} \sim W \sim r_0^\chi$. On the one hand this leads to the scaling relation $\beta = 1/\chi$. On the other hand we use the result established in a channel, (4.26), with this identification of a scale, and find $\dot{r}_0 = r_0^{\chi\mu}$. Comparing with (4.7) we find:

$$\beta = \frac{1}{(1-\chi\mu)} \ . \quad (4.32)$$

This result leads us to expect two dynamical regimes for our problem. Starting from smooth initial conditions, in relatively short times the roughness exponent remains close to unity. This is mainly since the typical scale $\mathcal{L}$ is not relevant yet, and most of the poles that are generated by noise merge into a few larger cusps. In later times the roughening exponent settles at its asymptotic value, and all the asymptotic scaling relations used above become valid. We thus expect $\beta$ to decrease from $1/(1-\mu)$ to an asymptotic value determined by $\chi = 1/\beta$ in (4.32):

$$\beta = 1 + \mu \approx 1.35 \pm 0.03 \ . \quad (4.33)$$

The expected value of $\chi$ is thus $\chi = 0.74 \pm 0.03$. We tested these predictions in numerical simulations. We integrated Eq.(4.24), and in Fig.4.2 we display the results for the growth velocity as a function of time. After a limited domain of exponential growth we observe a continuous reduction of the time dependent exponent. In the initial region we get $\beta = 1.65 \pm 0.1$ while in the final decade of the temporal range we find $\beta = 1.35 \pm 0.1$. We consider this a good agreement with (4.33). A second important test is provided by measuring the width of the system as a function of the radius, see Fig.4.2b. Again we observe a cross-over related to the initial dynamics; In the last temporal decade the exponent settles at $\chi = 0.75 \pm .1$. We conclude that at times large enough to observe the asymptotics our predictions are verified.

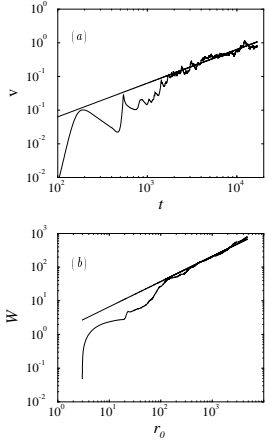

Figure 4.2: Panel a: a logarithmic plot of the velocity versus time for a radially evolving system. The parameters of the simulation are: $f = 10^{-8}$, $\gamma = 0.8$, $\nu = 1$. Panel b:Logarithmic plot of the width of the flame front as a function of the mean radius.

Finally, we stress some differences between radial and channel geometries. Fronts in a channel exhibit mainly one giant cusp which is only marginally disturbed by the small cusps that are introduced by noise. In the radial geometry, as can be concluded from the discussion above, there exist at any time cusps of all sizes from the smallest to the largest. This broad distribution of cusps (and scales) must influence correlation function in ways that differ qualitatively from correlation functions computed in channel geometries. To make the point clear we exhibit in Fig.4.3a the structure function

$$F(y) \equiv \sqrt{\langle |R(x+y) - R(x)|^2 \rangle} \tag{4.34}$$

computed for a typical radial front, with $x = R\theta$. To stress the scaling region we exhibit the second derivative of this function in Fig.4.3b. The low end of the graph can be fitted well by a power law $y^{-\alpha}$ with $\alpha \approx 0.6$. This indicates that $F(y) \approx Ay + By^{2-\alpha}$. In a channel geometry we get entirely different structure functions that do not exhibit such scaling functions at all. The way to understand this behavior in the radial geometry is to consider a distribution of cusps that remain distinct from each other but whose scales are distributed according to some distribution $P'(\ell)$. Let us define a distribution function $P(\ell) \sim \ell P'(\ell)$ which give us the probability that a point on the circuit with the mean radius lies on the basis of the cusp with the size $\ell$. For each of these cusps there is a contribution to the correlation function of the form $f(y, \ell) \approx \ell g(y/\ell)$ where $g(x)$ is a scaling function, $g(x) \approx x$ for $x < 1$ and $g(x) \approx$ constant for $x > 1$. The total correlation can be estimated (when the poles are distinct) as

$$F(y) \approx \sum_\ell P(\ell) \ell g(y/\ell) . \tag{4.35}$$

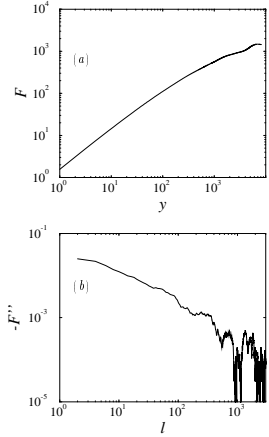

Figure 4.3: Panel a: a logarithmic plot of the correlation function F(y) of the interface versus the distance y between points. Panel b: Second derivative of the correlation function -F"(l) of the interface versus the distance l between points.

The first derivative leaves us with $\sum_\ell P(\ell)g'(y/\ell)$, and using the fact that $g'$ vanishes for $x > 1$ we estimate $F'(y) = \sum_{\ell=y}^{W} P(\ell)$. The second derivative yields $F''(y) \approx -P(y)$. Thus the structure function is determined by the scale distribution of cusps, and if the latter is a power law, this should be seen in the second derivative of $F(y)$ as demonstrated in Fig.4.3. The conclusion of this analysis is that the radial case exhibits a scaling function that characterizes the distributions of cusps, $P(\ell) \approx \ell^{-\alpha}$.

## 4.3 Conclusions

The main purpose of this chapter was to find exponents of the problem and find connections between them. Using the main result of the channel case (dependence of the velocity on the channel size) we can find the acceleration of the flame front in the radial case ($R_0(t) = (const + t)^\beta$)

$$\beta = 1 + \mu \ . \tag{4.36}$$

where $\mu$ is the exponent for the dependence of the velocity on the channel size found before, $\beta$ is the acceleration exponent and $R_0$ is a mean radius of the flame front.

Dependence of the width of the flame front in the radial case $W(t)$ on mean radius ($R_0(t) = (const + t)^\beta$) is

$$W(t) \sim R_0(t)^\chi \ . \tag{4.37}$$

$$\chi = 1/\beta \ . \tag{4.38}$$

In summary, we demonstrated that it is possible to use information about noisy channel dynamics to predict nontrivial features of the radial evolution, such as the acceleration and roughening exponents. It would be worthwhile to examine similar ideas in the context of Laplacian growth patterns.

# Chapter 5

# Laplacian Growth

## 5.1 Introduction

The problem of pattern formation is one of the most rapidly developing branches of nonlinear science today [1]. Of special interest is the study of the front dynamics between two phases (interface) that arises in a variety of nonequilibrium physical systems. If, as it usually happens, the motion of the interface is slow in comparison with the processes that take place in the bulk of both phases (such as heat transfer, diffusion, etc.), the scalar field governing the evolution of the interface is a harmonic function. It is natural then, to call the whole process *Laplacian growth*. Depending on the system, this harmonic scalar field is a temperature (in the freezing of a liquid or Stefan problem), a concentration (in solidification from a supersaturated solution), an electrostatic potential (in electrodeposition), a pressure (in flows through porous media), a probability (in diffusion-limited aggregation), etc.

The mathematical problem of Laplacian growth without surface tension exhibits a family of exact (analytic) solutions in terms of logarithmic poles in the complex plane. We show that this family of solutions has a remarkable property: generic initial conditions in channel geometry which begin with arbitrarily many features exhibit an inverse cascade into a single finger.

In the absence of surface tension, whose effect is to stabilize the short-wavelength perturbations of the interface, the problem of 2D Laplacian growth is described as follows

$$(\partial_x^2 + \partial_y^2)u = 0 \ . \tag{5.1}$$

$$u\mid_{\Gamma(t)} = 0 \ , \partial_n u\mid_{\Sigma} = 1 \ . \tag{5.2}$$

$$v_n = \partial_n u\mid_{\Gamma(t)} \ . \tag{5.3}$$

Here $u(x, y; t)$ is the scalar field mentioned, $\Gamma(t)$ is the moving interface, $\Sigma$ is a fixed external boundary, $\partial_n$ is a component of the gradient normal to the boundary (i.e. the normal derivative), and $v_n$ is a normal component of the velocity of the front.

We consider an infinitely long interface, obtained by a periodic continuation of the interface in the channel with periodic boundary conditions.Then We introduce a time-dependent conformal map $f$ from the lower half of a "mathematical" plane, $\xi \equiv \zeta + i\eta$, to the domain of the physical plane, $z \equiv x + iy$, where the Laplace equation 5.1 is defined as $\xi \xrightarrow{f} z$. We also require that $f(t, \xi) \approx \xi$ for $\xi \longrightarrow \zeta - i\infty$.

Thus the function $z = f(t, \zeta)$ describes the moving interface. From Eqs. (5.1), (5.2), (5.3) for function $f(t, \xi)$ we obtain the *Laplacian Growth Equation*

$$Im(\frac{\partial f(\xi,t)}{\partial \xi} \overline{\frac{\partial f(\xi,t)}{\partial t}}) = 1 \mid_{\xi=\zeta-i0}, f_\zeta \mid_{\zeta-i\infty} = 1 . \qquad (5.4)$$

Now we will extend these results obtained for periodic boundary conditions to the more physical "no-flux" boundary conditions (no flux across the lateral boundaries of the channel). This requires that the moving interface orthogonally intersects the walls of the channel. However, unlike the case of periodic boundary conditions, the end points at the two boundaries do not necessarily have the same horizontal coordinate. This is also a periodic problem where the period equals twice the width of the channel. The analysis is the same as before, but now only half of the strip should be considered as the physical channel, whereas the second half is the unphysical mirror image.

Let us look for a solution of Eq. (5.4) in the next form

$$f(\xi, t) = \lambda \xi - i\tau(t) - i \sum_{l=1}^{N} \alpha_l \log(e^{i\xi} - e^{i\xi_l(t)}),$$

$$\sum_{l=1}^{N} \alpha_l = 1 - \lambda, -1 < \lambda < 1 , \qquad (5.5)$$

where $\tau(t)$ is some real function of time, $\lambda$ is a real constant, $\alpha_l$ is a complex constant, $\xi_l = \zeta_l + i\eta_l$ denotes the position of the pole with the number $l$ and $N$ is the number of poles.

For the "no-flux" boundary condition we must add the condition that for every pole $\xi_l = \zeta_l + i\eta_l$ with $\alpha_l$ exists a pole $\xi_l = -\zeta_l + i\eta_l$ with $\overline{\alpha_l}$. So for the function $F(i\xi, t) = if(\xi, t)$

$$\overline{F(i\xi, t)} = F(\overline{i\xi}, t) \qquad (5.6)$$

We want to prove that the final state will be only one finger.

## 5.2 Asymptotic behavior of the poles in the mathematical plane

The main purpose of this chapter is to investigate the asymptotic behavior of the poles in the mathematical plane. We want to demonstrate that for time $t \mapsto \infty$, all poles go to a single point (or two points for no-flux boundary conditions). The equation for the interface is

$$f(\xi, t) = \lambda \xi - i\tau(t) - i \sum_{l=1}^{N} \alpha_l \log(e^{i\xi} - e^{i\xi_l(t)}),$$

$$\sum_{l=1}^{N} \alpha_l = 1 - \lambda, -1 < \lambda < 1 . \qquad (5.7)$$

By substitution of Eq. (5.7) in the *Laplacian Growth Equation*

$$Im(\frac{\partial f(\xi,t)}{\partial \xi} \overline{\frac{\partial f(\xi,t)}{\partial t}}) = 1 \mid_{\xi=\zeta-i0}, \tag{5.8}$$

we can find the equations of pole motion:

$$const = \tau(t) + (1 - \sum_{k=1}^{N} \overline{\alpha_k}) \log \frac{1}{a_l} + \sum_{k=1}^{N} \overline{\alpha_k} \log(\frac{1}{a_l} - \overline{a_k}) \tag{5.9}$$

and

$$\tau = t - \frac{1}{2} \sum_{k=1}^{N} \sum_{l=1}^{N} \overline{\alpha_k} \alpha_l \log(1 - \overline{a_k} a_l) + C_0, \tag{5.10}$$

where $a_l = e^{i\xi_l}$.

From eq. (5.9) we can find

$$C_1 = (1-\lambda)\tau - \sum_{l=1}^{N} \alpha_l \log a_l + \sum_{k=1}^{N} \sum_{l=1}^{N} \overline{\alpha_k} \alpha_l \log(1 - \overline{a_k} a_l). \tag{5.11}$$

From eqs. (5.10) and (5.11) we can obtain

$$Im(\sum_{l=1}^{N} \alpha_l \log a_l) = constant \tag{5.12}$$

and

$$t = (\frac{1+\lambda}{2})\tau + \frac{1}{2} Re(\sum_{l=1}^{N} \alpha_l \log a_l) + C_1/2, \tag{5.13}$$

where $\alpha_l$ is a constant, $\xi_l(t)$ is the position of the poles, $a_l = e^{i\xi_l(t)}$, and $\frac{\lambda+1}{2}$ is the portion of the channel occupied by the moving liquid. We will see that for $\tau \mapsto \infty$ we obtain one finger with wide $\frac{\lambda+1}{2}$.

In Appendix A we will prove from eq.(5.10) that $\tau \mapsto \infty$, if $t \mapsto \infty$ and if any finite time singularity does not exist.

The equations of pole motion are next from eq. (5.9)

$$const = \tau + i\overline{\xi_k} + \sum_{l} \alpha_l \log(1 - e^{i(\xi_l - \overline{\xi_k})}), \tag{5.14}$$

$$const = \zeta_k + \sum_{l}(\alpha_l'' \log |1 - e^{i(\xi_l - \overline{\xi_k})}| + \alpha_l' \arg(1 - e^{i(\xi_l - \overline{\xi_k})})), \tag{5.15}$$

$$const = \tau + \eta_k + \sum_{l}(\alpha_l' \log |1 - e^{i(\xi_l - \overline{\xi_k})}| - \alpha_l'' \arg(1 - e^{i(\xi_l - \overline{\xi_k})})), \tag{5.16}$$

$$\xi_l = \zeta_l + i\eta_l, \eta_l > 0. \tag{5.17}$$

$$\alpha_l = \alpha_l' + i\alpha_l'' \tag{5.18}$$

Let us examine

$$\arg(1 - e^{i(\xi_l - \overline{\xi_k})}) = \arg([1 - e^{i(\zeta_l - \zeta_k)} e^{-(\eta_l + \eta_k)}]) = \arg[1 - a_{lk} e^{i\varphi_{lk}}] \tag{5.19}$$

$$\varphi_{lk} = \zeta_l - \zeta_k, a_{lk} = e^{-(\eta_l+\eta_k)} \qquad (5.20)$$

$\arg[1 - a_{lk}e^{i\varphi_{lk}}]$ is a singlevalued function of $\varphi_{lk}$, i.e.

$$-\frac{\pi}{2} \leq \arg[1 - a_{lk}e^{i\varphi_{lk}}] \leq \frac{\pi}{2}. \qquad (5.21)$$

From the eq. (5.16) the only way to compensate for the divergence of term $\tau$ is that $\eta_k \mapsto 0$ for $\tau \mapsto \infty, 1 \leq k \leq N$.

We want to investigate asymptotic behavior of poles $\tau \mapsto \infty$. To eliminate the divergent term $\log|1 - e^{i(\xi_k - \overline{\xi_k})}|$ we multiply eq. (5.16) by $\alpha_k''$ and eq. (5.15) by $\alpha_k'$ and take difference

$$const = \alpha_k'\zeta_k - \alpha_k''\tau + \sum_{l\neq k}((\alpha_l''\alpha_k' - \alpha_k''\alpha_l')\log|1 - e^{i(\xi_l - \overline{\xi_k})}| +$$
$$(\alpha_l'\alpha_k' + \alpha_l''\alpha_k'')\arg(1 - e^{i(\xi_l - \overline{\xi_k})})). \qquad (5.22)$$

We have the divergent terms $\alpha_k''\tau$ in this equation. We may assume that for $t \mapsto \infty$, $N'$ groups of poles exist to eliminate the divergent terms ($\varphi_{lk} \mapsto 0$ for all members of a group). $N_l$ is the number of poles in each group, $1 < l < N'$. For each group by summation of eqs. (5.22) over all group poles we obtain

$$const = \alpha_k^{gr'}\zeta_k^{gr} - \alpha_k^{gr''}\tau + \sum_{l\neq k}((\alpha_l^{gr''}\alpha_k^{gr'} - \alpha_k^{gr''}\alpha_l^{gr'})\log|1 - e^{i(\xi_l^{gr} - \overline{\xi_k^{gr}})}| +$$
$$(\alpha_l^{gr'}\alpha_k^{gr'} + \alpha_l^{gr''}\alpha_k^{gr''})\arg(1 - e^{i(\xi_l^{gr} - \overline{\xi_k^{gr}})})). \qquad (5.23)$$

$$\alpha_l^{gr''} = \sum_k^{N_l} \alpha_k'', \qquad (5.24)$$

$$\alpha_l^{gr'} = \sum_k^{N_l} \alpha_k'. \qquad (5.25)$$

We have no merging of these groups for large $\tau$ and we investigate the motion of poles with this assumption

$$|\zeta_l^{gr} - \zeta_k^{gr}| \gg \eta_l^{gr} + \eta_k^{gr}, 1 \leq l, k \leq N. \qquad (5.26)$$

For $l \neq k$, $\eta_k^{gr} \mapsto 0$, $\varphi_{lk}^{gr} = \zeta_l^{gr} - \zeta_k^{gr}$ we obtain

$$\log|1 - e^{i(\xi_l^{gr} - \overline{\xi_k^{gr}})}| \approx \log|1 - e^{i(\zeta_l^{gr} - \zeta_k^{gr})}| = \log 2 + \frac{1}{2}\log\sin^2\frac{\varphi_{lk}^{gr}}{2} \qquad (5.27)$$

and

$$\arg(1 - e^{i(\xi_l^{gr} - \overline{\xi_k^{gr}})}) \approx \arg(1 - e^{i(\zeta_l^{gr} - \zeta_k^{gr})}) = \frac{\varphi_{lk}^{gr}}{2} + \pi n - \frac{\pi}{2}. \qquad (5.28)$$

We choose $n$ in Eq.(5.28) so that Eq.(5.21) is correct.
Substituting these results to the eqs. (5.23) we obtain

$$C_k = \alpha^{gr'}\zeta_k^{gr} - \alpha_k^{gr''}\tau + \sum_{l\neq k}[(\alpha_l^{gr''}\alpha_k^{gr'} - \alpha_k^{gr''}\alpha_l^{gr'})\log|\sin\frac{\varphi_{lk}^{gr}}{2}|$$
$$+(\alpha_l^{gr'}\alpha_k^{gr'} + \alpha_l^{gr''}\alpha_k^{gr''})\frac{\varphi_{lk}^{gr}}{2}] \qquad (5.29)$$

## 5.3 Theorem about coalescence of the poles

From eqs. (5.29) we can conclude

(i) By summation of eqs.(5.29) (or exactly from eq. (5.12)) we obtain

$$\sum_k \alpha_k^{gr\prime} \zeta_k^{gr} = const . \tag{5.30}$$

(ii) For $| \varphi_{lk}^{gr} | \mapsto 0, 2\pi$, we obtain $\log | \sin \frac{\varphi_{lk}^{gr}}{2} | \mapsto \infty$, meaning that the poles can not pass off each other;

(iii) From (ii) we conclude that $0 < | \varphi_{lk}^{gr} | < 2\pi$

(iv) From (i) and (iii), $\zeta_k^{gr} \mapsto \infty$ is impossible;

(v) In eq.(5.29) we must compensate the second divergent term. From (iv) and (iii) we can do it only if $\alpha_l^{gr\prime\prime} = \sum_k^{N_l} \alpha_k'' = 0$ for all $l$.

So from eq. (5.29) we obtain

$$\sum_k^{N_l} \alpha_k'' = 0 , \tag{5.31}$$

$$\dot{\varphi}_{lk}^{gr} = 0 , \tag{5.32}$$

$$\varphi_{lk}^{gr} \neq 0 , \tag{5.33}$$

$$\dot{\zeta}_k^{gr} = 0 . \tag{5.34}$$

For the asymptotic motion of poles in the group $N_m$ we obtain from eqs. (5.31), (5.32), (5.33), (5.34) taking leadind terms in eqs. (5.13), (5.14)

$$\tau = \frac{2}{\lambda+1} t , \tag{5.35}$$

$$0 = \dot{\tau} + \sum_l^{N_m} \alpha_l \frac{\dot{\eta}_k + \dot{\eta}_l + i(\dot{\zeta}_k - \dot{\zeta}_l)}{\eta_k + \eta_l + i(\zeta_k - \zeta_l)} . \tag{5.36}$$

The solution to these equations is

$$\eta_k = \eta_k^0 e^{-\frac{1}{\alpha_m^{gr\prime}} \frac{2}{1+\lambda} t} , \tag{5.37}$$

$$\varphi_{lk} = \varphi_{lk}{}^0 e^{-\frac{1}{\alpha_m^{gr\prime}} \frac{2}{1+\lambda} t} , \tag{5.38}$$

$$\dot{\zeta}_k = 0 . \tag{5.39}$$

So we may conclude that for eliminating the divergent term we need

$$\alpha_l^{gr\prime\prime} = \sum_k^{N_l} \alpha_k'' = 0, \tag{5.40}$$

$$\alpha_l^{gr\prime}(1+\lambda) > 0 \tag{5.41}$$

for all $l$.

## 5.4 Conclusions

With the periodic boundary condition, eq.(5.40) is correct for all poles, so we obtain $N' = 1$, $m = 1$ and $N_m = N$.

Therefore the unique solution is

$$\eta_k = \eta_k^0 e^{-\frac{2}{(1-\lambda^2)}t} , \tag{5.42}$$

$$\varphi_{lk} = \varphi_{lk}{}^0 e^{-\frac{2}{(1-\lambda^2)}t} , \tag{5.43}$$

$$\dot{\zeta}_k = 0 . \tag{5.44}$$

$$1 - \lambda^2 > 0 \tag{5.45}$$

With the no-flux boundary condition we have a pair of the poles whose condition of eq. (5.40) is correct so all these pairs must merge. Because of the symmetry of the problem these poles can merge only on the boundaries of the channel $\zeta = 0, \pm\pi$. Therefore we obtain two groups of the poles on boundaries $N' = 2$, $m = 1, 2$, $N_1 + N_2 = N$, $\alpha_1^{gr'} + \alpha_2^{gr'} = 1 - \lambda$.

Consequently we obtain the solution (on two boundaries):

$$\eta_k^{(1)} = \eta_k^{(1),0} e^{-\frac{1}{\alpha_1^{gr'}}\frac{2}{1+\lambda}t} , \tag{5.46}$$

$$\varphi_{lk}^{(1)} = \varphi_{lk}{}^{(1),0} e^{-\frac{1}{\alpha_1^{gr'}}\frac{2}{1+\lambda}t} , \tag{5.47}$$

$$\zeta_k^{(1)} = 0 ; \tag{5.48}$$

$$\eta_k^{(2)} = \eta_k^{(2),0} e^{-\frac{1}{\alpha_2^{gr'}}\frac{2}{1+\lambda}t} , \tag{5.49}$$

$$\varphi_{lk}^{(2)} = \varphi_{lk}{}^{(2),0} e^{-\frac{1}{\alpha_2^{gr'}}\frac{2}{1+\lambda}t} , \tag{5.50}$$

$$\zeta_k^{(2)} = \pm\pi ; \tag{5.51}$$

$$\alpha_1^{gr'}(1 + \lambda) > 0, \tag{5.52}$$

$$\alpha_2^{gr'}(1 + \lambda) > 0. \tag{5.53}$$

## 5.5 Appendix A

We need to prove that $\tau \mapsto \infty$, if $t \mapsto \infty$ and if any finite time singularity does not exist. This is apparent if the second term in the next formula for $\tau$ is greater than zero:

$$\tau = t + [-\frac{1}{2}\sum_{k=1}^{N}\sum_{l=1}^{N}\overline{\alpha_k}\alpha_l \log(1 - \overline{a_k}a_l)] + C_0 , \tag{5.54}$$

where $\mid a_l \mid < 1$ for all $l$. Let us prove it.

$$-\frac{1}{2}\sum_{k=1}^{N}\sum_{l=1}^{N}\overline{\alpha_k}\alpha_l \log(1-\overline{a_k}a_l) = -\frac{1}{2}\sum_{k=1}^{N}\sum_{l=1}^{N}\overline{\alpha_k}\alpha_l \sum_{n=1}^{\infty}(-\frac{(\overline{a_k}a_l)^n}{n})$$

$$\frac{1}{2}\sum_{n=1}^{\infty}\frac{1}{n}(\sum_{k=1}^{N}\overline{\alpha_k}(\overline{a_k})^n)(\sum_{l=1}^{N}\alpha_l(a_l)^n)$$

$$= \frac{1}{2}\sum_{n=1}^{\infty}\frac{1}{n}(\overline{\sum_{l=1}^{N}\alpha_l(a_l)^n})(\sum_{l=1}^{N}\alpha_l(a_l)^n) > 0 \qquad (5.55)$$

# Chapter 6

# Summary

The problem of flame propagation is studied as an example of unstable fronts that wrinkle on many scales. The analytic tool of pole expansion in the complex plane is employed to address the interaction of the unstable growth process with random initial conditions and perturbations. We argue that the effect of random noise is immense and that it can never be neglected in sufficiently large systems. We present simulations that lead to scaling laws for the velocity and acceleration of the front as a function of the system size and the level of noise, and analytic arguments that explain these results in terms of the noisy pole dynamics.

We consider flame front propagation in channel geometries. The steady state solution in this problem is space dependent, and therefore the linear stability analysis is described by a partial integro-differential equation with a space dependent coefficient. Accordingly it involves complicated eigenfunctions. We show that the analysis can be performed to required detail using a finite order dynamical system in terms of the dynamics of singularities in the complex plane, yielding detailed understanding of the physics of the eigenfunctions and eigenvalues.

The roughening of expanding flame fronts by the accretion of cusp-like singularities is a fascinating example of the interplay between instability, noise and nonlinear dynamics that is reminiscent of self-fractalization in Laplacian growth patterns. The nonlinear integro-differential equation that describes the dynamics of expanding flame fronts is amenable to analytic investigations using pole decomposition. This powerful technique allows the development of a satisfactory understanding of the qualitative and some quantitative aspects of the complex geometry that develops in expanding flame fronts.

Flame Propagation is used as a prototypical example of expanding fronts that wrinkle without limit in radial geometries but reach a simple shape in channel geometry. We show that the relevant scaling laws that govern the radial growth can be inferred once the simpler channel geometry is understood in detail. In radial geometries (in contrast to channel geometries) the effect of external noise is crucial in accelerating and wrinkling the fronts. Nevertheless, once the interrelations between system size, velocity of propagation and noise level are understood in channel geometry, the scaling laws for radial growth follow.

The mathematical problem of Laplacian growth without surface tension exhibits a family of exact (analytic) solutions in terms of logarithmic poles in the complex plane. We show that this family of solutions has a remarkable property: generic initial conditions in channel geometry which begin with arbitrarily many features exhibit an inverse cascade into a single finger.

**Тезисы ведущие
к степени Ph.D**

# Процессы Роста Поверхностей

Купервассер Олег Юрьевич

**Под руководством: Итамара Прокаччо**

*Отделение Химической Физики
Научно-исследовательского Института имени Вейцмана
76100 Реховот, Израиль*

*Ноябрь 1999*

# Оглавление

# Глава 1

# Введение.

Задачи роста поверхности раздела недавно получили большое внимание [1–3]. Так, например, это диффузионно-ограниченная агрегация (DLA) [4], случайная последовательная адсорбция (RSA) [5], Лапласовский рост [6–8] или распространение фронта пламени [9]. Мы главным образом обратим внимание в этих Тезисах на численное и аналитическое исследование последних двух задач.

В дополнение к тому факту, что распространение фронта пламени - интересная физическая задача, мы чувствуем, что мы можем также объяснить экспериментальные результаты на основе теоретических исследований. Также существует возможность использовать методы, найденные для распространения фронта пламени, в различных областях, где подобные задачи появляются, такие как важная модель Лапласовского роста.

Заранее перемешанное пламя - самоподдерживающаяся волна экзотермической химической реакции - является одним из основных проявлений газообразного сгорания. Это хорошо установлено, однако, что самая простая вообразимая конфигурация пламени - неограниченное плоское пламя, свободно распространяющееся через первоначально неподвижную гомогенную горючую смесь – существенно неустойчива и спонтанно принимает характерные двух или трехмерные структурные формы.

В недавней статье Гостинцева, Истратова и Шуленина [10] представлен интересный обзор экспериментальных исследований распространяющегося сферического и цилиндрического пламени направленного наружу в режиме хорошо развивающейся гидродинамический (Ландау-Дерри) неустойчивости. Доступные данные ясно указывают, что свободно расширяющееся неровное пламя обладает двумя существенными особенностями:

1. Структура фронта пламени в виде многих «квази-изломов». (Фронт пламени состоит из большого количества квази-изломов, то есть, точек излома с округленными вершинами).

2. Заметное ускорение фронта пламени.

Кроме того, временная зависимость радиуса пламени почти идентична для всех обсужденных предсмесей и коррелирует хорошо с простым отношением:

$$R_0(t) = At^{3/2} + B \qquad (1.1)$$

Здесь $R_0(t)$ эффективный (средний) радиус неровного пламени, а A, B - эмпирические константы.

В этих Тезисах мы изучаем пространственное и временное поведение нелинейных моделей континуума (то есть, моделей, которые обладают бесконечным числом степеней свободы). Они включают все особенности, которые считались существенными для заранее перемешанных систем с пламенем; а именно, дисперсность, нелинейность и линейную неустойчивость. Севашинский, Фильянд и Френкель [11] недавно получили уравнение, обозначаемое как SFF в дальнейшем, чтобы описать, как двумерные изгибы цилиндрического заранее перемешанного пламени возникают как следствие из известной Ландау-Дерри гидродинамической неустойчивости. Уравнение SFF читается следующим образом:

$$\frac{\partial R}{\partial t} = \frac{U_b}{2R_0^2(t)}\left(\frac{\partial R}{\partial \theta}\right)^2 + \frac{D_M}{R_0^2(t)}\frac{\partial^2 R}{\partial \theta^2} + \frac{\gamma U_b}{2R_0(t)}I\{R\} + U_b \ . \quad (1.2)$$

где $0 < \theta < 2\pi$ являются углом; $R(\theta, t)$ модуль радиус-вектора на поверхности раздела пламени; $U_b, D_M, \gamma$ являются константами.

$$I(R) = \frac{1}{\pi}\sum_{n=1}^{\infty} n \int_0^{2\pi} cos[n(\theta - \theta^*)]R(\theta^*, t)d\theta^* =$$
$$= -\frac{1}{\pi}P\int_{-\infty}^{+\infty} \frac{\frac{\partial R(\theta^*,t)}{\partial \theta^*}}{\theta^* - \theta}d\theta^* \quad (1.3)$$

$$R_0(t) = \frac{1}{2\pi}\int_0^{2\pi} R(\theta, t)d\theta \ . \quad (1.4)$$

Севашинский, Фильянд и Френкель [11] провели прямое числовое моделирование этого нелинейного уравнения для эволюции цилиндрической динамики поверхности пламени. Результат показал, что оба упомянутых экспериментальных эффекта имеют место. Кроме того, оцененный порядок ускорения весьма совместим со степенным законом, данным уравнением (1.1). Для сравнения численное моделирование свободно распространяющегося диффузионно-нестабильного пламени были представлено также. В этом случае никакая тенденция к ускорению не наблюдалась.

В отсутствии поверхностного натяжения, действие которого должно стабилизировать возмущения короткой длины волны поверхности, задача двумерного Лапласовского роста описана следующим образом

$$(\partial_x^2 + \partial_y^2)u = 0 \ . \quad (1.5)$$

$$u\mid_{\Gamma(t)} = 0 \ , \partial_n u\mid_{\Sigma} = 1 \ . \quad (1.6)$$

$$v_n = \partial_n u\mid_{\Gamma(t)} \ . \quad (1.7)$$

Здесь $u(x, y; t)$ - скалярное поле, $\Gamma(t)$ движущаяся поверхность раздела, $\Sigma$ фиксированная внешняя граница, $\partial_n$ компонент градиента, нормальная к границе (то есть производная по нормали), и $v_n$ нормальная компонента скорости фронта.

Чтобы получить результаты для радиального роста пламени, необходимо вначале исследовать случай канала. Версия уравнения в канале для распространения фронта пламени - так называемое уравнение Михельсона-Севашинского [12, 13]

$$\frac{\partial H}{\partial t} = \frac{1}{2}\left(\frac{\partial H}{\partial x}\right)^2 + \nu\frac{\partial^2 H}{\partial x^2} + I\{H\} \ . \quad (1.8)$$

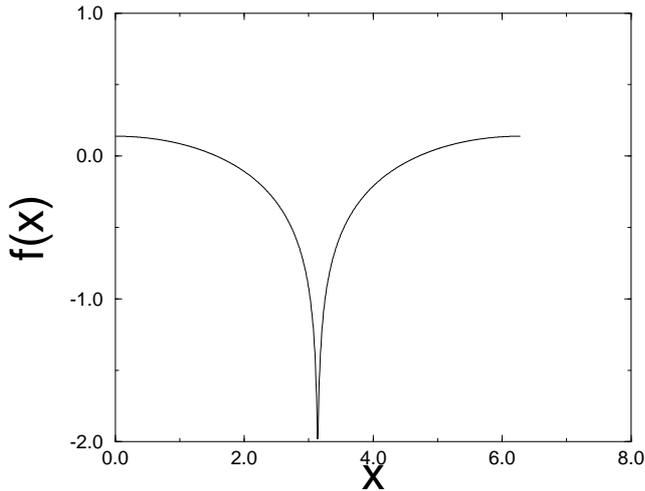

Рис. 1.1: Решение в виде гигантского излома

$$I(H) = -\frac{1}{\pi} P \int_{-\infty}^{+\infty} \frac{\frac{\partial H(x^*,t)}{\partial x^*}}{x^* - x} dx^* \,. \qquad (1.9)$$

с периодическими граничными условиями на интервале x [0, L], где L - размер системы. $\nu$ является постоянным, $\nu > 0$. H - высота точки фронта пламени, $P \int$ обычный сингулярный интеграл.

У уравнений для распространения фронта пламени и Лапласового роста с нулевым поверхностным растяжением есть замечательное свойство: эти уравнения могут быть решены в терминах полюсов в комплексной плоскости [6, 12, 14–16]. Таким образом, мы получаем ряд обычных дифференциальных уравнений для координат этих полюсов. Число полюсов - постоянная величина в системе, но чтобы объяснить такой эффект как рост скорости фронта пламени, мы должны рассмотреть небольшой шум, который является источником новых полюсов. Таким образом, мы должны решить задачу взаимодействия случайных флуктуаций и движений полюсов.

Самый простой случай - геометрия канала. Главные результаты для этого случая - существование решения [12] (рис. 7.1) в виде гигантского излома, которое представлено в пространстве конфигураций полюсами, которые организованы на линии, параллельной мнимой оси. Это полюсное решение - точка притяжения для всех полюсов.

Полный анализ этого установившегося решения был сначала представлен в работе [12], и главные полученные результаты формулируются следующим образом:

1. Есть только одно устойчивое стационарное решение, которое геометрически представлено гигантским изломом (или эквивалентно одним пальцем) и аналитически $N(L)$ полюсами, которые выстраиваются на одной линии, параллельной мнимой оси. Существование этого решения объясняется в следующих пунктах.

2. Существует притяжение между полюсами вдоль действительной линии. В получающейся динамике происходит слияние всех $x$ положений полюсов, $y$-положения которых остается конечным.

3. $y$ положения отличны, и полюса выравниваются один выше другого в положениях $y_{j-1} < y_j < y_{j+1}$ с неким максимальным $y_{N(L)}$. Это может быть понято из уравнений для движения полюсов. Для них взаимодействие является отталкиванием на коротких диапазонах, но изменяет знак на более длинных диапазонах.

4. Если мы прибавим дополнительный полюс к такому решению, то этот полюс (или другой) будет выкинут в бесконечность вдоль мнимой оси. Если у системы есть меньше чем $N(L)$ полюсов, то она неустойчива к добавлению полюсов, и любой шум будет вести систему к этому единственному стационарному состоянию. Число $N(L)$ описывается

$$N(L) = \left[\frac{1}{2}\left(\frac{L}{\nu} + 1\right)\right], \quad (1.10)$$

где $\left[\ldots\right]$ - целая часть числа и $2\pi L$ - размер системы. Чтобы увидеть это, рассмотрим систему с $N$ полюсами, таким образом, что все значения $y_j$ удовлетворяют условию $0 < y_j < y_{max}$. Прибавьте теперь один дополнительный полюс, координаты которого $z_a \equiv (x_a, y_a)$ с $y_a \gg y_{max}$. Из уравнений движения для $y_a$, мы видим, что члены в сумме - все порядка единицы, также как и $\cot(y_a)$ член. Таким образом, уравнение движения $y_a$ приблизительно

$$\frac{dy_a}{dt} \approx \nu \frac{2N+1}{L^2} - \frac{1}{L}. \quad (1.11)$$

Судьба этого полюса зависит от числа других полюсов. Если $N$ будет слишком большим, то полюс убежит в бесконечность, тогда как, если $N$ является небольшим, полюс будет притянут к действительной оси. Условие для того, чтобы убежать в бесконечность состоит в том, что $N > N(L)$, где $N(L)$ дан (1.10). С другой стороны $y$ координата полюсов не может достичь нуля. Ноль - отталкивающая линия, и полюса отталкиваются от нуля с бесконечной скоростью. Чтобы увидеть это рассмотрим полюс, чей $y_j$ приближается к нулю. Для любого конечного $L$ член $\coth(y_j)$ растет неограниченно, в то время как все другие члены в уравнении для движения полюсов остаются ограниченными.

5. Высота излома - пропорциональна $L$. Распределение положений полюсов вдоль линии с постоянным $x$ было найдено в [12].

Мы будем ссылаться на решение со всеми этими свойствами как на Чуал-Фриш-Хенон (TFH) - решение в виде гигантского излома.

Главные результаты нашей собственной работы следующие.

Традиционный линейный анализ был сделан для этого решения в виде гигантского излома. Этот анализ демонстрирует существование отрицательных собственных значений, которые стремятся к нулю, когда размер системы стремится в бесконечность.

1. Существует очевидная мода Голдстоуна или трансляционная мода с собственным значением $\lambda_0 = 0$. Этот собственная мода происходит от Галилеевой инвариантности уравнения движения.

2. Перемасштабированные собственные значения ($L^2\lambda_i$) периодически колеблются между значениями, которые $L$-независимы в этом представлении. Другими словами, до колебательного поведения собственные значения зависят от $L$ как $L^{-2}$.

3. Собственные значения $\lambda_1$ и $\lambda_2$ периодически обнуляются. Функциональная зависимость в этом представлении кажется почти кусочно-линейной.

4. Более высокие собственные значения также показывают подобное качественное поведение, но не достигая ноля. Мы отмечаем, что решение становится незначительно устойчивым для каждого значения $L$, для которого собственные значения $\lambda_1$ и $\lambda_2$ обнуляются. $L^{-2}$ зависимость спектра указывает, что решение становится более чувствительным к шуму, когда $L$ увеличивается.

Было доказано, что произвольные начальные условия могут быть написаны в виде набора полюсов в комплексной плоскости. Обратный каскадный процесс формирования гигантского излома был исследован в численной и аналитической форме. Зависимости ширины фронта пламени и средней скорости были найдены. Следующим шагом в исследовании случая канальной геометрии было исследование влияния случайного шума на динамику полюсов. Главный эффект внешнего шума - появление новых полюсов в максимумах фронта пламени и слияния этих полюсов с гигантским изломом. Зависимость средней скорости фронта пламени от шума и размера системы была найдена. Скорость почти независима от шума, пока шум не достигает некоторого критического значения. При исследовании зависимости скорости от размера системы мы видим рост скорости с некоторым показателем степени, до тех пор, пока скорость не достигает некоторого значения насыщения.

Обозначим $v$ как скорость фронта пламени и L как размер системы:

1. Мы можем видеть два различных режима поведения средней скорость $v$ как функции шума $f$ для фиксированного размера системы L. Для шума $f$ меньшего, чем некоторое фиксированное значение $f_{cr}$

$$v \sim f^{\xi} . \qquad (1.12)$$

Для этих значений $f$ эта зависимость очень слабая, и $\xi \approx 0.02$. Для больших значений $f$ зависимость намного более сильная

2. Мы можем видеть рост средней скорости $v$ как функции L. После некоторых значений L мы можем видеть насыщение скорости. Для режима $f < f_{cr}$ рост скорости может быть записан как

$$v \sim L^{\mu}, \quad \mu \approx 0.35 \pm 0.03 . \qquad (1.13)$$

Зависимость числа полюсов в системе и числа полюсов, которые появляются в системе в единицу времени, была исследована в численной форме как функция шума и параметров системы. Время жизни полюса было найдена в численой форме. Теоретическое обсуждение влияния шума на динамику полюсов и среднюю скорость было сделано [17].

Динамика полюсов может быть использована также, чтобы проанализировать небольшое возмущение фронта пламени и сделать полный анализ стабильности гигантского излома. Два вида мод были найдены. Первый – собственные колебания полюсов в гигантском изломе. Второй - моды, связанные с появлением новых полюсов в системе. Собственные значения этих мод были найдены. Результаты находятся в хорошем соглашении с традиционным анализом стабильности [18].

Результаты, найденные для случая канала, могут использоваться, чтобы проанализировать распространение фронта пламени в радиальном случае [19, 20]. Главная особенность этого случая - соревнование между притяжением полюсов и расширением фронта пламени. Так в этом случае мы получаем не только один гигантский излом, но набор изломов. Новые полюса, которые появляются в системе из-за шума - это изломы. На основе уравнения движения полюсов мы можем найти связь между ускорением фронта пламени и шириной поверхности раздела. На основе результата для средней скорости в случае канала может быть найдено ускорение фронта пламени. Таким образом, мы получаем полную картину распространения фронта пламени в радиальном случае.

Следующий шаг в исследовании задачи рассматривает Лапласовский рост с нулевым поверхностным натяжением, у которого также есть полюсные решения. В случае Лапласовского роста мы получаем результат, который походит на слияние полюсов в случае распространения фронта пламени в канале: все полюса срастаются в один полюс в случае периодического граничного условия или два полюса на границах в случае граничных условий без потоков. Этот результат может быть доказан теоретически [21].

В статьях [22–26] рассматривают само-ускорение без привлечения внешнего принуждения. Никакое само-ускорение не существует для конечного числа полюсов. Таким образом, мы можем объяснить как само-ускорение, так и появление новых полюсов либо шумом, либо "дождем" полюсов из "облака" в бесконечности. Действительно, любое данное начальное условие может быть написано как сумма бесконечного числа полюсов (Раздел 8.4.1).

Позвольте нам рассмотреть один полюс, который появляется из "облака" в бесконечности. Мы пренебрегаем силой отталкивания от остальной части полюсов в системе и рассматриваем только силу притяжения в уравнениях 4.13 $-\frac{\gamma}{2r_0}$ Для $r_0$, мы можем написать в случае само-ускорения $r_0(\tau) = (a+\tau)^\beta, r_0(0) = a^\beta, \beta > 1$. Так от $\tau = 0$ до $\tau = \infty$ полюс снижается на расстояние $\Delta y = \int_0^\infty \frac{\gamma}{2r_0(\tau)}d\tau = \frac{\gamma}{2}\frac{1}{\beta-1}\frac{1}{r_0(0)^{\frac{\beta-1}{\beta}}}$. Поэтому "дождь" снижается на конечное расстояние после бесконечного времени. Кроме того это расстояние сходится к нулю, если $r_0(0) \mapsto \infty$! Таким образом, мы думаем, что появление новых полюсов из бесконечности может быть объяснено только внешним шумом.

Характерный размер излома в системе $\mathcal{L} \sim r_0^{\frac{1}{\beta}}$. Из Рис. 8.23 шум $f \sim \frac{1}{\mathcal{L}^5} \sim \frac{1}{r_0^{\frac{5}{\beta}}}$ необходим для появления новых изломов в системе. Если шум больше чем это значение, зависимость от шума является очень медленной ($f^{0.2}$ для режима II и $f^{0.02}$ для режима III). Этот результат объясняет почти полную неизменность результатов численного моделирования при уменьшении уровня шума ( [22], Рис. 2).

Жоулин и др. [27–30] используют очень похожий подход для анализа роста пламени в канале и в радиальной геометрии. Но главное внимание в нашей работе сосредоточено на скорости фронта пламени (само-ускорение для радиального случая) и ширине фронта пламени. Главное внимание в случае канала в работе Жоулина было уделено исследованию среднего интервала между изломами (гребнями). Для радиального случая только линейная зависимость радиуса от времени (без само-ускорения) рассматривается в работе Жоулина.

Наши работы очень хорошо дополняют друг друга, но не конкурируют друг с другом. Например, для вычисления расстояния между изломами (или, что то же самое, среднее значение размера излома) без всякого доказательства мы использовали уравнение (8.70). Но Рис. 9 из работы ( [28]) дает нам превосходное обоснование для этого уравнения.

Структура этих Тезисов следующая.

Глава 7 - это Введение.

В Главе 8 мы получаем главные результаты для случая распространения фронта пламени в канале. Мы приводим результаты для стационарного решения, представляем традиционный линейный анализ задачи и занимаемся аналитическим и численным исследованием влияния шума на среднюю скорость, динамику полюсов и фронта.

В Главе 9 мы получаем результаты линейного анализа стабильности с помощью полюсных решений.

В Главе 10 мы используем результат, полученный для случая канала для анализа распространения фронта пламени в радиальном случае

В Главе 11 мы исследуем асимптотическое поведение полюсов в комплексной плоскости для Лапласовского роста с нулевым поверхностным натяжением в случае периодического и граничного условия без потоков.

Глава 12 – это резюме.

# Глава 2

# Динамика полюсов при нестабильном распространении фронта пламени: случай геометрии канала

## 2.1 Введение

Цель этой главы состоит в том, чтобы исследовать роль случайных колебаний на динамику роста неровных поверхностей, которая описывается нелинейными уравнениями движения. Мы интересуемся теми примерами, для которых рост плоских или гладких поверхностей характеризуется нестабильностью. Известный пример таких явлений роста предоставлен в [1] картинами Лапласовского роста [1–3]. Экспериментальная реализация таких картин происходит, например, в ячейках Хеле-Шоу [1], в которых воздух или другая жидкость низкой вязкости перемещают нефть или некоторую другую жидкость высокой вязкости. При нормальных условиях продвигающиеся фронты не остаются плоскими; в плоском канале они формируют в асимптотике устойчивый палец, ширина которого определяется тонкими эффектами, которые являются результатом существования поверхностного натяжения. В радиальной геометрии рост поверхности приводит к искаженной и ветвящейся фрактальной форме. Аналогичное явление было изучено в модельном уравнении для распространения пламени, которое имеет те же самые свойства линейные устойчивости, что проблема Лапласовского роста [9]. Физическая проблема в этом случае - проблема распространения пламени в заранее перемешанных смесях, которое существует как самоподдерживающийся фронт экзотермических химических реакций при газообразном сгорании. Эксперименты [10] о распространении пламени в радиальной геометрии демонстрируют, что фронт пламени ускоряет со временем и характеризуется характерными экспонентами, описывающими его неровности. Оба наблюдения не получали надлежащие теоретические объяснения прежде Известно, что движение в плоском канале и радиальное движение являются заметно различными. Первый приводит к единственному гигантскому острому излому на двигающемся фронте. Тогда как во втором случае появляется очень много острых изломов, которые появляются в сложной иерархической последовательности на двигающемся фронте пламени ( [11, 19] и глава 10). Доступны аналитические методики для изучения таких процессов [38]. В контексте распространения пламени [12, 15, 19, 39], и в контексте Лапласовского роста с нуле-вым поверхностным натяжением [6,35,36] можно исследовать решения, которые описаны в терминах движения полюсов в комплексной плоскости. Это описание очень полезно и описывается конечным числом обычных дифференциальных уравнений для положений полюсов, из чего уже можно получить геометрию развивающегося фронта чрезвычайно экономичным и эффективным способом. К сожалению, это описание не доступно в случае Лапласовского роста с нулевым поверхностным натяжением, и это делает проблему распространения пламени очень привлекательной. Однако это описание страдает от одного фундаментального недостатка. Для уравнения динамики полюсов без внешнего шума всегда сохраняется число полюсов, которые существовали в начальных условиях. В результате есть лишь конечная степень разветвления, которая предоставлена каждым набором начальных условий даже в радиальной геометрии, и не очевидно, как описать продолжающийся самоподобный рост, который замечен в экспериментальных условиях или при численном моделировании. Кроме того, как упомянуто ранее, по крайней мере, в случае распространения пламени наблюдается [10] *ускорение* фронта со временем. Такое явление невозможно, когда число полюсов сохраняется. Поэтому заманчиво предположить, что шум может быть важная роль в воздействии на фактические явления роста, которые наблюдаются в таких системах. Фактически, воздействие шума на нестабильную динамику фронта соответственно не освещено в литературе. С точки зрения аналитических методик шум может, конечно, генерировать новые полюса, даже если у начальных условий было конечное число полюсов. Тема динамики полюсов в присутствии белого шума, и взаимодействие между случайными колебаниями и детерминированным распространением фронта - основные проблемы этой главы.

Мы решили изучить пример распространения пламени, а не Лапласовского роста, просто потому, что у него есть аналитическое описание в терминах полюсов также и в экспериментально соответствующем случае конечной вязкости. Мы хотим начать наше исследование с геометрии плоского канала. Причина - то, что в радиальной геометрии более трудно разделить эффекты внешнего шума и начальных условий. В конце концов, первоначально система может содержать бесконечно много полюсов, очень далеко в бесконечности в комплексной плоскости (и поэтому имеющие бесконечно маленькое влияние на форму поверхности). Так как рост радиуса изменяет стабильность системы, все больше этих полюсов могло бы падать к оси действительных значений и стать заметным. В геометрии плоского канала анализ эффекта начальных условий является относительно прямолинейным, и можно понять его перед концентрацией на (более интересных) эффектах от внешнего шума [12]. Основная причина для этого - то, что в этой геометрии бесшумное устойчивое стационарное решение для растущего фронта известно аналитически. Как описано в Разделе II, в канале ширины $L$ стационарное решение дано в терминах $N(L)$ полюсов, которые группируются на линии, параллельной мнимой оси. Можно показать, что для любого числа полюсов в начальных условиях это единственный аттрактор динамики полюсов. После установления этого стационарного состояния мы можем начать систематически исследовать эффекты внешнего шума на это решение. Как заявлено прежде, в радиальной геометрии нет никакого устойчивого стационарного состояния с конечным числом полюсов, и разъединение влияния начальных условий и внешних воздействий является менее прямым ( [19] и глава 10). Мы покажем позднее, что идеи, предоставленные в этой главе, имеют значение и для радиального роста, и поэтому также будут обсуждаться впоследствии. У нас есть ряд целей в этой главе. Во-первых, после описания декомпозиции решения на полюса, описания динамики полюсов и основного стационарного состояния, мы представим анализ стабильности решений проблемы распространения пламени в геометрии плоского канала. Он покажет, что решение в виде гигантского излома хоть линейно устойчиво, но нелинейно является неустойчивым. Эти результаты, которые описаны в Разделе III, могут быть получены или линеаризацией динамики вокруг решения в виде гигантского излома, чтобы изучить собственные значения стабильности, или исследуя возмущения в форме полюсов в комплексной плоскости. Основной результат Раздела III состоит в том, что там существует одна мода Голдстоуна и две моды, собственные значения которых периодически попадают на ось действительных значений, когда размер системы $L$ увеличивается. Таким образом, система незначительно устойчива при определенных значениях $L$, и всегда неустойчива нелинейно, позволяя конечным возмущениям ввести новые полюса в систему. Эта идея позволяет нам понять соотношение между размером системы и влиянием шумов. В Разделе IV мы обсуждаем динамику релаксации, которая следует после старта системы с "маленькими" начальными условиями. Мы изучаем процесс огрубления, который ведет к конечному решению в виде гигантского острого излома, и понять типичные временные рамки, которые существуют в нашей динамике. Мы предлагаем в этом Разделе некоторые результаты численного моделирования, которые интерпретируются в более поздних разделах. В Разделе V мы сосредотачиваемся на явлении ускорения фронта пламени и его происхождения из существования шума. В условиях без шума скорость фронта пламени в конечном канале ограничена [12]. Это можно показать или при использовании динамики полюсов или непосредственно из уравнения движения. Мы представим результаты числового моделирования, где шум можно менять, и покажем, как скорость фронта пламени зависит от уровня шума и размера системы.

Основные результаты таковы: (i) Шум ответственен за появление новых полюсов системе; (ii) Для низких уровней шума скорость фронта пламени масштабируется как размер системы с характерной экспонентой; (iii) есть фазовый переход при определенном значение шумового уровня (зависящем от размера системы), после которого поведение системы меняется качественно; (iv) После фазового перехода скорость движения фронта пламени меняется очень быстро с изменением уровня шума. В последнем Разделе мы отмечаем значение этих наблюдений для поведения масштабирования проблемы роста в радиальной геометрии, и представляем резюме и заключения.

## 2.2 Уравнения движения и декомпозиции на полюса в геометрии плоского канала

Известно, что плоское пламя, свободно перемещающееся через первоначально неподвижные гомогенные горючие смеси, существенно нестабильно. Было известно, что такое пламя развивает на фронте характерные структуры, которые включают острые изломы, и что при обычных экспериментальных условиях фронт пламени ускоряется с течением времени. Модель в $1+1$ размерностях, которая описывает распространение фронта пламени в каналах ширины $\tilde{L}$, была предложена в [9]. Оно записано в терминах позиции $h(x,t)$ фронта пламени выше $x$-оси. После соответствующего перемасштабирования оно принимает форму:

$$\frac{\partial h(x,t)}{\partial t} = \frac{1}{2}\left[\frac{\partial h(x,t)}{\partial x}\right]^2 + \nu \frac{\partial^2 h(x,t)}{\partial x^2} + I\{h(x,t)\} + 1 \ . \qquad (2.1)$$

Область описания – домен $0 < x < \tilde{L}$, $\nu$ - параметр, и мы используем периодические граничные условия. Функционал $I[h(x,t)]$ - преобразование Гильберта, которое удобно определить в терминах пространственного Фурье преобразования:

$$h(x,t) = \int_{-\infty}^{\infty} e^{ikx} \hat{h}(k,t) dk \qquad (2.2)$$

$$I[h(k,t)] = |k|\hat{h}(k,t) \qquad (2.3)$$

С целью представления декомпозиции на полюса удобно повторно масштабировать домен описания к $0 < \theta < 2\pi$. Выполняя это перемасштабирование и переобозначая получающиеся величины теми же самыми обозначениями, мы имеем:

$$\frac{\partial h(\theta,t)}{\partial t} = \frac{1}{2L^2}\left[\frac{\partial h(\theta,t)}{\partial \theta}\right]^2 + \frac{\nu}{L^2}\frac{\partial^2 h(\theta,t)}{\partial \theta^2}$$
$$+ \frac{1}{L}I\{h(\theta,t)\} + 1 \ . \qquad (2.4)$$

В этом уравнении $L = \tilde{L}/2\pi$. Затем мы изменяем переменные на $u(\theta,t) \equiv \partial h(\theta,t)/\partial \theta$. Мы находим:

$$\frac{\partial u(\theta,t)}{\partial t} = \frac{u(\theta,t)}{L^2}\frac{\partial u(\theta,t)}{\partial \theta} + \frac{\nu}{L^2}\frac{\partial^2 u(\theta,t)}{\partial \theta^2} + \frac{1}{L}I\{u(\theta,t)\} \ . \qquad (2.5)$$

Хорошо известно, что плоское решение этого уравнения линейно нестабильно. Линейный спектр в $k$ - представлении является:

$$\omega_k = |k|/L - \nu k^2/L^2 \ . \qquad (2.6)$$

Существует типичный масштаб $k_{max}$, который является последней нестабильной модой:

$$k_{max} = \frac{L}{\nu} \ . \qquad (2.7)$$

Нелинейные эффекты стабилизируют новое стационарное состояние, которое обсуждается ниже. Выдающаяся особенность решений этого уравнения - появление структур в виде квази-острых изломов на распространяющемся фронте. Поэтому представление в терминах Фурье разложения очень неэффективно. Скорее, кажется очень стоящим представить такие решения в терминах сумм функций с полюсами в комплексной плоскости. Как будет показано ниже, позиция острого излома на фронте определена действительной координатой полюса, тогда как высота острого излома определяется мнимой координатой. Кроме того, будет замечено, что динамика распространения фронта может быть хорошо описана в терминах динамики полюсов. Следуя [12, 19, 38, 39] мы разворачиваем решения $u(\theta,t)$ в функциях, которые зависят от $N$ полюсов, чья позиция $z_j(t) \equiv x_j(t) + iy_j(t)$ в комплексной плоскости зависит от времени следующим образом:

$$u(\theta,t) = \nu \sum_{j=1}^{N} \cot\left[\frac{\theta - z_j(t)}{2}\right] + c.c.$$
$$= \nu \sum_{j=1}^{N} \frac{2\sin[\theta - x_j(t)]}{\cosh[y_j(t)] - \cos[\theta - x_j(t)]} \ , \qquad (2.8)$$

$$h(\theta, t) = 2\nu \sum_{j=1}^{N} \ln\left[\cosh(y_j(t)) - \cos(\theta - x_j(t))\right] + C(t) \ . \qquad (2.9)$$

В (2.9) $C(t)$ является функцией времени. Функция (2.9) является суперпозицией квази-острых изломов (то есть острые изломы, которые округлены в точке излома). Вещественная часть позиции полюса (то есть $x_j$) является координатой (в домене $[0, 2\pi]$) максимума квази-острого излома, а мнимая части позиции полюса (то есть $y_j$) связана с глубиной квази-острого излома. Когда $y_j$ уменьшается, глубина острого излома увеличивается. Когда $y_j \to 0$, глубина излома стремится к бесконечности. Наоборот, когда $y_j \to \infty$ глубина уменьшается и стремится к нулю. Основное преимущество этого представления состоит в том, что распространение и неровности фронта могут быть описаны через динамику полюсов. Подставляя (2.8) в (2.5) мы получаем следующие обыкновенные дифференциальные уравнения для позиций полюсов:

$$-L^2 \frac{dz_j}{dt} = \left[\nu \sum_{k=1, k\neq j}^{2N} \cot\left(\frac{z_j - z_k}{2}\right) + i\frac{L}{2}sign[Im(z_j)]\right]. \qquad (2.10)$$

Отметим, что в (2.8), из-за комплексного сопряжения, мы имеем $2N$ полюса, которые упорядочены в парах таким образом, что для $j < N$ $z_{j+N} = \bar{z}_j$. Во второй сумме в (2.8) каждая пара полюсов вносит вклад в один член. В уравнении (2.10) мы снова используем $2N$ полюса, так как все они взаимодействуют. Мы можем написать динамику полюса в терминах вещественных и мнимых частей $x_j$ и $y_j$. Из-за организации в пары достаточно написать уравнение либо для $y_j > 0$ либо для $y_j < 0$. Мы выбираем первое. Уравнения для позиций полюсов записываются:

$$-L^2 \frac{dx_j}{dt} = \nu \sum_{k=1,k\neq j}^{N} \sin(x_j - x_k)\Big[[\cosh(y_j - y_k) \qquad (2.11)$$
$$- \cos(x_j - x_k)]^{-1} + [\cosh(y_j + y_k) - \cos(x_j - x_k)]^{-1}\Big]$$

$$L^2 \frac{dy_j}{dt} = \nu \sum_{k=1,k\neq j}^{N} \Big(\frac{\sinh(y_j - y_k)}{\cosh(y_j - y_k) - \cos(x_j - x_k)}$$
$$+ \frac{\sinh(y_j + y_k)}{\cosh(y_j + y_k) - \cos(x_j - x_k)}\Big) + \nu \coth(y_j) - L. \qquad (2.12)$$

Отметим, что если начальные условия дифференциального уравнения (2.5) выражены в конечном числе полюсов, уравнения движения сохраняют это число как функцию времени. С другой стороны, это может быть неустойчивым решением для частичного дифференциального уравнения, и шум может изменить число полюсов. Эта проблема будет исследована подробно в Разделе 2.5.

## 2.3 Линейный анализ стабильности в канальной геометрии

В этом разделе мы обсуждаем линейную стабильность решения - TFH острого излома. С этой целью мы сначала используем Eq. (2.8), чтобы написать стаци-онарное решение $u_s(\theta)$ в форме:

$$u_s(\theta) = \nu \sum_{j=1}^{N} \frac{2\sin[\theta - x_s]}{\cosh[y_j] - \cos[\theta - x_s]} , \qquad (2.13)$$

где $x_s$ вещественная (общая) компонента стационарных полюсов и $y_j$ их постоянная мнимая компонента. Чтобы изучить стабильность этого решения, мы должны определить фактические позиции $y_j$. Это сделано в численной форме, интегрируя уравнения движения для полюсов, начиная с $N$ полюсов в начальных позициях и ожидающих релаксацию. Затем это решение возмущается маленьким возмущением $\phi(\theta, t)$: $u(\theta, t) = u_s(\theta) + \phi(\theta, t)$. Линеаризация динамики для малых $\phi$ будет иметь результатом уравнения движения

$$\frac{\partial \phi(\theta, t)}{\partial t} = \frac{1}{L^2}\left[\partial_\theta[u_s(\theta)\phi(\theta,t)] + \nu \partial_\theta^2 \phi(\theta,t)\right]$$
$$+ \frac{1}{L}I(\phi(\theta,t)) . \qquad (2.14)$$

### 2.3.1 Разложение Фурье и собственные значения

Линейное уравнение может быть разложено по Фурье модам согласно:

$$\phi(\theta, t) = \sum_{k=-\infty}^{\infty} \hat{\phi}_k(t) e^{ik\theta} \qquad (2.15)$$

$$u_s(\theta) = -2\nu i \sum_{k=-\infty}^{\infty} \sum_{j=1}^{N} \text{sign}(k) e^{-|k|y_j} e^{ik\theta} \qquad (2.16)$$

В этих суммах дискретное $k$ значение пробегает все целые числа. Подставляя в (2.14) мы получаем уравнения:

$$\frac{d\hat{\phi}_k(t))}{dt} = \sum_n a_{kn} \hat{\phi}_n(t) , \qquad (2.17)$$

Где $a_{kn}$, бесконечная матрица, чьи элементы даются следующим выражением

$$a_{kk} = \frac{|k|}{L} - \frac{\nu}{L^2}k^2 \qquad (2.18)$$

$$a_{kn} = \frac{k}{L^2}\text{sign}(k-n)(2\nu \sum_{j=1}^{N} e^{-|k-n|y_j}) \quad k \neq n . \qquad (2.19)$$

Чтобы найти собственные значения этой матрицы, мы должны обрезать ее на некотором $k$-векторе $k^*$. Выбор $k^*$ может быть основан на линейном анализе стабильности плоского фронта. Масштаб $k_{max}$, смотри (2.7), является наибольшим $k$, который все еще соответствует линейной нестабильности. Мы должны выбрать $k^* > k_{max}$ и проверить выбор сходимостью собственных значений. Выбранное значение $k^*$ в наших численных данных составляло $4k_{max}$. Результаты для собственных значений низшего порядка матрицы $a_{kn}$, которые были получены из сходившихся численных расчетов, представлены на Рис. 8.1.

Собственные значения умножены $L^2$ и изображаются как функция $L$. Мы упорядочиваем собственные значения в порядке убывания и обозначаем их как $|\lambda_0| \leq |\lambda_1| \leq |\lambda_2| \ldots$.

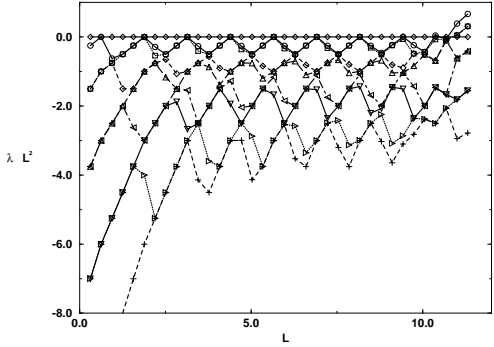

Рис. 2.1: Первые 10 самых высоких собственных значений матрицы стабильности с $\nu = \pi/5$, умноженные на квадрат размера системы $L^2$ как функция размера системы $L$. Отметим, что все собственные значения колеблются вокруг фиксированных значений в этом представлении, и что самые высокие два собственных значения периодически обнуляются.

Рис. 8.1 содержит странный результат - положительные собственные значения при большом L. Один из методов, чтобы проверить некоторый числовой результат, - это сделать аналитическое исследование. Например, в Главе 3 мы делаем детализированное аналитическое исследование для числового результата на рис. 8.1 и получаем, что все собственные значения не положительны. Действительно, два типа мод существует. Первый тип связан со смещением полюсов в гигантском остром изломе. Из-за притяжения полюсов гигантский острый излом устойчив относительно продольного смещения полюсов и таким образом соответствующие собственные значения не позитивны. Для поперечного смещения существует функция Ляпунова и таким образом гигантский острый излом устойчив относительно поперечного смещения, и соответствующие собственные значения не положительны. Второй тип мод связан с дополнительными полюсами. Эти полюса стремятся к бесконечности из-за отталкивания от полюсов гигантских острого излома $N(L)$. Таким образом, соответствующие собственные значения также не позитивны. Соответственно, позитивные собственные значения при большом L - численный артифакт. Рисунок демонстрирует ряд качественных явлений:

1. Существует очевидная Голдстоуновская или трансляционная мода $u'_s(\theta)$ с собственным значением $\lambda_0 = 0$, которая показана ромбами на Рис. 8.1. Это собственная мода происходит из-за галилевской инвариантности уравнений движения.

2. Собственные значения периодически колеблются между значениями, которые являются независимыми от $L$ в представлении, где они умножаются на $L^2$. Другими словами, с точностью до колебательного поведения собственные значения зависят от $L$ как $L^{-2}$.

3. Собственные значения $\lambda_1$ и $\lambda_2$, которые изображаются квадратными и круглыми символами на Рис. 8.1, периодически пересекают нулевую ось. Функциональная зависимость в этом представлении кажется почти кусочно-линейной

4. , более высокие собственные значения также показывают подобное качественное поведение, но не достигают при этом нуля. Отметим, что решение становится очень слабоустойчивым для тех значений $L$, для которых собственные значения $\lambda_1$ и $\lambda_2$ обнуляются. Вид $L^{-2}$ зависимости спектра от $L$ указывает, что решение становится более чувствительным к шуму, когда $L$ увеличивается.

### 2.3.2 Качественное понимание, используя анализ с помощью разложения на полюса.

Самые интересные качественные аспекты – это перечисленные выше пункты 2 и 3. Чтобы понять их, полезно возвратиться к описанию с помощью полюсов, и сосредоточиться на Eq. (1.11). Это уравнение описывает динамику единственного далекого полюса. Заметим, прежде всего, что это уравнение демонстрирует, что для *фиксированной $L$* постоянное число полюсов – это целочисленная часть (1.10). Определим теперь число $\alpha$, $0 \leq \alpha \leq 1$, согласно

$$\alpha = \left[\frac{1}{2}\left(\frac{L}{\nu}+1\right)\right] - \frac{1}{2}\left(\frac{L}{\nu}-1\right) . \qquad (2.20)$$

Используя это число мы переписываем уравнение(1.11) как

$$\frac{dy_a}{dt} \approx \frac{2\nu}{L^2}\alpha . \qquad (2.21)$$

Когда $L$ увеличивается, $\alpha$ становится кусочно-линейной и периодически колеблющейся между нулем и единицей. Это показывает, что некий отдаленный полюс, который добавляется к решению в виде гигантского острого излома, обычно выталкивается в бесконечность кроме тех случаев, когда $\alpha$ точно нуль и система становятся слабоустойчивой к добавлению нового полюса.

Чтобы связать это с анализом линейной стабильности, мы замечаем из Eq. (2.8), что решение для одиночного удаленного полюса (то есть с очень большим $y$) может быть написано как

$$u(\theta, t) = 4\nu e^{-y(t)}\sin(\theta - x(t)) . \qquad (2.22)$$

Предположим, что мы добавляем к нашему решению в виде гигантского острого излома возмущение такой функциональной формы. Из Eq. (2.21) мы знаем, что $y$ растет линейно во времени, и, следовательно, это решение затухает экспоненциально во времени. Скорость затухания – это собственное значение, получаемое из решения проблемы линейной стабильности, и из Eq. (2.21) мы находим как $1/L^2$ зависимость, так и периодически возникающую слабую устойчивость. Мы должны отметить, что эти доводы дают нам большую часть зависимости собственных значений от $L$, но не всю. Переменная $\alpha$ повышается от нуля до единицы периодически, но после достижения единицы падает в ноль немедленно. Соответственно, если бы самое высокое не нулевое собственное значение было полностью определено анализом с помощью полюсов, то мы ожидали бы, что это собственное значение будет вести себя как сплошная линия, показанная на Рис. 8.2.

Фактическое самое высокое собственное значение, вычисленное из матрицы стабильности, показано ромбами, соединенными пунктиром. Ясно, что полюсный анализ дает нам, большое качественное и количественное понимание, но не все особенности.

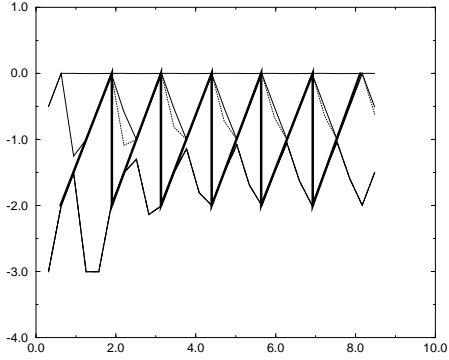

Рис. 2.2: Сравнение четырех самых высоких собственных значений матрицы стабильности, полученных численно, с предсказанием полюсного анализа. Собственные значения матрицы стабильности : $\lambda_0$, $\lambda_1$, $\lambda_2$ и $\lambda_3$. Полюсный анализ (сплошная линия) обеспечивает качественное понимание стабильности, и перекрывается с самым высоким собственным значением на одной половине диапазона, и с четвертым собственным значением по другой половине.

### 2.3.3 Динамика вблизи слабой устойчивости

Открытие слабой устойчивости при изолированных значениях $L$ поднимает вопросы относительно судьбы полюсов, которые добавлены с очень большом $y$-ом в определенных $x$-позициях. Можно показать, что, когда система становится незначительно устойчивой, новый полюс может быть добавлен к тем, которые уже существуют в решении в виде гигантского острого излома. Мы помним, что у этих полюсов есть общая $\theta$ позиция, которую мы обозначаем как $\theta = \theta_c$. Судьба нового полюса, добавляемого в бесконечности, зависит от его $\theta$ позиции. В то же время, она минимальна, когда $\theta_a - \theta_c = \pi$. Это следует из того факта, что косинусный член имеет значение $+1$, когда $\theta_a = \theta_c$, и значение $-1$, когда $\theta_a - \theta_c = \pi$. Для большого $y$ разницы между членами в сумме принимают свое минимальное значение, когда cos член равен $-1$, и максимальные значения при $+1$. Для бесконечно больших $y_a$ уравнениями движения являются (1.11), которые независимы от $\theta_a$. Так как правая часть этого уравнения становится нулем в точках со слабой устойчивостью, мы заключаем, что для очень большого, но конечного $y_a$ $dy_a/dt$ изменяет знак с положительного на отрицательный, когда $\theta_a - \theta_c$ изменяется от нуля до $\pi$. Значение этого наблюдения в том, что самые нестабильные точки в системе – это те точки, которые дальше всего удалены от гигантского острого излома. Интересно обсудить судьбу полюса, который добавлен к системе в такой позиции. С точки зрения динамики полюса $\theta = \theta_c + \pi$ - неустойчивая точка для движения вдоль $\theta$ оси. Притяжение к гигантскому острому излому исчезает точно в этой точке. Если мы запустим полюс с очень большой $y_a$ близко к этому значению $\theta$, то падение этого полюса вдоль $y$ координаты будет быстрее, чем боковое движение к гигантскому острому излому. Мы ожидаем увидеть, поэтому, создание маленького острого излома при $\theta$ значении близко к $\pi$, который предшествует более поздней стадии движения, в котором маленький острый излом двигается до слияния с гигантским острым выступом. После слияния нового полюса с гигантским острым изломом все существующие полюса продвинутся и самый далекий полюс в $y_{max}$ будет выброшен в бесконечность. Мы позже объясним, что этот тип динамики происходит и в устойчивых системах, которые подвергаются шуму. Искажение генерирует далекие полюса

### 2.3.4 Возмущенная Система.

Интуиция, полученная из рассмотрения выше, может использоваться, чтобы обсудить проблему стабильности устойчивой системы к *большим* возмущениям. Другими словами, мы можем захотеть добавить к системе новые полюса при конечных значениях $y$ и спросить об их судьбе. Мы сначала показываем в этом подразделе, что полюса, начальная величина которых $y$ имеет значение ниже $y_{max} \sim \log(L^2/\nu^2)$, будут притянуты к вещественной оси. Сценарий подобен тому, который был описан в последнем параграфе.

Предположите, что мы создаем устойчивую систему с гигантским острым изломом в $\theta_c = 0$ с полюсами, распределенными вдоль $y$ ось до $y_{max}$. Мы знаем, что сумма всех сил, которые действуют на верхний полюс, является нулем. Рассмотрим теперь дополнительный полюс, вставленный в позицию $(\pi, y_{max})$. Очевидно из Eq. (2.12), что силы, действующие на этот полюс, переместят его вниз. С другой стороны, если его начальная позиция будет значительно выше $y_{max}$, то сила, действующая на него, будет отталкивающей по направлению к бесконечности. Мы видим, что этот простой параметр идентифицирует $y_{max}$ как типичный масштаб для нелинейной неустойчивости.

Затем мы оцениваем $y_{max}$ и интерпретируем свой результат в терминах *амплитуды* возмущения фронта пламени. Мы объяснили выше, что высшая позиция полюса, вносящего неустойчивость, колеблется между минимальным значением и бесконечностью при изменении $L$. Мы хотим оценить характерный масштаб минимального значения $y_{max}(L)$. С этой целью мы используем результат ref. [12] относительно устойчивого распределения позиций полюсов в устойчивой большой системе. Параметризация [12] отличается от нашей; чтобы перейти от нашей параметризации в Eq. (2.5) к их, мы должны повторно масштабировать $u$ с помощью $L^{-1}$ и $t$ с помощью $L$. Параметр $\nu$ в их параметризации - это $\nu/L$ в нашей. Согласно [12] число полюсов между $y$ и $y + dy$ дано $\rho(y)dy$, где плотность $\rho(y)$ определяется из:

$$\rho(y) = \frac{L}{\pi^2 \nu} \ln[\coth(|y|/4)] \ . \qquad (2.23)$$

Чтобы оценить минимальное значение $y_{max}$, мы требуем, чтобы хвост распределения $\rho(y)$, проинтегрированный между этим значением и бесконечностью, позволил только один полюс. Другими словами,

$$\int_{y_{max}}^{\infty} dy \rho(y) \approx 1 \ . \qquad (2.24)$$

Раскладывая (2.23) для большого $y$ и интегрируя явно результат в (2.24) мы получаем оценку

$$y_{max} \approx 2 \ln \left[ \frac{4L}{\pi^2 \nu} \right] \qquad (2.25)$$

Для большого $L$ результатом является $y_{max} \approx \ln(\frac{L^2}{\nu^2})$. Если мы теперь добавляем дополнительный полюс в позиции $(\theta, y_{max})$ это эквивалентно возмущению решения $u(\theta, t)$ функцией $\nu e^{-y_{max}} \sin(\theta)$, как это можно видеть непосредственно из

(2.8). Мы, таким образом, заключаем, что система неустойчива к возмущению *большему* чем

$$u(\theta) \sim \nu^3 \sin(\theta)/L^2 \ . \tag{2.26}$$

Это указывает на очень сильную зависимость чувствительности решения (в виде гигантского острого излома) по отношению к внешнему возмущению от размера системы. Этот факт будет важным компонентом в нашем обсуждении систем с шумом.

## 2.4 Начальные Условия, Разложение на Полюса и Огрубление.

В этом разделе мы показываем сначала, что любые начальные условия могут быть приближены декомпозицией на полюса. Позже, мы показываем, что динамика достаточно гладких начальных условий может быть хорошо понята благодаря декомпозиции на полюса. Наконец мы используем эту картину, чтобы описать *обратный каскад* острых изломов внутри гигантского острого излома, который является конечным устойчивым состоянием. Под обратным каскадом мы подразумеваем нелинейный процесс огрубления, в котором мелкие масштабы объединяются в пользу больших масштабов, и, наконец, система насыщается в наибольшем возможном масштабе [40].

### 2.4.1 Разложение на полюса: Общие комментарии

Фундаментальный вопрос - то, сколько полюсов необходимо, чтобы описать любое данное начальное условие. Ответ, конечно, зависит от того, насколько гладкими являются начальные условия. Предположим также, что у нас есть начальная функция $u(\theta, t=0)$, которая является $2\pi$-периодической и которая в момент времени $t = 0$ допускает представление Фурье

$$u(\theta) = \sum_{k=1}^{\infty} A_k \sin\left(k\theta + \phi_k\right) \ , \tag{2.27}$$

с $A_k > 0$ для всех $k$. Предположим, что мы хотим найти представление декомпозиции на полюса $u_p(\theta)$ таким образом, что

$$|u_p(\theta) - u(\theta)| \leq \epsilon \qquad \theta \ , \tag{2.28}$$

где $\epsilon$ это данная требуемая точность. Если $u(\theta)$ дифференцируемо, мы можем обрезать разложение Фурье при некотором конечном $k = K$, зная, что остаток меньше чем, скажем, $\epsilon/2$. Выберем теперь большое число $M$ и маленькое число $\Delta \ll 1/M$ и напишем представление в виде полюсов для $u_p(\theta)$ как

$$u_p(\theta) = \sum_{k=1}^{K} \sum_{p=0}^{M-1} \frac{2k \sin\left(k\theta + \phi_k\right)}{\cosh\left[k(y_k + p\Delta)\right] - \cos(k\theta + \phi_k)} \ . \tag{2.29}$$

Для того, чтобы увидеть, что это представление является частной формой общей формулы (2.8) мы используем следующие два равенства

$$\sum_{k=0}^{\infty} e^{-kt} \sin xk = \frac{1}{2} \frac{\sin x}{\cosh t - \cos x} \ , \tag{2.30}$$

$$\sum_{k=0}^{K-1} \sin(x+ky) = \sin(x+\frac{K-1}{2}y)\sin\frac{Ky}{2}\operatorname{cosec}\frac{y}{2}\,. \tag{2.31}$$

Из них следует третье равенство

$$\sum_{j=0}^{K-1} \frac{2\sin\left(x-\frac{2\pi j}{K}+\phi\right)}{\cosh y - \cos\left(x-\frac{2\pi j}{K}+\phi\right)}$$
$$= \frac{2K\sin\left(Kx+\phi\right)}{\cosh Ky - \cos\left(Kx+\phi\right)}\,. \tag{2.32}$$

Заметим, что левая часть (2.32) имеет форму (2.8) с $K$ полюсами, позиции которых - все на линии $y_j = y$ и чьи $x_j$ находятся в узлах Затем мы используем (2.30), чтобы переписать (2.29) в форме

$$u_p(\theta) = \sum_{k=1}^{K}\sum_{p=0}^{M-1}\sum_{n=1}^{\infty} 4k e^{-nk(y_k+p\Delta)}\sin\left(nk\theta+n\phi_k\right)\,. \tag{2.33}$$

Меняя порядок суммирования по $n$ и по $p$, мы можем использовать формулу для суммы геометрической прогрессии для суммы по $p$. Обозначая

$$b_{n,k} \equiv \sum_{p=0}^{M-1} e^{-nkp\Delta} = \frac{1-e^{-Mkn\Delta}}{1-e^{-kn\Delta}}\,, \tag{2.34}$$

мы находим

$$\begin{aligned} u_p(\theta) &= \sum_{k=1}^{K}\sum_{n=1}^{\infty} 4k b_{n,k} e^{-nky_k}\sin\left(nk\theta+n\phi_k\right) \\ &= \sum_{k=1}^{K}\sum_{n=2}^{\infty} 4k b_{n,k} e^{-nky_k}\sin\left(nk\theta+n\phi_k\right) \\ &+ \sum_{k=1}^{K} 4k b_{1,k} e^{-ky_k}\sin\left(k\theta+\phi_k\right)\,. \end{aligned} \tag{2.35}$$

Сравнивая теперь второй член правой части (2.35) с (2.27) мы можем получить равенство

$$e^{-ky_k} = \frac{A_k}{4k b_{1,k}} \tag{2.36}$$

Первый член может быть затем ограничен сверху как

$$\Big|\sum_{k=1}^{K}\sum_{n=2}^{\infty} 4k b_{n,k} e^{-nky_k}\sin\left(nk\theta+n\phi_k\right)\Big| \tag{2.37}$$
$$\leq \sum_{k=1}^{K}\sum_{n=2}^{\infty}\Big|4k b_{n,k}\left[\frac{A_k}{4k b_{1,k}}\right]^n\sin\left(nk\theta+n\phi_k\right)\Big|.$$

Синусная функция и фактор $(4K)^{1-n}$ могут быть заменены на единицу, и мы можем ограничить правую часть (2.37) как следующее

$$\sum_{k=1}^{K}\sum_{n=2}^{\infty}\left[\frac{A_k}{b_{1,k}}\right]^n b_{n,k} \leq \sum_{k=1}^{K} A_k \sum_{n=1}^{\infty}\left[\frac{A_k}{b_{1,k}}\right]^n\,, \tag{2.38}$$

где мы использовали факт, что $b_{n,k} \leq b_{1,k}$, который следует напрямую из (2.34). Используем теперь тот факт, что $b_{1,K} \leq b_{1,k}$ для каждого $k \leq K$ и что $A_k$ ограничен некоторым конечным $C$, так как это коэффициент Фурье. Соответственно мы можем ограничить (2.38) величиной $C^2 K/(b_{1,K} - C)$. Мы можем свободно выбрать параметры $\Delta$ и $M$. Следовательно, мы можем сделать $b_{1,K}$ такой величины, как мы хотим. Таким образом, мы можем сделать остаток ряда меньше по абсолютной величине, чем $\epsilon/2$.

Вывод из этих умозаключений следующий - любое начальное условие, которое может быть представлено в виде ряда Фурье, может быть хоть и приближено, но с любой желаемой точностью декомпозировано на полюса. Число необходимых полюсов имеет порядок $K^2 \times M$. Конечно, число полюсов в таким образом сгенерированных начальных условиях может превысить число $N(L)$, найденный в Eq. (1.10). В таком случае лишние полюса будут двигаться в бесконечность и станут несущественными для динамики. Таким образом, меньшее число полюсов может быть необходимо, чтобы описать состояние в асимптотике по времени. Следует отметить, что декомпозиция на полюса избыточна; например, если есть точно один полюс в $t = 0$, и мы использовали бы вышеупомянутую методику, чтобы получить декомпозиции на полюса, то мы получили бы большое количество полюсов в нашем представлении.

### 2.4.2 Начальные стадии эволюции фронта: экспоненциальная стадия и обратный каскад

В этом разделе мы используем связь между разложением Фурье и декомпозицией на полюса, чтобы понять начальную экспоненциальную стадию эволюции фронта пламени с маленькими начальными данными $u(\theta, t = 0)$. Затем мы используем свое знание о взаимодействии полюсов, чтобы объяснить медленную динамику огрубления, дающую в итоге устойчивое стационарное решение.

Предположим, что первоначально разложение (2.27) доступно со всеми коэффициентами $A_k \ll 1$. Мы знаем из линейной неустойчивости плоского фронта пламени, что каждая компонента Фурье меняется экспоненциально во времени согласно линейному спектру (2.6). Компоненты с волновым вектором, большим чем (2.7) уменьшаются, в то время как компоненты с меньшим волновым вектором увеличиваются. Наиболее быстро растущий режим соответствует $k_c = L/2\nu$. В линейной стадии роста эта мода будет доминантной и будет деформировать плоский фронт пламени, то есть:

$$u(\theta, t) \approx A_{k_c} e^{\omega_{k_c} t} \sin(k_c \theta) . \tag{2.39}$$

Используя Eq. (2.32) для большого значения $y$ (который эквивалентен маленькому $A_{k_c}$) мы видим, что порядок $O(A_{k_c}^2)$ (2.39) может быть представлен как сумма по $L/2\nu$ полюсам, распределенными периодически вдоль $\theta$ оси. Другие нестабильные режимы также внесут подобные массивы полюсов, но для много более высоких значениях $y$, так как их амплитуда экспоненциально меньше. Кроме того, у нас есть нелинейные коррекции на представление мод в терминах полюсов. Эти коррекции могут быть снова разложены в члены, отвечающие Фурье-модам, и могут быть снова идентифицированы с полюсами, которые еще дальше поедут в направлении бесконечности вдоль $y$ оси, и с более высокими частотами. Чтобы видеть это, можно использовать Eq.(2.35), вычесть из $u_p(\theta)$ главные члены, получающиеся из разложения на полюса, и повторно разложить в ряд Фурье. Затем мы идентифицируем ведущий порядок с двойным числом полюсов, которые расположены в два раза дальше вдоль $y$ оси.

Заметим, что, даже когда все неустойчивые моды представлены, число полюсов для первого порядка декомпозиции конечно для конечного $L$. Это вытекает из факта, что существует только $L/\nu$ неустойчивых мод. Считая число полюсов, которые вводит каждая мода, мы получаем общее количество $\left(L/\nu\right)^2$ полюсов. Число $L/2\nu$ полюсов, которые связаны с самой неустойчивой модой, является точно числом, позволенным в устойчивом стационарном решении, сравнивая с (1.10). Когда полюса приближаются к вещественной оси, и острые изломы начинают развиваться, линейный анализ больше не применим, но описание в виде полюсов работает.

Опишем теперь качественный сценарий установления стационарного состояния. Во-первых, мы понимаем, что все полюса, которые принадлежат менее стабильным режимам, будут выброшены в бесконечность. Чтобы увидеть это, представим систему на данном этапе как массив несвязанных систем с масштабом порядка единицы. У каждой такой системы будет характерное значение $y$. Как мы уже обсуждали выше, полюса, которые находятся выше вдоль $y$ оси, будет выброшены в бесконечность. Поэтому система останется с $L/2\nu$ полюсами самого неустойчивого режима. Результирующее влияние полюсов, принадлежащих (нелинейно) устойчивым модам, должно разрушить совершенную периодичность полюсов неустойчивого режима. Чтобы увидеть действие этих коррекций более высокого порядка на динамику полюсов, мы повторно вспоминаем, что они тоже могут быть представлены как набор полюсов с более высокими частотами, динамика которых подобна менее неустойчивым модам, которые были только что обсуждены. Но с течением времени их влияние не увеличивается. Как только полюса устойчивых режимов становятся достаточно далекими от вещественной оси, динамика остающихся полюсов начнет развиваться согласно взаимодействиям, которые направлены вдоль вещественной оси. Эти взаимодействия намного более слабы, и результирующая динамика происходит на намного больших временных масштабах. Качественная картина представляет собой обратные каскады объединений $\theta$ позиций полюсов. Отметим, что у системы есть ряд неустойчивых фиксированных точек, которые являются 'ячеистыми решениями'. Они описываются периодическим расположением полюсов вдоль вещественной оси с частотой $k$. Эти фиксированные точки неустойчивы. Они коллапсируют под влиянием возмущения, с характерными временными масштабами (который зависит от $k$) к следующему неустойчивому решению с фиксированным числом точек на частоте $k' = k/2$. Этот процесс продолжается до $k \sim 1/L$, то есть мы достигаем гигантского острого излома, который является стационарным устойчивым решением [40]. Этот сценарий ясно виден при численных расчетах. На рисунке 8.3 мы видим временную эволюцию фронта пламени, начинающуюся из нулевых начальных условий возмущенных малым белым шумом. Нижняя кривая относится к самому раннему времени на этой картине, сразу после быстрого экспоненциального роста, и ясно виден периодический массив формирующихся острых изломов. Последовательные изображения показывают продвижение фронта пламени во времени, и демонстрирует формирование больших масштабов с более глубокими острыми изломами, которые представляют частичное объединение полюсов на одной и той же $\theta$ позиции. На Рис. 8.4

Мы показываем, что ширину и скорость этого фронта как функция времени. Распознается экспоненциальная стадия роста, в которой $L/2\nu$ полюсов приближаются $\theta$ оси, а затем виден ясный переход к намного более медленной динамике, в которой эффективный масштаб в системе растет с более медлен-

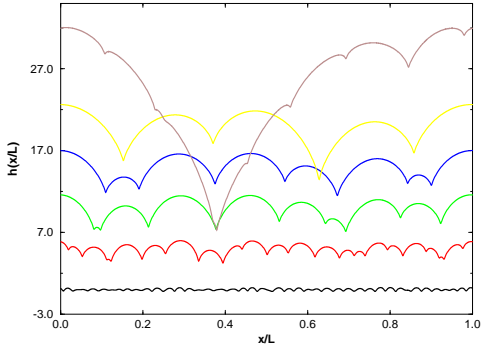

Рис. 2.3: Процесс обратного каскада огрубления, который происходит после приготовления системы со случайными, маленькими начальными условиями. Можно видеть, что в последовательные моменты времени типичный масштаб увеличивается вплоть до величины гигантского острого излома, и притягивает все другие лежащие в стороне полюса. Воздействие дополнительного численного шума приводит к появлению новых полюсов, которые появляются как побочные острые изломы, которые непрерывно притягиваются к гигантскому острому излому. Этот эффект заметен для глаза только после того, как типичный масштаб является достаточно большим, т.е. лишь на финальной стадии (смотри текст для дальнейших подробностей).

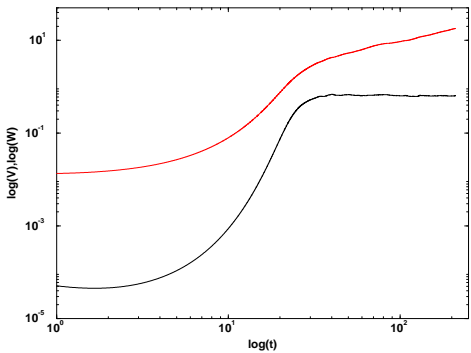

Рис. 2.4: log-log изображение скорости фронта (более низкая кривая) и ширина (верхняя кривая) как функции времени в процессе обратного каскада изображены на Рис. 8.3 в системе размера 2000 и $\nu = 1$. Обе величины демонстрируют вначале экспоненциальный рост, который переходит затем в степенной рост (после того, как $t \approx 30$). Скорость становится постоянной после этого времени, а ширина увеличивается как $t^\zeta$. Отметим, что в самое раннее время есть небольшое уменьшение в скорости; это происходит из-за затухания линейно устойчивых режимов, которые существуют в случайных начальных условиях.

ной скоростью. Медленная стадия динамики может быть понята качественно, используя предыдущую интерпретацию в виде каскада следующим образом. Если начальное число полюсов, принадлежащих неустойчивой моде, равно $L/2\nu$, то начальный эффективный линейный масштаб - это $2\nu$. Таким образом, первый шаг обратного каскада будет закончен во временном масштабе порядка $2\nu$. В этот момент эффективный линейный масштаб удваивается до $4\nu$, и второй шаг будет закончен после такого же временного масштаба. Хотелось бы знать, какой типичный масштаб длины $l_t$ наблюдается в системе во время $t$. Определением ширины фронта является $l_t = \sqrt{\frac{1}{L}\int_0^{\tilde{L}}[h(x,t)-\bar{h}]^2 dx}, \bar{h} = \frac{1}{L}\int_0^{\tilde{L}} h(x,t)dx$. Типичная ширина системы на этом этапе будет пропорциональна этому масштабу. Обозначим число шагов каскада, которые имело место, до тех пор, пока этот масштаб не был достигнут, выражением $s_l$. Полное протекшее время $t(l_t)$ является суммой

$$t(l_t) \sim \sum_{i=1}^{s_l} 2^i \ . \tag{2.40}$$

Сумма геометрической прогрессии определяется ее наибольшим членом, и мы поэтому оцениваем $t(l_t) \sim l_t$. Мы заключаем, что масштаб и ширина линейны интервалу времени, прошедшему с начального момента ($l_t \sim t^\zeta$, $\zeta = 1$). В численном моделировании без шума мы находим (см. Рис. 8.4) значение $\zeta$, который равняется $\zeta \approx 0.95 \pm 0.1$.

### 2.4.3 Обратный каскад в присутствии шума

Интересное следствие обсуждения в последнем разделе – это то, что процесс обратного каскада - это эффективный "тактовый генератор", который измеряет типичные временные рамки в этой системе. Для дальнейших целей мы должны знать типичные временные масштабы в случае, когда динамика возмущена белым шумом. С этой целью мы выполняем моделирование обратного каскада в *присутствии* внешнего шума. Основной результат, который будет использоваться в дальнейших аргументах, состоит в том, что теперь появление типичного масштаба $l_t$ происходит не после времени $t$, а скорее согласно

$$l_t \sim t^\zeta \ , \quad \zeta \approx 1.2 \pm 0.1 \ . \tag{2.41}$$

Численное подтверждение этого закона представлено на Рис. 8.5 .

Мы также находим, что скорость фронта в этом случае растет со временем согласно

$$v \sim t^\gamma \ , \quad \gamma \approx 0.48 \pm 0.05 \ . \tag{2.42}$$

Этот результат будет связан с ускорением фронта пламени при численном моделировании с учетом шума, как будет видно в следующем Разделе.

## 2.5 Ускорение Фронта Пламени, Динамика Полюсов и Шум

Главная мотивация для написания этого Раздела - наблюдение, что в радиальной геометрии то же самое уравнение движения показывает ускорение фронта пламени. Цель этого раздела состоит в том, чтобы продемонстрировать, что это явление вызвано появлением новых полюсов, сгенерированных шумом. Кроме того, мы убеждены, что многое об ускорении в радиальной геометрии мы можем

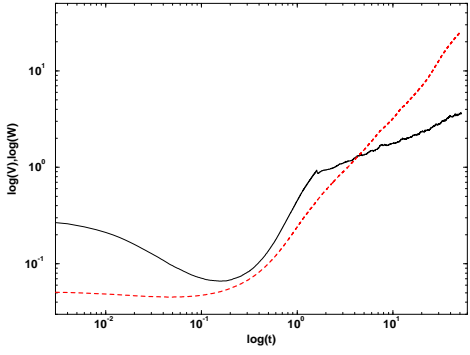

Рис. 2.5: То же самое что и на Рис. 8.4, но с совокупным белым шумом для системы размера 1000, $\nu = 0.1$ и $f = 10^{-13}$. Скорость не насыщается теперь, и экспонента $\zeta$, характеризующая увеличение ширины со временем, изменяется на $\zeta = 1.2 \pm 0.1$. Скорость увеличивается во времени как $t^\gamma$ с $\gamma \approx 0.48 \pm 0.04$.

понять, просто рассматривая воздействие шума на развитие фронта пламени в геометрии плоского канала. В работе [12] показано, что любое начальное условие, которое представимо в полюсах, приводит к единственному стационарному решению. Это решение является гигантским острым изломом, который движется с постоянной скоростью $v = 1/2$ с точностью до маленьких коррекций порядка $1/L$. В свете нашего обсуждения в последнем разделе мы ожидаем, что любое достаточно гладкое начальное условие, приводит к тому же самому стационарному решению. Таким образом, если нет никакого шума в динамике конечного канала, никакое ускорение фронта пламени не возможно. Что случается, когда мы добавляем шум к системе? Для конкретности мы вводим дополнительный член белого шума $\eta(\theta, t)$ в уравнение движения (2.5) где

$$\eta(\theta, t) = \sum_k \eta_k(t) \exp\left(ik\theta\right), \quad (2.43)$$

и амплитуды Фурье $\eta_k$ коррелированны согласно

$$<\eta_k(t)\eta_{k'}^*(t')> = \frac{f}{L}\delta_{k,k'}\delta(t-t') . \quad (2.44)$$

Мы сначала исследуем результат численного моделирования динамики, управляемой шумом, а позднее возвращаемся к теоретическому анализу.

### 2.5.1 Численное Моделирование с Шумом

Предыдущие попытки численного моделирования [11, 13] не вводили явный шум, которым можно управлять. Мы увидим в дальнейшем, что некоторые из явлений, с которыми сталкиваются при этом моделировании, могут быть приписаны неуправляемому численному шуму. Мы осуществили численное моделирование уравнения (2.5, используя псевдо-спектральный метод. Была выбрана шагающая по времени схема Адамса - Бешфорса с 2-ым порядком точности во времени. Совокупный белый шум был сгенерирован в Фурье-пространстве, выбирая $\eta_k$ для каждого $k$ из однородного плоского распределения в интервале $[-\sqrt{2\frac{f}{L}}, \sqrt{2\frac{f}{L}}]$. Мы исследовали среднюю скорость стационарного состояния фронта как функцию $L$ для фиксированного $f$ и как функцию $f$ для фиксированной $L$. Мы нашли интересные явления, которые перечислены здесь:

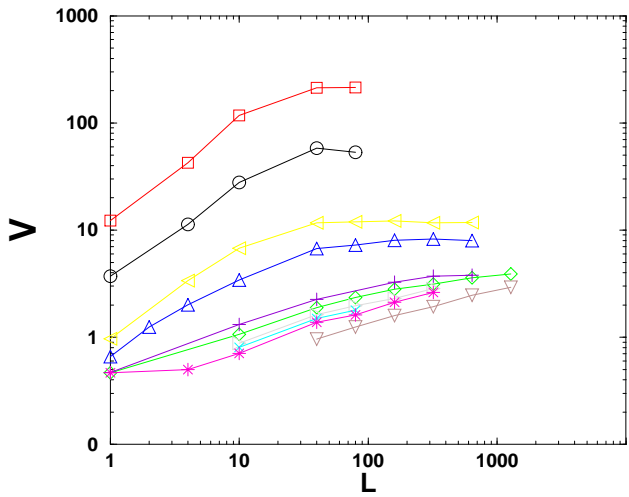

Рис. 2.6: Зависимость средней скорости $v$ от размера системы $L$ для $f^{0.5} = 0, 2.7 \times 10^{-6}, 2.7 \times 10^{-5}, 2.7 \times 10^{-4}, 2.7 \times 10^{-3}, 2.7 \times 10^{-2}, 2.7 \times 10^{-1}, 0.5, 1.3, 2.7$.

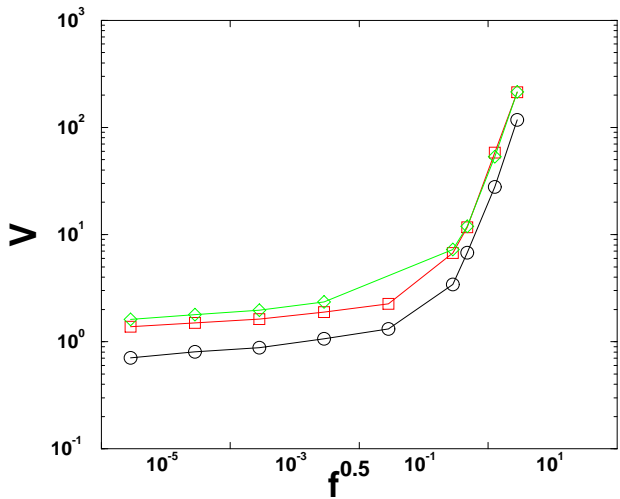

Рис. 2.7: Зависимость средней скорости $v$ от шума $f^{0.5}$ для L=10, 40, 80.

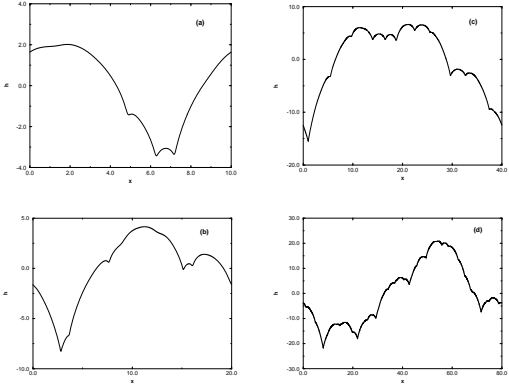

Рис. 2.8: Типичные фронты пламени для $f < f_{cr}$, где система является достаточно маленькой, чтобы не быть сильно искаженной шумом. Эффект шума в этом режиме должен добавить дополнительные маленькие острые изломы к гигантскому острому излому. На рисунках a-d мы представляем фронты для растущих размеров системы $\tilde{L} = 10, 20, 40$ и $80$ соответственно, $\nu = 0.1$. Можно заметить, что, когда размер системы растет, есть больше острых изломов с более сложной структурой.

1. На Рис. 8.7 мы видим два различных режима поведения средней скорости $v$ как функции шума $f^{0.5}$ для фиксированного размера системы L. Для шума $f$ меньшего, чем такая же фиксированная величина $f_{cr}$

$$v \sim f^\xi . \tag{2.45}$$

   Для этих значений $f$ эта зависимость очень слаба, и $\xi \approx 0.02$. Для больших значений $f$ зависимость намного более сильна.

2. На Рис. 8.6 мы можем видеть рост средней скорости $v$ как функции размера системы L. После некоторых значений L мы можем видеть насыщение скорости. Для режима $f < f_{cr}$ рост скорости может быть записан как

$$v \sim L^\mu, \quad \mu \approx 0.35 \pm 0.03 . \tag{2.46}$$

3. На Рис. 8.8 и рис. 8.9 мы можем видеть фронт пламени для $f < f_{cr}$ и $f > f_{cr}$.

### 2.5.2 Расчет Числа Полюсов в Системе

Интересная проблема, которую мы хотели бы решить здесь, чтобы лучше понять динамику полюсов, состоит в том, чтобы определить те полюса, которые существуют в нашей системе вне гигантского острого выступа. Это можно сделать, вычисляя число острых изломов (точек минимума или точек перегиба) и их позиции на интервале $\theta : [0, 2\pi]$ в каждый момент времени и изображением позиций острых изломов как функций времени, см. Рис. 8.10. На этом изображении мы можем видеть x-позиции всех острых изломов в системе как функцию времени.

Мы предположили, что наша система находится в "квази-устойчивом" состоянии большую часть времени, то есть каждый новый острый излом, который появляется в системе, включает только один полюс. Используя изображения, полученные таким образом, мы можем найти:

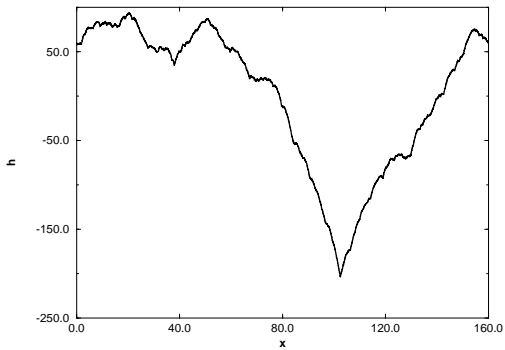

Рис. 2.9: Типичный фронт пламени для $f > f_{cr}$. Размер системы 160. Этого достаточно, чтобы вызвать качественное изменение в поведении фронта пламени: шум вводит существенные уровни структуры мелких масштабов в дополнение к острым изломам.

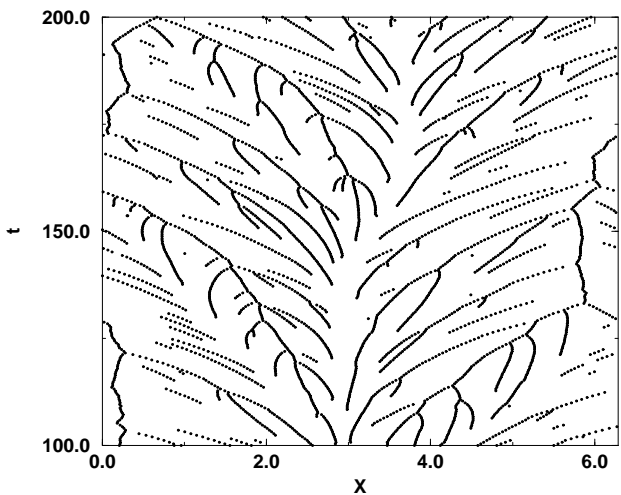

Рис. 2.10: Зависимость позиций изломов от времени. $L = 80$ $\nu = 0.1$ $f = 9 \times 10^{-6}$

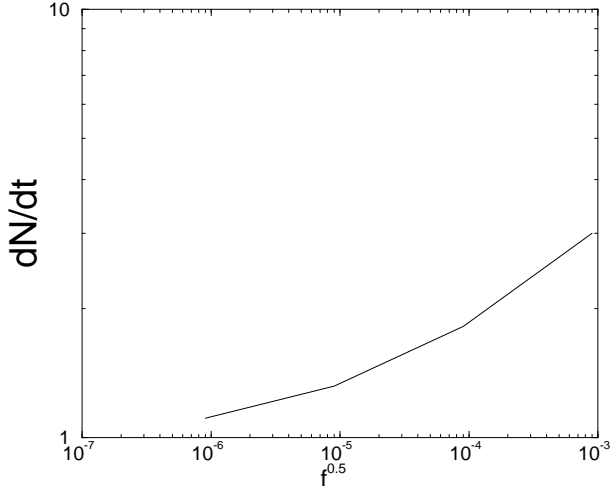

Рис. 2.11: Зависимость числа полюсов, появляющихся за единицу времени $dN/dt$ от шума $f^{0.5}$. $\nu = 0.1$ $L = 80$

1. Среднее число полюсов в системе. Вычисляя число острых изломов через некоторый момент времени и исследуя хронологию каждого острого выступа (кроме гигантского острого выступа), то есть сколько начальных острых изломов принимает участие в форматировании этого острого излома, и после усреднения числа полюсов, найденных относительно различных моментов времени, мы можем найти среднее число полюсов, которые существуют в нашей системе вне гигантского острого излома. Позвольте нам обозначать это число $\delta N$. Есть четыре режима, которые могут быть определены в зависимости от вида зависимости этого числа от шума$f$:

(i) Режим I: Такой маленький шум, что не существует ни одного излома вне гигантского излома

(ii) Режим II: Сильная зависимость числа полюсов $\delta N$ от шума $f$;

(iii) Режим III: Насыщение числа полюсов $\delta N$ от уровня шума $f$, так что это число зависит слабо от шума (Рис. 8.12);

$$\delta N \sim f^{0.03} \tag{2.47}$$

Величина насыщения $\delta N$ определяется следующей формулой (Рис. 8.14, Рис. 8.16)

$$\delta N \approx N(L)/2 \approx \frac{1}{4}\frac{L}{\nu} \tag{2.48}$$

где $N(L) \approx \frac{1}{2}\frac{L}{\nu}$ число полюсов в гигантском изломе.

(iv) Режим IV: Мы опять видим сильную зависимость числа полюсов $\delta N$ От шума $f$ (Рис. 8.12);

$$\delta N \sim f^{0.1} \tag{2.49}$$

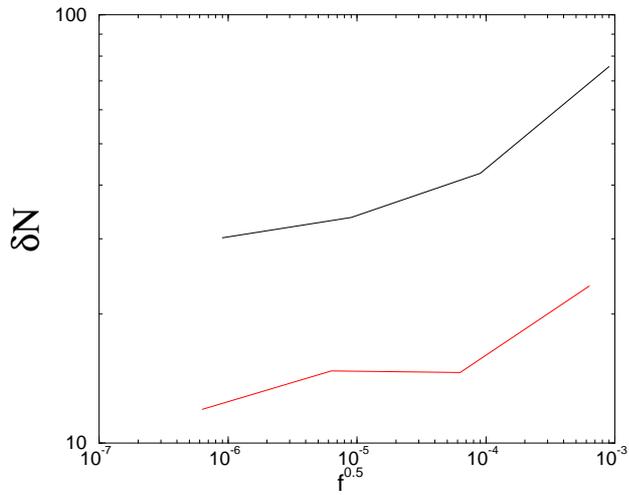

Рис. 2.12: Зависимость дополнительного числа полюсов $\delta N$ от шума $f^{0.5}$. $\nu = 0.1$ $L = 40, 80$.

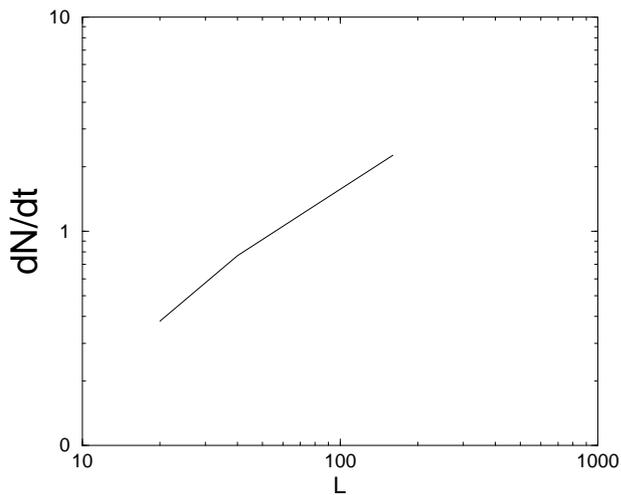

Рис. 2.13: Зависимость числа полюсов, появляющихся за единицу времени $dN/dt$ от размеров системы L. $\nu = 0.1$ $f^{0.5} = 9 \times 10^{-6}$.

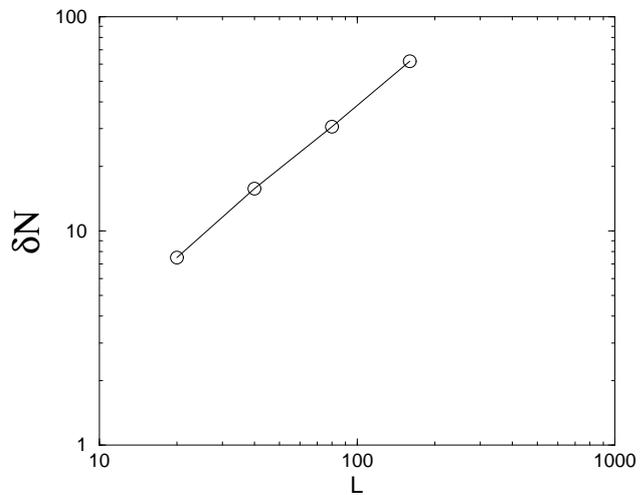

Рис. 2.14: Зависимость дополнительного числа полюсов $\delta N$ от размеров системы L. $\nu = 0.1$ $f^{0.5} = 9 \times 10^{-6}$

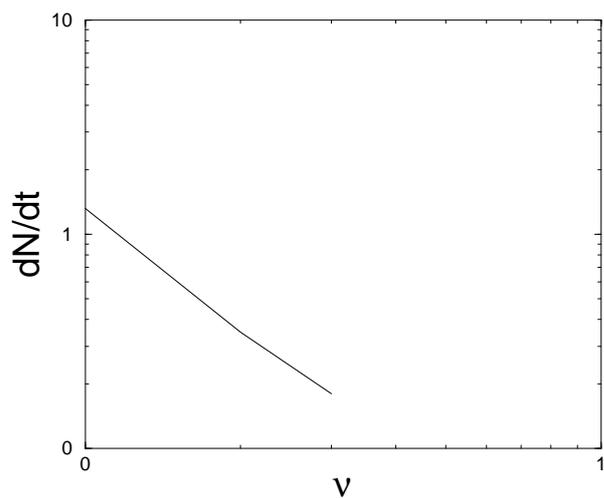

Рис. 2.15: Зависимость числа полюсов, появляющихся за единицу времени $dN/dt$ от параметра $\nu$. $L = 80$ $\nu = 0.1$.

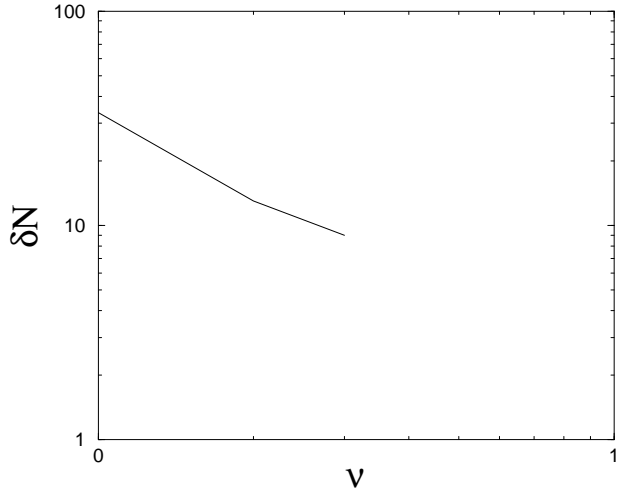

Рис. 2.16: Зависимость дополнительного числа полюсов $\delta N$ от параметра $\nu$. $L = 80$ $\nu = 0.1$.

Из-за численного шума мы можем видеть в большинстве численных расчетов только режимы III и IV. В дальнейшем, если не оговорено иное, мы будем обсуждать режим III.

2. Вычисляя число новых острых изломов, которые появляются в системе за единицу времени, мы можем найти число полюсов, которые появляются в системе в единицу времени $\frac{dN}{dt}$. В режиме III (рис. 8.11)

$$\frac{dN}{dt} \sim f^{0.03} \qquad (2.50)$$

Зависимость от $L$ и $\nu$ определяется (Рис. 8.13 и Рис. 8.15)

$$\frac{dN}{dt} \sim L^{0.8} \qquad (2.51)$$

$$\frac{dN}{dt} \sim \frac{1}{\nu^2} \qquad (2.52)$$

В режиме IV зависимость от шума определяется как следующее (Рис. 8.11):

$$\frac{dN}{dt} \sim f^{0.1} \qquad (2.53)$$

### 2.5.3 Теоретическое Обсуждение Влияния Шума

**Пороговая Нестабильность к Добавленному Шуму. Переход от Режима I к Режиму II**

Вначале мы представляем теоретические аргументы, которые объясняют чувствительность гигантского решения в виде острого излома к действию добавленного шума. Эта чувствительность увеличивается драматично с увеличением размера системы $L$. Чтобы видеть это, мы используем снова отношения между линейным анализом стабильности и динамикой полюсов.

Наш добавленный шум вводит возмущение для всех $k$-векторов. Мы показали ранее, что самая неустойчивая мода $k = 1$ с компонентой $A_1 \sin(\theta)$. Таким образом, самым эффективным шумовым возмущением является $\eta_1 \sin(\theta)$, которое может потенциально привести к росту самой непостоянной моды. Будет ли этот режим расти, зависит от амплитуды шума. Чтобы увидеть это ясно, мы возвращаемся к описанию в виде полюсов. Для маленьких значений амплитуды $A_1$ мы представляем $A_1 \sin(\theta)$ как решение для одиночного полюса функциональной формы $\nu e^{-y} \sin \theta$. $y$ позиция определяется из $y = -\log |A_1|/\nu$, и $\theta$-позиция $\theta = \pi$ - для положительного $A_1$, и $\theta = 0$ - для отрицательного $A_1$. Из анализа в Разделе III мы знаем, что для очень маленького $A_1$ полюс должен убежать в бесконечность, независимо от его $\theta$ позиции; динамика симметрична для $A_1 \to -A_1$, когда $y$ является достаточно большим.

С другой стороны, когда значение $A_1$ увеличивается и симметрия нарушена, то $\theta$-позиция и знак $A_1$ становятся очень важными. Если $A_1 > 0$, то существует пороговое значение $y$, ниже которого полюс падает вниз. С другой стороны, если $A_1 < 0$, и $\theta = 0$ отталкивание от полюсов гигантского острого излома растет с уменьшением $y$. Мы, таким образом, понимаем, что, качественно говоря, динамика $A_1$ характеризуется асимметричным "потенциалом" согласно с

$$\dot{A}_1 = -\frac{\partial V(A_1)}{\partial A_1}, \qquad (2.54)$$

$$V(A_1) = \lambda A_1^2 - a A_1^3 + \dots . \qquad (2.55)$$

Из анализа линейной стабильности мы знаем, что $\lambda \approx \nu/L^2$, сравни с уравнением (1.11). Мы знаем для дальнейшего, что порог для линейной нестабильности определяется как $A_1 \approx \nu^3/L^2$, сравни с уравнением (2.26). Этим определяется величина коэффициента $a \approx 2/3\nu^2$. Амплитуда «потенциала» в максимуме определяется из

$$V(A_{max}) \approx \nu^7/L^6 . \qquad (2.56)$$

Воздействие шума на развитие моды $A_1 \sin \theta$ может быть понято из следующего стохастического уравнения

$$\dot{A}_1 = -\frac{\partial V(A_1)}{\partial A_1} + \eta_1(t) . \qquad (2.57)$$

Хорошо известно [41], что для такой динамики скорость истечения $R$ через потенциальный барьер для малого шума пропорциональна

$$R \sim \frac{\nu}{L^2} \exp^{-\nu^7/fL^5} . \qquad (2.58)$$

Заключаем отсюда, что любой произвольно маленький шум становится эффективным, когда размер системы увеличивается и когда $\nu$ уменьшается. Если мы вводим в систему шум с амплитудой $\frac{f}{L}$, система может всегда быть чувствительной к этому шуму, когда ее размер превышает критическое значение $L_c$, которое определено выражением $f/L_c \sim \nu^7/L_c^6$. Эта формула определяет переход от режима I (нет новых острых изломов) к режиму II. Для $L > L_c$ шум введет новые полюса в систему. Даже численный шум при моделировании, включающем системы большого размера, может иметь макроскопическое влияние.

Появление новых полюсов должно увеличить скорость фронта. Действительно, скорость пропорциональна среднему от $(u/L)^2$. Новые полюса искажают гигантский острый излом дополнительными меньшими острыми изломами

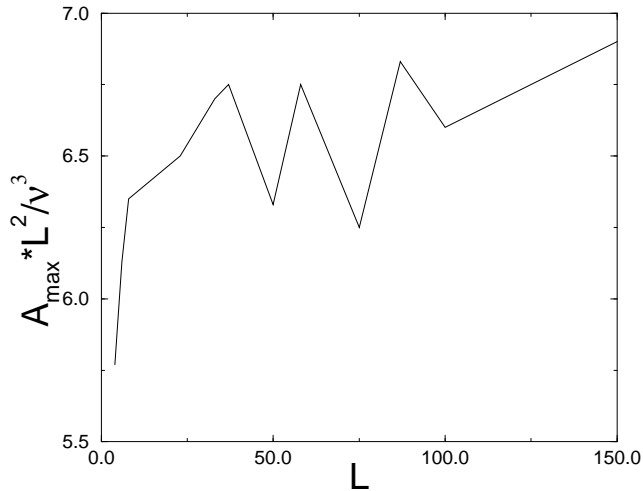

Рис. 2.17: Зависимость нормализованной амплитуды $A_{max}L^2/\nu^3$ от размера системы $L$.

на «крыльях» гигантского острого излома, увеличивая $u^2$. После увеличения амплитуды шума все больше и больше маленьких острых изломов появляется на фронте, что неизбежно ведет к увеличению скорости. Это явление обсуждается количественно в Разделе 2.5.

**Численная проверка ассиметричной «потенциальной» формы и зависимости шума от $L_c$**

Из уравнений движения для полюсов мы можем найти распределение полюсов в гигантском остром изломе [12]. Если мы знаем распределение полюсов в гигантском остром изломе, мы можем тогда найти форму "потенциала" и проверить в цифровой форме выражения для значений $\lambda$, $A_{max}$ и $\frac{\partial V(A_1)}{\partial A_1}$, обсужденных ранее. Связь между амплитудой $A_1$ и позицией полюса $y$ определена как $A_1 = 4\nu e^{-y}$. В то же время, связь между потенциальной функцией $\frac{\partial V(A_1)}{\partial A_1}$ и позицией полюса $y$ определены формулой $\frac{\partial V(A_1)}{\partial A_1} = 4\nu \frac{dy}{dt} e^{-y}$. Производная $\frac{dy}{dt}$ может быть определена из уравнений движения для полюсов. Мы можем найти $A_{max}$ как нулевая точка $\frac{\partial V(A_1)}{\partial A_1}$ и $\lambda$ может быть найден как $\frac{1}{2}\frac{\partial^2 V(A_1)}{\partial A_1^2}$ для $A_1 = 0$. Числовые измерения были сделаны для набора значений $L = 2n\nu$, где $n$ целое число и $n > 2$. Для наших численных измерений мы используем константу $\nu = 0.005$ и переменную $L$, где $L$ изменяется в интервале [1,150], или же переменную $\nu$, которая изменяется в интервале [0.005,0.05] и постоянную $L = 1$. Получены следующие результаты:

1. $\frac{A_{max}L^2}{\nu^3}$ как функция $L$ является почти постоянной. (Рис. 8.17)

2. $\frac{A_{max}L^2}{\nu^3}$ как функция $\nu$ является почти постоянной. (Рис. 8.18)

3. $\frac{A_{max}}{A_{N(L)}}$ как функция $L$ является почти постоянной. ($A_{N(L)}$ определяется позицией верхнего полюса.) (Рис. 8.19)

4. $\frac{A_{max}}{A_{N(L)}}$ как функция $\nu$ является почти постоянной.

   (Рис. 8.20)

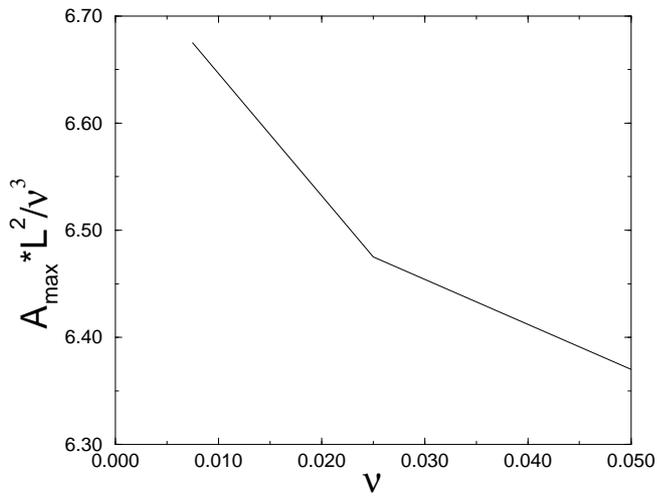

Рис. 2.18: Зависимость нормализованной амплитуды $A_{max}L^2/\nu^3$ от параметра $\nu$.

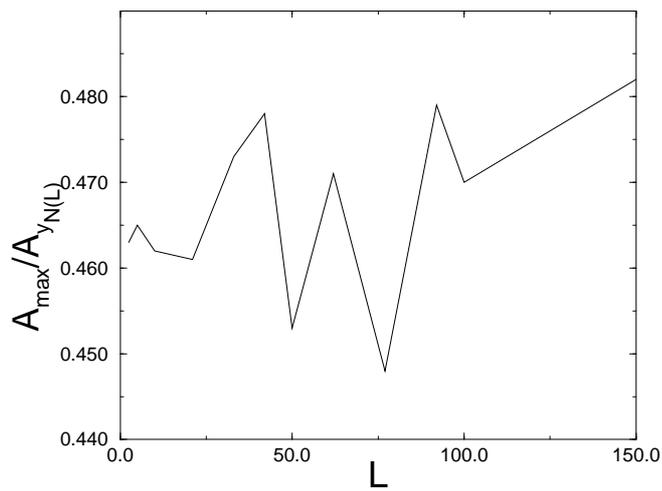

Рис. 2.19: Связь между амплитудой, определенной минимумом потенциала $A_{max}$ и амплитудой, определенной позицией верхнего полюса $A_{N(L)}$ как функция размера системы $L$.

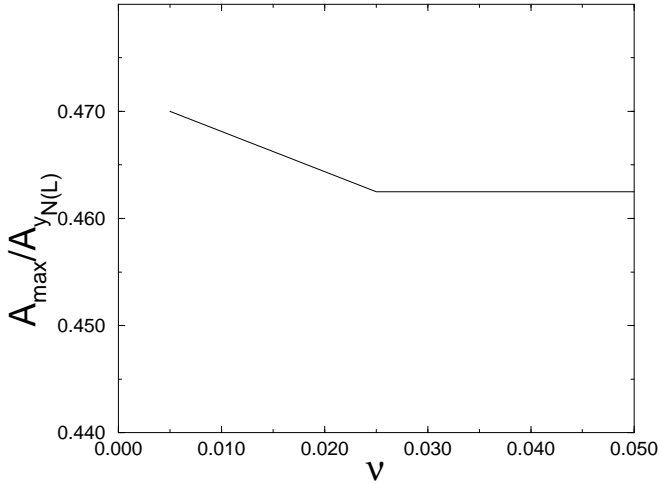

Рис. 2.20: Связь между амплитудой, определенной минимумом потенциала $A_{max}$ и амплитудой, определенной позицией верхнего полюса $A_{N(L)}$ как функция параметра $\nu$.

5. Величина $\frac{\lambda L^2}{\nu}$ как функция $L$ является постоянной. (Рис. 8.21).

6. Величина $\frac{\lambda L^2}{\nu}$ как функция $\nu$ является постоянной ( Рис. 8.22 ).

Мы также проверяем границу между Режимом I (никаких новых острых изломов), и Режимом II (новые острые выступы появляются). Рис. 8.23 показывает зависимости $\frac{f}{L_c}$ от $L_c$. Мы можем видеть что $f/L_c \sim 1/L_c^6$. Эти результаты находятся в хорошем согласии с теорией.

**Стационарное состояние в присутствии шума и его разрушение при увеличении шума или размера системы**

В этом подразделе мы обсуждаем реакцию решения в виде гигантского острого излома на уровень шума, который в состоянии ввести большое количество дополнительных полюсов кроме тех, которые уже существуют в гигантском остром изломе. Мы обозначим дополнительное число полюсов как $\delta N$. Первый вопрос, к которому мы обращаемся, насколько это трудно, вставить еще один дополнительный полюс, когда уже существует избыток $\delta N$. С этой целью мы оцениваем эффективный потенциал $V_{\delta N}(A_1)$, который подобен (2.55), но принимает во внимание существование дополнительного числа полюсов. Основное приближение, которое мы используем это то, что фундаментальная форма решения в виде гигантского острого излома серьезно не изменена существованием дополнительного числа полюсов. Конечно, это приближение количественно уже искажается с введением уже одного лишнего полюсом. Качественно, однако, это выполняется хорошо до тех пор, пока число дополнительных полюсов не имеет порядок характерного числа $N(L)$ для решения в виде гигантского острого излома. Другое приближение – это то, что остальная часть линейных мод не играет серьезной роли в этом случае. С этого момента мы ограничиваем обсуждение следовательно ситуацией $\delta N \ll N(L)$ (режим II).

Чтобы оценить параметр $\lambda$ в эффективном потенциале, мы рассматриваем динамику одного полюса, чья $y$ позиция $y_a$ сильно выше $y_{max}$. Соответственно

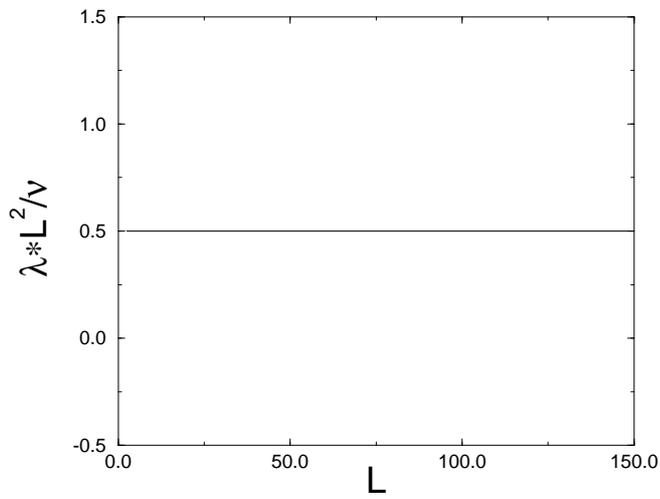

Рис. 2.21: Зависимость нормализованного параметра $\lambda L^2/\nu$ от размера системы $L$.

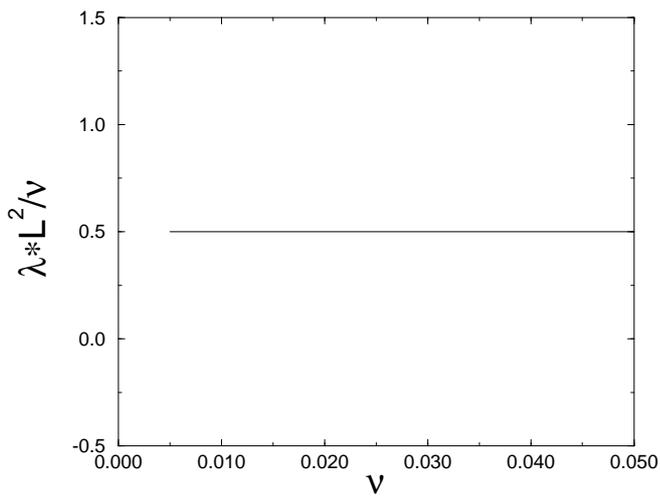

Рис. 2.22: Зависимость нормализованного параметра $\lambda L^2/\nu$ от параметра $\nu$.

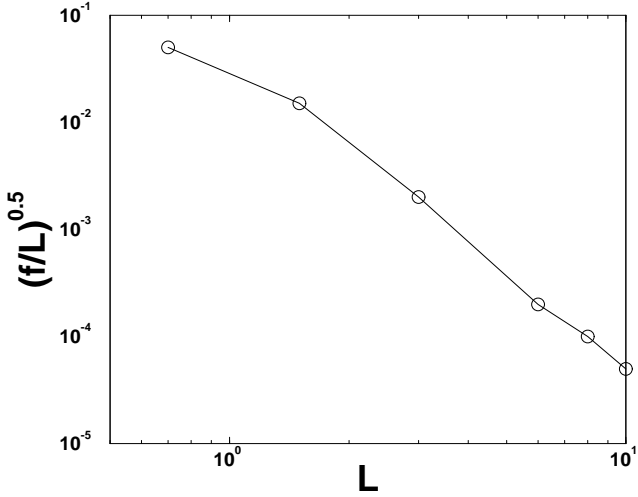

Рис. 2.23: Зависимость критического шума от размера ситемы.

в уравнении (1.11) динамика дает

$$\frac{dy_a}{dt} \approx \frac{2\nu(N(L)+\delta N)}{L^2} - \frac{1}{L} \qquad (2.59)$$

Так как член $N(L)$ пренебрежимо мал в сравнении с членом $L^{-1}$ (сравни с Разделом II A), мы остаемся с отталкивающим членом, который в эффективном потенциале транслируется в

$$\lambda = \frac{\nu\delta N}{L^2} \ . \qquad (2.60)$$

Затем мы оцениваем значение потенциала в точке равновесия между областью притяжения и отталкивания. В последнем подразделе мы видели, что новый полюс должен быть вставлен ниже $y_{max}$, чтобы быть притянутым к вещественной оси. Теперь мы должны поместить новый полюс ниже позиции существующего полюса, индекс которого $N(L) - \delta N$. Эта позиция оценена в Разделе III C, используя функцию распределения TFH (2.23). Мы находим

$$y_{\delta N} \approx 2\ln\left[\frac{4L}{\pi^2\nu\delta N}\right] \ . \qquad (2.61)$$

Как и прежде, это подразумевает пороговое значение амплитуды однополюсного решения $A_{max}\sin\theta$, которое получено из равенства $A_{max} = \nu e^{-y_{\delta N}}$. Мы, таким образом, находим, что в данном случае $A_{max} \sim \nu^3(\delta N)^2/L^2$. Используя снова кубическое представление для эффективного потенциала, мы находим $a = 2/(3\nu^2\delta N)$ и

$$V(A_{max}) = \frac{1}{3}\frac{\nu^7(\delta N)^5}{L^6} \ . \qquad (2.62)$$

Повторяя вычисление скорости истечения через потенциальный барьер, мы находим в данном случае

$$R \sim \frac{\nu\delta N}{L^2}\exp^{-\nu^7(\delta N)^5/fL^5} \ . \qquad (2.63)$$

Для данной амплитуды шума $f$ всегда есть значение $L$ и $\nu$ для которых скорость истечения имеет величину $O(1)$ до тех пор, пока $\delta N$ не является слишком большим. Когда $\delta N$ увеличивается, скорость истечения уменьшается, и в конечном счете никакие дополнительные полюса не могут попасть в систему. Типичное число $\delta N$ для фиксированных значений параметров оценено путем приравнивания к единице аргумента экспоненты.

$$\delta N \approx \left(fL^5/\nu^7\right)^{1/5} . \qquad (2.64)$$

Мы можем видеть, что $\delta N$ сильно зависит от шума $f$, в отличие от режима III. Позвольте нам находить условия перехода от режима II к III, где мы видим насыщение $\delta N$ относительно уровня шума $f$.

(i) Мы используем выражение $A_{max} = 4\nu e^{-y_{\delta N}}$ для амплитуды полюсного решения, которое равняется $\frac{2\nu \sin \theta}{\cosh(y_{\delta N}) - \cos \theta}$; однако, это правильно только для большого числа $y_{\delta N} < 1$. Когда $y_{\delta N} = 1$, лучшее приближение $A_{max} = 4\nu e^{-y_{\delta N}}$. Из уравнения (2.61) мы находим, что граничная величина $y_{\delta N} = 1$ соответствует $\delta N \approx N(L)/2$.

(ii) Мы используем выражение $y_{\delta N} \approx 2 \ln\left[\frac{4L}{\pi^2 \nu \delta N}\right]$, но для большого значения $\delta N$ лучшее приближение, которое может быть найдено тем же самым путем $y_{\delta N} \approx \frac{\pi^2 \nu}{2L}(N(L) - \delta N) \ln\left[\frac{8eL}{\pi^2 \nu (N(L) - \delta N)}\right]$ [12]. Эти выражения дают нам почти тот же самый результат для $\delta N \approx N(L)/2$. Из (i) и (ii) мы можем сделать следующие заключения:

(a) Переход от режима II к режиму III происходит для $\delta N \approx N(L)/2$;

(b) Используя новые выражения в (i) и (ii) для амплитуды $A_{max}$ и $y_{\delta N}$, мы можем определить шум $\frac{f}{L}$ в режиме III выражением

$$\frac{f}{L} \sim V(A_{max}) \sim \lambda A_{max}^2 \sim \frac{\nu \delta N}{L^2}\left(\frac{4\nu}{y_{\delta N}^2}\right)^2 \sim \frac{L^2}{\nu} \frac{\delta N}{(N(L) - \delta N)^4} \qquad (2.65)$$

Это выражение определяет очень слабую зависимость $\delta N$ от шума $f$ для $\delta N > N(L)/2$, который объясняет насыщение $\delta N$ от уровня шума для режима III.

(c) Формой решения в виде гигантского острого излома управляют полюса, которые расположены близко к нулю относительно $y$. Для режима III, $N(L)/2$ полюса, у которых есть позиции $y < y_{\delta N = N(L)/2} = 1$ остаются в этой позиции. Этот результат объясняет, почему решение в виде гигантского острого излома не может быть серьезно изменено для режима III. Из ур. (2.64) используя условие

$$\delta N \approx N(L)/2 \qquad (2.66)$$

Пороговый шум $f_b$ между режимами II и III может быть найден как

$$f_b \sim \nu^2 . \qquad (2.67)$$

Базовое уравнение, описывающее полюсную динамику следующее

$$\frac{dN}{dt} = \frac{\delta N}{T} , \qquad (2.68)$$

где $\frac{dN}{dt}$ - число полюсов, которые появляются в единицу времени в нашей системе, $\delta N$ - добавочное число полюсов, и T - среднее время жизни полюса (между появлением и объединением с гигантским острым изломом). Используя результат численного моделирования для $\frac{dN}{dt}$ и (2.66) мы можем найти для $T$

$$T = \frac{\delta N}{\frac{dN}{dt}} \sim \nu L^{0.2} . \qquad (2.69)$$

Таким образом, время жизни пропорционально $\nu$ и зависит от размера системы $L$ очень слабо. Кроме того, время жизни полюса определяется временем жизни полюсов, которые находятся в остром изломе. Из точки максимума линейной части ур. (2.1), мы можем найти средний характерный размер (Рис. 9 ( [28]))

$$\lambda_m \sim \nu \tag{2.70}$$

который определяет характерный размер наших изломов. Среднее число полюсов в изломе

$$n_{big} \approx \frac{\lambda_m}{2\nu} \sim const \tag{2.71}$$

не зависит от $L$ и $\nu$. Среднее число изломов

$$N_{big} \sim \frac{\delta N}{n_{big}} \sim \frac{L}{\nu} \ . \tag{2.72}$$

Предположим, что некий острый излом существует в основном минимуме системы. Время жизни полюса в таком остром изломе определяется тремя компонентами.

(I) Время формирования острого излома. Это время пропорционально размеру острого излома (с ln-коррекциями) и числу полюсов в остром изломе (из уравнений движения полюсов)

$$T_1 \sim \lambda_m n_{big} \sim \nu \tag{2.73}$$

(II) Время, которое острый излом находится в окрестности минимума. Это время определено выражением

$$T_2 \sim \frac{a}{v} \tag{2.74}$$

где $a$ окрестность минимума такая, что сила от гигантского острого излома меньше, чем сила, связанная с колебаниями числа избыточных полюсов $\delta N$; $v$ является скоростью полюса в этой окрестности. Колебания числа избыточных полюсов $\delta N$ выражается как

$$N_{fl} = \sqrt{\delta N} \ . \tag{2.75}$$

Из этого результата и уравнения движения полюсов мы находим что

$$v \sim \frac{\nu}{L} N_{fl} \sim \frac{\nu}{L} \sqrt{\frac{L}{\nu}} \sim \sqrt{\frac{\nu}{L}} \ . \tag{2.76}$$

Скорость от гигантского острого излома определена

$$v \sim \frac{\nu}{L} N(L) \frac{a}{L} \sim \frac{a}{L} \ . \tag{2.77}$$

Приравнивая эти два уравнения, получаем

$$a \sim \sqrt{\nu L} \ . \tag{2.78}$$

Таким образом, для времени $T_2$ мы получаем

$$T_2 \sim \frac{a}{v} \sim L \ . \tag{2.79}$$

(III) Время притяжения к гигантскому острому излому. Из уравнений движения для полюсов мы получаем

$$T_3 \sim L \ln(\frac{L}{a}) \sim L \ln \sqrt{L} \sim L \ . \qquad (2.80)$$

Для исследованной области размеров системы было найдено

$$T_1 \gg T_2, T_3 \qquad (2.81)$$

Следовательно, полное время жизни

$$T = T_1 + T_2 + T_3 \sim \nu + sL \ , \qquad (2.82)$$

где $s$ - постоянная и

$$0 < s \ll 1 \ . \qquad (2.83)$$

Этот результат качественно и частично количественно объясняет зависимость (2.69). Из (2.69), (2.68) и (2.66) мы можем видеть, что в режиме III $\frac{dN}{dt}$ насыщается как функция размера системы $L$.

### 2.5.4 Ускорение фронта пламени из-за шума

В этом разделе мы оцениваем экспоненты масштабирования, которые характеризуют скорость фронта пламени как функцию размера системы. Чтобы оценить скорость фронта пламени, мы должны создать уравнение для среднего от $< dh/dt >$ даваемого произвольным числом $N$ полюсов в системе. Это уравнение следует непосредственно из (2.4)

$$\left\langle \frac{dh}{dt} \right\rangle = \frac{1}{L^2} \frac{1}{2\pi} \int_0^{2\pi} u^2 d\theta \ . \qquad (2.84)$$

После подстановки (2.8) в (2.84) мы получаем, используя (2.11) и (2.12)

$$\left\langle \frac{dh}{dt} \right\rangle = 2\nu \sum_{k=1}^{N} \frac{dy_k}{dt} + 2 \left( \frac{\nu N}{L} - \frac{\nu^2 N^2}{L^2} \right) \ . \qquad (2.85)$$

Оценка второго и третьего членов в этом уравнении является прямолинейной. Записывая $N = N(L) + \delta N(L)$ и помня, что $N(L) \sim L/\nu$ и $\delta N(L) \sim N(L)/2$, мы находим, что эти члены вносят вклад порядка $O(1)$. Первый член вносит вклад только, когда поток полюсов асимметричен. Шум вводит полюса на конечной величине $y_{min}$, тогда как отклоненные полюса утекают в бесконечность и исчезают на границе нелинейности, определенной позицией самого высокого полюса как

$$y_{max} \approx 2 \ln \left[ \frac{4L}{\pi^2 \nu} \right] \ . \qquad (2.86)$$

Таким образом, у нас есть асимметрия, которая вносит вклад в скорость фронта. Чтобы оценить первый член позвольте нам определять:

$$d(\sum \frac{dy_k}{dt}) = \sum_{l}^{l+dl} \frac{dy_k}{dt} \ , \qquad (2.87)$$

где $\sum_{l}^{l+dl} \frac{dy_k}{dt}$ - сумма по полюсам, которые находятся на интервале $y : [l, l+dl]$. Мы можем написать

$$d(\sum \frac{dy_k}{dt}) = d(\sum \frac{dy_k}{dt})_{up} + d(\sum \frac{dy_k}{dt})_{down} , \qquad (2.88)$$

где $d(\sum \frac{dy_k}{dt})_{up}$ является потоком полюсов, двигающихся вверх, и $d(\sum \frac{dy_k}{dt})_{down}$ является потоком полюсов, двигающихся вниз.

Для этих потоков мы можем написать

$$d(\sum \frac{dy_k}{dt})_{up}, -d(\sum \frac{dy_k}{dt})_{down} \leq \frac{dN}{dt}dl . \qquad (2.89)$$

Поэтому для первого члена

$$\begin{aligned}
0 \leq \sum_{k=1}^{N} \frac{dy_k}{dt} &= \int_{y_{min}}^{y_{max}} \frac{d(\sum \frac{dy_k}{dt})}{dl} dl \qquad (2.90) \\
&= \int_{y_{min}}^{y_{max}} \frac{d(\sum \frac{dy_k}{dt})_{up} + d(\sum \frac{dy_k}{dt})_{down}}{dl} dl \\
&\leq \frac{dN}{dt}(y_{max} - y_{min}) \\
&\leq \frac{dN}{dt} y_{max}
\end{aligned}$$

Из-за небольшой (ln) зависимости $y_{max}$ от $L$ и $\nu$, $\frac{dN}{dt}$ член определяет порядок нелинейности для первого члена в ур. (2.85). Этот член равняется нулю для симметричного потока полюсов и достигает максимума для максимально асимметричного потока полюсов. Сравнение $v \sim L^{0.42} f^{0.02}$ and $\frac{dN}{dt} \sim L^{0.8} f^{0.03}$ подтверждает это вычисление.

## 2.6 Краткое заключение и выводы

Главное два посыла этой главы: (i) есть важное взаимосвязь между неустойчивостью распространяющихся фронтов и белым шумом; (ii) Это взаимосвязь и ее значение может быть поняты качественно, а и иногда и количественно, используя описание в терминах многих полюсов. Описание в форме полюсов является естественным в этом контексте в первую очередь потому, что это обеспечивает точное (и эффективное) представление стационарного состояния без шума. Как только мы преуспеваем в описании также и *возмущения* этого стационарного состояния в терминах полюсов, мы получаем очень ясный язык для исследования взаимодействия между шумом и неустойчивостью. Этот язык также позволяет нам описывать в качественных и полуколичественных терминах процесс обратного каскада увеличения типичных длин, когда система релаксирует к стационарному состоянию из малых и случайных начальных условий.

Основные концептуальные шаги в этой главе следующие: во-первых очевидно, что стационарное решение, которое характеризуется $N(L)$ полюсами, выровненными вдоль мнимой оси, слабо устойчиво по отношению к шуму в периодическом массиве $L$ значений. Для всех значений $L$ стационарное состояние нелинейно неустойчиво по отношению к шуму. Основное и главное воздействие шума данной амплитуды $f$ введение дополнительного числа полюсов $\delta N(L, f)$ в систему. Существование этого избыточного числа полюсов ответственно как за дополнительную волновую деформацию переднего фронта пламени на вершине гигантского острого излома, так и за наблюдаемое ускорение фронта пламени. Рассматривая появление новых полюсов из-за шума, мы описываем наблюдаемые законы масштабирования как функции амплитуды шума и размера системы. Теоретически мы поэтому концентрируемся на нахождении оценки для $\delta N(L, f)$. Заметим, что некоторые из наших рассмотрений являются только качественными. Например, мы оценивали $\delta N(L, f)$, предполагая, что решение в виде гигантского острого излома серьезно не возмущено. С другой стороны мы находим поток полюсов, идущих в бесконечность из-за введения шумом полюсов на конечных величинах $y$. Существование этого распределения полюсов между $y_{max}$ и бесконечностью *является* существенным возмущением решения в виде гигантского острого излома. Таким образом, сравнение между измеренными и предсказанными экспонентами масштабирования должно быть сделано с осторожностью; мы не можем гарантировать, что даже в тех случаях, в которых наше предсказание близко к измерению, теория является количественно однозначно верной. Однако мы полагаем, что наше рассмотрение демонстрирует основные черты правильной теории.

"Фазовая диаграмма" как функция $L$ и $f$ в этой системе состоит из четырех режимов. В первом, обсуждаемом в Разделе 2.5.3, шум является слишком маленьким, чтобы иметь какое-либо воздействие на решение в виде гигантского острого излома. Во втором, шум вводит добавочные полюса, которые украшают гигантский острый излом побочными острыми изломами. В этом режиме мы находим законы масштабирования для скорости как функция $L$ и $f$, и мы имеем успех в понимании экспонент масштабирования. В третьем режиме происходит насыщение числа полюсов $\delta N$ от уровня шума $f$, так что это число зависит слабо от шума. В четвертом режиме шум является достаточно большим, чтобы создать мелкомасштабные структуры, которые не интерпретируются в терминах индивидуальных полюсов. Кажется из наших численных данных, что в этом режиме огрубление фронта пламени получает вклад от мелкомасштабной структуры путем, подобным *устойчивым* КПЗ (Кардар-Паризи-Занг) моделям роста, управляемых шумом.

Наша важнейшая мотивация для этого исследования была понять явления, наблюдаемые в радиальной геометрии с расширяющимися фронтами пламени. Отметим, что многие из идей, предлагаемых выше, немедленно переносится на эту проблему. Действительно, в радиальной геометрии фронт пламени ускоряется, и острые изломы множатся и формируют иерархическую структуру на протяжении времени. Так как радиус (и типичный масштаб) увеличиваются в этой системе все время, новые полюса будут добавляться к системе даже при незначительно слабом шуме. Слабая стабильность, найденная выше, существует также и в этом случае, и система позволит введение дополнительных полюсов в результате шума. Результаты, обсуждаемые в работе [19], могут быть объединены с изложенными идеями, чтобы обеспечить теорию радиального роста (глава 10).

Наконец, успех этого подхода в случае распространения пламени дает надежду, что к картинам Лапласовского роста можно применить подобные идеи. Проблема непосредственного интереса – Лапласовский рост в канале, в котором «пальцевидное» стационарное решение, как известно, существует. Известно, что стабильность такого «пальцевидного» решения к шуму уменьшается быстро с увеличением ширины канала. Кроме того, ясно, что шум вызывает дополнительные геометрические особенности на вершине пальца. Есть достаточно многие общие черты здесь, чтобы указать, что осторожный анализ аналитической теории может пролить много света на эту проблему.

# Глава 3

# Использование динамики полюсов для анализа стабильности фронтов пламени: Приближение динамическими системами в комплексной плоскости.

## 3.1 Введение

В этой главе мы обсуждаем стабильность стационарных фронтов пламени в канальной геометрии. Мы писали коротко об этой теме в главе 2 (Раздел 2.3), и мы хотим рассмотреть ее подробно в этой главе. Традиционно [1–3] при изучении стабильности, рассматривается линейный оператор, полученный линеаризацией уравнения движения вокруг устойчивого решения. Полученные собственные функции *делокализованы*, и в определенных случаях не просты для интерпретации. В случае фронта пламени стационарное решение пространственно зависимо. Поэтому собственные функции очень отличаются от простых Фурье мод. Мы показываем в этой главе, что хорошее понимание природы собственного спектра и собственных мод может быть получено, идя путем почти противоположным традиционному анализу стабильности, то есть, изучая *ограниченную* динамику сингулярностей в комплексной плоскости. Уменьшая анализ стабильности до исследования динамической системы конечной размерности можно получить значительное интуитивное понимание природы проблем стабильности. Анализ основан на понимании, что для данной ширины канала $L$ стационарное решение для фронта пламени дано в терминах $N(L)$ полюсов, которые организованы на линии, параллельной мнимой оси [12]. Стабильность этого решения можно тогда рассмотреть за два шага. На первом шаге мы исследуем реакцию этого набора $N(L)$ полюсов на возмущение их позиций. Эта процедура приводит к важной части спектра стабильности. Во втором шаге мы исследуем общие возмущения, которые могут также быть описаны добавлением дополнительных полюсов к системе из $N(L)$ полюсов. Ответ на эти возмущения дает нам остальную часть спектра стабильности; комбинация этих двух шагов рационализирует все качественные особенности, найденные традиционным анализом стабильности. В Разделе 2 мы представляем результаты традиционного линейного анализа стабильности, и показываем собственным значениям и собственные функции, которые мы хотим интерпретировать при использова-нии декомпозиции на полюса. Раздел 3 представляет этот анализ в терминах сингулярностей на комплексной плоскости в два шага, как обсуждалось выше. Резюме и обсуждение представлены в Разделе 4.

## 3.2 Анализ линейной стабильности в канальной геометрии

Стандартная методика, чтобы изучить линейную стабильность стационарного решения должна внести в него малое возмущение $\phi(\theta,t)$: $u(\theta,t) = u_s(\theta) + \phi(\theta,t)$. Линеаризация динамики для маленького $\phi$ приводит к следующему уравнению движения

$$\begin{aligned}\frac{\partial \phi(\theta,t)}{\partial t} &= \frac{1}{L^2}\Big[\partial_\theta[u_s(\theta)\phi(\theta,t)] \\ &+ \nu\partial_\theta^2 \phi(\theta,t)\Big] + \frac{1}{L}I(\phi(\theta,t)) \ . \end{aligned} \quad (3.1)$$

где линейный оператор, содержит $u_s(\theta)$ как коэффициент. Соответственно простые Фурье моды не диагонализируют его. Однако, мы осуществляем разложение $\phi(x)$ на Фурье моды согласно:

$$\phi(\theta,t) = \sum_{k=-\infty}^{\infty} \hat{\phi}_k(t) e^{ik\theta} \quad (3.2)$$

$$u_s(\theta) = -2\nu i \sum_{k=-\infty}^{\infty} \sum_{j=1}^{N} sign(k) e^{-|k|y_j} e^{ik\theta} \quad (3.3)$$

Последнее уравнение следует из (2.13), путем разложения в ряд по $\sin k\theta$. В этих суммах дискретные $k$ значения пробегают все целые числа. Подставляя в ур. (3.1) мы получаем:

$$\frac{d\hat{\phi}_k(t))}{dt} = \sum_n a_{kn} \hat{\phi}_n(t) \ , \quad (3.4)$$

где $a_{kn}$ элементы бесконечной матрицы:

$$a_{kk} = \frac{|k|}{L} - \frac{\nu}{L^2}k^2 \ , \quad (3.5)$$

$$a_{kn} = \frac{k}{L^2}sign(k-n)(2\nu \sum_{j=1}^{N} e^{-|k-n|y_j}) \quad k \neq n \ . \quad (3.6)$$

Чтобы найти собственные значения этой матрицы, мы должны усечь ее на некотором $k$-векторе $k^*$. Масштаб $k^*$ может быть выбран на основе ур. (3.5). Из этого уравнения мы видим, что наибольшее значение $k$, для которого $a_{kk} \geq 0$ это масштаб, который мы обозначаем как $k_{max}$. Он является целой частью $L/\nu$. Мы должны выбрать $k^* > k_{max}$ и проверить выбор сходимостью собственных значений. Выбранное значение $k^*$ в наших численных данных составляло $4k_{max}$. Нужно обратить внимание, что это обрезание ограничивает число собственных значений, которое должно быть бесконечным. Однако более низкие собственные значения будут хорошо представлены. Результаты для собственных значений низкого порядка матрицы $a_{kn}$, которые были получены из сходящегося численного вычисления, представлены на Рис. 9.1 Собственные значения умножены

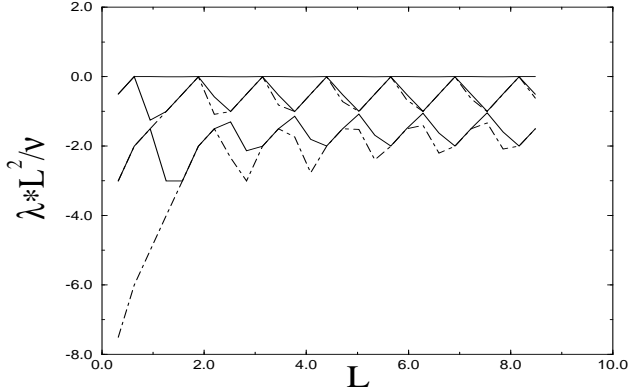

Рис. 3.1: Изображение первых пяти собственных значений, полученных диагонализацией матрицы, полученной традиционным анализом стабильности, как функций размера системы. Собственные значения нормализованы $L^2/\nu$. Наибольшее собственное значение - нуль, который является режимом Голдстоуна. Все другие собственные значения - отрицательны за исключением второго и третьего, которые касаются нуля периодически. Вторые и четвертые собственные значения представлены сплошной линией, и третьи и пятые собственные значения представлены точечной пунктирной линией.

на $L^2/\nu$ и чертятся как функция $L$. Мы упорядочиваем собственные значения в порядке их убывания и обозначаем их как $\lambda_0 \geq \lambda_1 \geq \lambda_2 \ldots$. В дополнение к собственным значениям обрезанная матрица также имеет собственные вектора, которые мы обозначаем как $A^{(\ell)}$. У каждого такого вектора есть порог $k^*$, и мы можем вычислить собственную функцию $f^{(\ell)}(\theta)$ линейного оператора (3.1), используя (3.2), как

$$f^{(\ell)}(\theta) \equiv \sum_{-k^*}^{k^*} e^{ik\theta} A_k^{(\ell)} . \qquad (3.7)$$

Уравнение (3.1), разделяет четные и нечетные решения по $\theta$, как может быть проверено напрямую. Следовательно, доступные решения имеют четный или нечетный тип, разложимый или в cos или в sin функции. Первые две нетривиальные собственные функции $f^{(1)}(\theta)$ и $f^{(2)}(\theta)$ показаны на Рис. 9.2 и 9.3. а Очевидно, что функция на Рис. 9.2 нечетна вокруг ноля, тогда как на Рис. 9.3 функция является четной. Так же мы можем в числено получить любую другую собственную функцию линейного оператора, но мы не понимаем ни физического значения этих собственных функций, ни $L$ зависимость их соответствующих собственных значений, показанных на Рис. 9.1. В следующем разделе, мы продемонстрируем, как приближение динамических систем в терминах особенностей в комплексной плоскости дает нам необходимую интуицию для понимания этих проблем.

## 3.3 Линейная стабильность в терминах комплексных сингулярностей

Так как дифференциальное уравнение в частных производных непрерывно у него существует бесконечное число мод. Чтобы понять это в терминах динамики

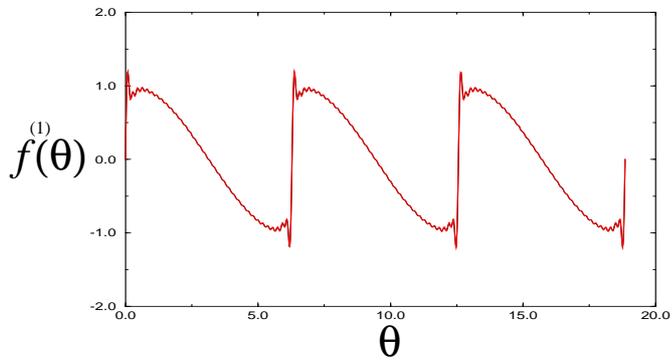

Рис. 3.2: Первая нечетная собственная функция, полученная традиционным анализом стабильности.

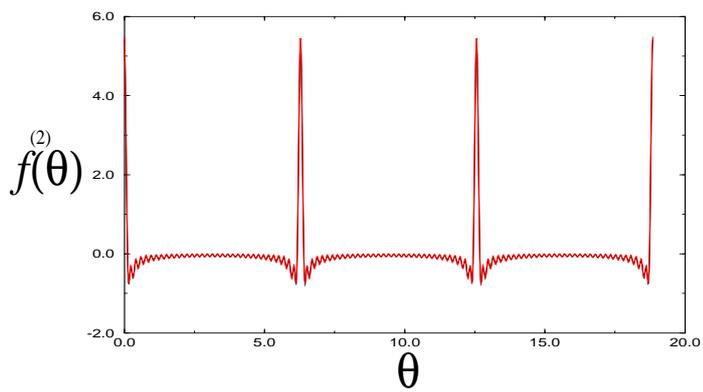

Рис. 3.3: Первая четная собственная функция, полученная традиционным анализом стабильности.

полюсов, мы рассматриваем задачу в два шага: Во-первых, мы рассматриваем $2N(L)$ мод, ассоциированных с динамикой $N(L)$ полюсов гигантского излома. На втором шаге мы объясняем, что все дополнительные моды следуют из введения дополнительных полюсов, включая реакцию $N(L)$ полюсов гигантского излома на новые полюса. После этих двух шагов мы будем в состоянии идентифицировать все линейные моды, которые были найдены путем диагонализации матрицы стабильности в предыдущем разделе.

### 3.3.1 Моды, ассоциированные с гигантским изломом

В стационарном решении все полюса занимают устойчивые равновесные положения. Силы, воздействующие на любой данный полюс, точно обнуляются, и мы можем написать матричные уравнения для небольших возмущений для положений полюса $\delta y_i$ and $\delta x_i$. Следуя [12] мы переписываем уравнения движения (2.12) используя функции Ляпунова $U$:

$$L\dot{y_i} = \frac{\partial U}{\partial y_i} \qquad (3.8)$$

где $i = 1, ..., N$ и

$$U = \frac{\nu}{L}[\sum_i \ln\sinh y_i \; + \; 2\sum_{i<k}(\ln\sinh\frac{y_k - y_i}{2} \\ + \; \ln\sinh\frac{y_k + y_i}{2})] - \sum_i y_i \qquad (3.9)$$

Линеаризованные уравнения движения для $\delta y_i$:

$$L\dot{\delta y_i} = \sum_k \frac{\partial^2 U}{\partial y_i \partial y_k} \delta y_k \; . \qquad (3.10)$$

Матрица $\partial^2 U/\partial y_i \partial y_k$ действительная и симметричная ранга $N$. Мы таким образом ожидаем найти $N$ действительных собственных значений and N ортогональных собственных функций. Для отклонений $\delta x_i$ в $x$ позициях мы находим следующие линеаризованные уравнения движения:

$$L\dot{\delta x_j} = -\frac{\nu}{L}\delta x_j \sum_{k=1,k\neq j}^{N}(\frac{1}{\cosh(y_j - y_k) - 1} \\ +\frac{1}{\cosh(y_j + y_k) - 1}) \\ +\frac{\nu}{L}\sum_{k=1,k\neq j}^{N} \delta x_k(\frac{1}{\cosh(y_j - y_k) - 1} + \frac{1}{\cosh(y_j + y_k) - 1}) \qquad (3.11)$$

Сокращенно:

$$L\frac{d\delta x_i}{dt} = V_{ik}\delta x_k \; . \qquad (3.12)$$

Матрица $V$ также действительна и симметрична. Таким образом, $V$ и $\partial^2 U/\partial y_i \partial y_k$ вместе поставляют $2N(L)$ действительных собственных значений и $2N(L)$ ортогональных собственных векторов. Явная форма матриц $V$ и $\partial^2 U/\partial y_i \partial y_k$ следующая:

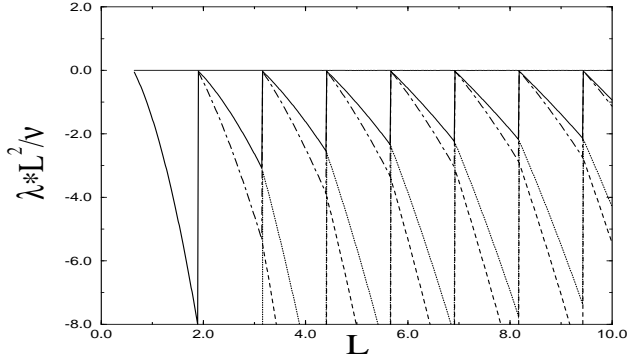

Рис. 3.4: Собственные значения, ассоциированные с возмущением положений полюсов, которые создают гигантский излом. Наибольшее собственное значение - ноль. Вторые, третьи, четвертые и пятые собственные значения представлены сплошной линией, точечной пунктирной линией, пунктиром и пунктирной линией, соответственно.

Для $i \neq k$:

$$\frac{\partial^2 U}{\partial y_i \partial y_k} = \frac{\nu}{L}[\frac{1/2}{\sinh^2(\frac{y_k-y_i}{2})} - \frac{1/2}{\sinh^2(\frac{y_k+y_i}{2})}] \qquad (3.13)$$

$$V_{ik} = \frac{\nu}{L}(\frac{1}{\cosh(y_i-y_k)-1} + \frac{1}{\cosh(y_i+y_k)-1}) \qquad (3.14)$$

и для $i = k$ получаем:

$$\frac{\partial^2 U}{\partial y_i^2} = - \frac{\nu}{L}[\sum_{k \neq i}^N \left(\frac{1}{2\sinh^2(\frac{y_k-y_i}{2})} + \frac{1}{2\sinh^2(\frac{y_k+y_i}{2})}\right)$$
$$+ \frac{1}{\sinh^2(y_i)}] \qquad (3.15)$$

$$V_{ii} = \sum_{k \neq i}^N [-\frac{\nu}{L}(\frac{1}{\cosh(y_i-y_k)-1} + \frac{1}{\cosh(y_i+y_k)-1})] \qquad (3.16)$$

Используя известные решения для стационарного состояния $y_i$ для любого данного $L$, мы можем диагонализовать $N(L) \times N(L)$ матриц численно. На Рис. 9.4 мы представляем собственные значения мод самого низкого порядка получены из этой процедуры. Наименее отрицательное собственное значение периодически касается нуля. Это собственное значение может быть полностью идентифицировано с движением самого высокого полюса $y_{N(L)}$ в гигантском изломе. При изолированных значениях $L$ положение этого полюса уходит в бесконечность, и, соответственно, ряд и столбец в наших матрицах, которые содержат $y_{N(L)}$, обращаются в нуль тождественно, приводя к нулевому собственному значению. Остальная часть верхних собственных значений соответствует полностью половине наблюдаемых собственных значений на Рис. 9.1. Другими словами, собственные значения, наблюдаемые здесь, отлично согласуются с изображенными на Рис. 9.1 собственными значениями до их разрывного увеличения от минимальных точек. "Вторая половина"колебания в собственных значениях

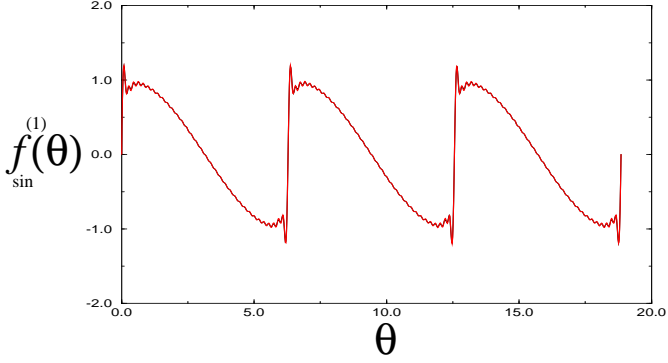

Рис. 3.5: Первая нечетная собственная функция, ассоциированная с возмущением положений полюсов в гигантском изломе.

как функция $L$ не содержится в этом спектре $N(L)$ полюсов гигантского излома. Чтобы понять остальную часть спектра, мы должны рассмотреть возмущение гигантского излома дополнительными полюсами. Собственные функции могут быть найдены, используя знание собственных векторов этих матриц. Позвольте нам обозначать собственные векторы $\partial^2 U/\partial y_i \partial y_k$ и $a^{(\ell)}$ and $b^{(\ell)}$, соответственно. Возмущенное решение явно дано как (взято для $x_s = 0$):

$$u_s(\theta) + \delta u = 2\nu \sum_{i=1}^{N} \frac{\sin(\theta - \delta x_i)}{\cosh(y_i + \delta y_i) - \cos(\theta - \delta x_i)} \quad (3.17)$$

где $\delta u$ находится из

$$\begin{aligned}\delta u = &- 4\nu \sum_{i=1}^{N} \sum_{k=1}^{\infty} \delta y_i k e^{-k y_i} \sin k\theta \\ &- 4\nu \sum_{i=1}^{N} \sum_{k=1}^{\infty} \delta x_i k e^{-k y_i} \cos k\theta \end{aligned} \quad (3.18)$$

Поэтому, зная собственные вектора $a^{(\ell)}$ и $b^{(\ell)}$, мы можем оценить собственные вектора $f^{(\ell)}(\theta)$ of (3.7):

$$f_{\sin}^{(\ell)}(\theta) = -4\nu \sum_{i=1}^{N} \sum_{k=1}^{\infty} a_i^{(j)} k e^{-k y_i} \sin k\theta \ , \quad j=1,...,N \quad (3.19)$$

или

$$f_{\cos}^{(\ell)}(\theta) = -4\nu \sum_{i=1}^{N} \sum_{k=1}^{\infty} b_i^{(j)} k e^{-k y_i} \cos k\theta \ , \quad j=1,...,N \quad (3.20)$$

где мы показываем отдельно sin разложение и cos разложение. Для случая $j = 1$, собственное значение - ноль, и одинаковый перенос полюсов на любую дистанцию $\delta x_i$ имеет результатом Голдстоуновскую моду. Это охарактеризовано собственным вектором $b_i^{(1)} = 1$ для всех $i$. Собственные векторы $f^{(\ell)}$ (Рис. 9.5,9.6) вычисляющие этот путь, идентичны следующему Численная точность с ними показана на Рис. 9.2, 9.3 и наблюдается согласие.

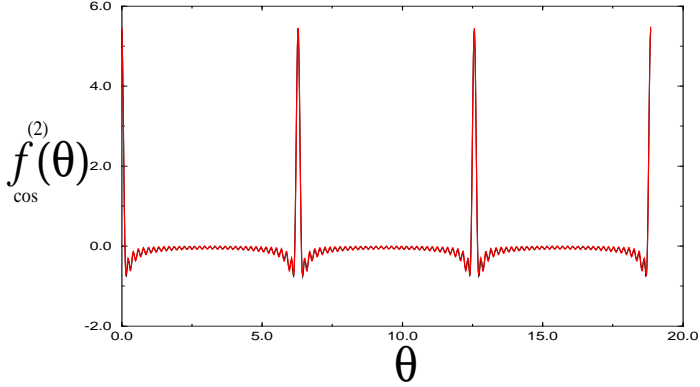

Рис. 3.6: Первая четная собственная функция, ассоциированная с возмущением положений полюсов в гигантском изломе.

### 3.3.2 Моды, связанные с дополнительными полюсами

В этом подразделе мы идентифицируем остальную часть мод, которые не были найдены в предыдущем подразделе. С этой целью мы изучаем реакцию решения TFH на введение дополнительных полюсов. Мы хотим добавить $M$ новых полюсов, помещенных на одну и ту же мнимую координату $y_p \ll y_{max}$, но распределенных на равных расстояниях вдоль действительной оси $\{x_j = x_0 + (2\pi/M)j\}_{j=1}^M$. Для $x_0 = 0$ мы используем (2.8) и разложение Фурье, чтобы получить возмущение формы

$$\delta u(\theta, t) \simeq 4\nu M e^{-M y_p(t)} \sin M\theta \qquad (3.21)$$

Для $x_0 = -\pi/2M$ мы получаем

$$\delta u(\theta, t) \simeq 4\nu M e^{-M y_p(t)} \cos M\theta \qquad (3.22)$$

в обоих случаях уравнения для динамики $y_p$ следует из ур.(2.11)-(2.12):

$$\frac{dy_p}{dt} \simeq 2\frac{\nu}{L^2}\alpha(M) \ , \qquad (3.23)$$

где $\alpha(M)$ дается как:

$$\alpha(M) = [\frac{1}{2}(\frac{L}{\nu} + 1)] - \frac{1}{2}(\frac{L}{\nu} - M) \qquad (3.24)$$

Так как (3.23) линейно, мы можем решить его и подставить в уравнения (3.21) - (3.22). Ища форму $\delta u(\theta, t) \sim exp(-\lambda(M)t)$ мы находим, что собственное значение $\lambda(M)$ является

$$\lambda(M) = 2M\frac{\nu}{L^2}\alpha(M) \qquad (3.25)$$

Эта собственная величина изображена на Рис. 9.7 В этот момент мы рассмотрим динамику полюсов в гигантском изломе под влиянием дополнительных $M$ полюсов. Из уравнений (3.10), (3.12), (2.11), (2.12) мы получаем, после некоторой очевидной алгебры,

$$L\dot{\delta y_i} = \sum_j \frac{\partial^2 U}{\partial y_i \partial y_j}\delta y_j - 4\frac{\nu}{L}M e^{-M y_p(t)}\sinh(M y_i) \qquad (3.26)$$

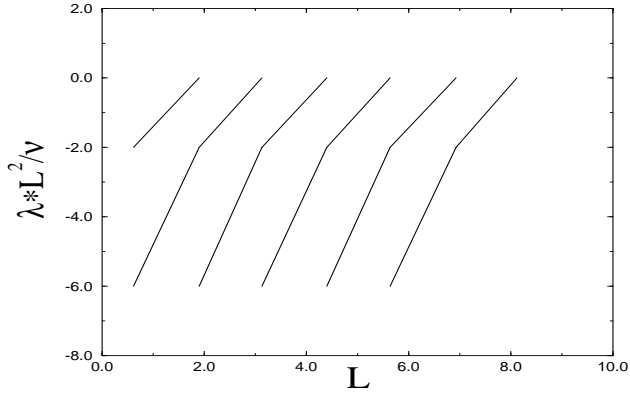

Рис. 3.7: Спектр собственных значений, ассоциированных с реакцией полюсов в гигантском изломе на добавление новых полюсов.

или

$$L\dot{\delta x_i} = \sum_j V_{ij}\delta x_j - 4\frac{\nu}{L}Me^{-My_p(t)}\cosh(My_i) \quad (3.27)$$

Удобно теперь преобразовать из базиса $\delta y_i$ к естественному базису $w_i$, который получен, используя линейное преобразование $w = A^{-1}\delta y$. Здесь у матрицы $A$ есть столбцы, которые являются собственными векторами $\partial^2 U/\partial y_i\partial y_j$, которые были вычислены ранее. Так как матрица была действительной симметричной, матрица $A$ является ортогональной, и $A^{-1} = A^T$. Определим $C = 4\frac{\nu}{L^2}Me^{-My_p(0)}$ и напишем

$$\dot{w_i} = -\lambda_i w_i - Ce^{-\lambda(M)t}\xi_i , \quad (3.28)$$

ult $-\lambda_i$ является собственными значениями, ассоциированным со столбцами $A$, и

$$\xi_i = \sum_j A_{ji}\sinh My_j . \quad (3.29)$$

Мы теперь ищем решение, которое затухает по экспоненте на скорости $\lambda(M)$:

$$w_i(t) = w_i(0)e^{-\lambda(M)t} \quad (3.30)$$

Подставляя желаемое решение в (3.28) мы находим условие на начальное значение $w_i$:

$$w_i(0) = -\frac{C}{\lambda_i - \lambda(M)}\xi_i \quad (3.31)$$

Трансформируя обратно к $\delta y_i$ мы получаем

$$\begin{aligned}\delta y_i(0) &= \sum_k A_{ik}w_k(0) = -\sum_k A_{ik}\frac{C}{\lambda_k - \lambda(M)}\sum_l A_{lk}\sinh My_l\\ &= -C\sum_l \sinh My_l \sum_k \frac{A_{ik}A_{lk}}{\lambda_k - \lambda_p^M}\end{aligned} \quad (3.32)$$

Мы можем получить собственные функции линейного оператора, как ранее, используя ур.(3.18), (3.21), (3.22), (3.32). Мы получаем

$$f_{\sin}^{(M)}(\theta) = 4C\nu \sum_{i=1}^{N(L)} \sum_{k=1}^{\infty} (\sum_{l} \sinh M y_l \sum_{m} \frac{A_{im}A_{lm}}{\lambda_m - \lambda(M)})$$
$$\times k e^{-k y_i} \sin k\theta + L^2 C \sin M\theta \qquad (3.33)$$

Расчеты идентичные расчетам, проведенным для ур. (3.28) могут быть проведены для отклонений $\delta x_i$. Итоговый результат дает

$$f_{\cos}^{(M)}(\theta) = 4C\nu \sum_{i=1}^{N(L)} \sum_{k=1}^{\infty} (\sum_{l} \cosh M y_l \sum_{m} \frac{\tilde{A}_{im}\tilde{A}_{lm}}{\tilde{\lambda}_m - \lambda(M)})$$
$$\times k e^{-k y_i} \cos k\theta + L^2 C \cos M\theta , \qquad (3.34)$$

где $\tilde{A}$ является матрицей, чьи столбцы являются собственными векторами $V$, и $-\tilde{\lambda}_i$ ее собственные величины.

Мы теперь в состоянии объяснить весь линейный спектр, используя знание, что мы получили. Спектр состоит из двух отдельных типов вкладов. Первый тип имеет $2N$ моды, которые принадлежат динамике возмущенных $N(L)$ полюсов в гигантском изломе. Вторая часть, которая является большей частью спектра, построена из мод второго типа, так как $M$ может пойти в бесконечность. Эта структура видна на Рис. 9.4 и Рис. 9.7.

Мы можем утверждать, что множество собственных функций, полученных выше, является полным и исчерпывающим. Чтобы сделать это, мы покажем, что любая произвольная периодическая функция $\theta$ может быть разложена на эти собственные функции. Начнем со стандартного ряда Фурье в виде *sin* и *cos* функций. В этот момент решаем для $\sin k\theta$ и $\cos k\theta$ из уравнений (3.33-3.34). Подставляем результат в ряд Фурье. У нас теперь есть разложение в терминах собственных мод $f^{(M)}$ и в терминах тройных сумм. Тройные суммы, однако, могут быть разложены, используя уравнения (3.19-3.20), в терминах собственных функций $f^{(\ell)}$ Мы можем таким образом анализировать любую функцию в терминах собственных функций $f^{(M)}$ и $f^{(\ell)}$.

## 3.4 Выводы

Мы обсуждали стабильность фронтов пламени в геометрии канала, используя представление решений в терминах сингулярностей в комплексной плоскости. На этом языке стационарное решение, которое является гигантским изломом в конфигурационном пространстве, представлено $N(L)$ полюсами, которые организованы на линии, параллельной мнимой оси. Мы показали, что задача о стабильности может быть понята в терминах двух типов возмущений.

Первый тип - возмущение в положениях полюсов, которые составляют гигантский излом. Продольные движения полюсов дают начало нечетным модам, тогда как поперечные движения - четным. Собственные значения, ассоциированные с этими модами, являются собственными значениями конечных, действительных и симметричных матриц, сравни уравнения (3.13), (3.14), (3.15), (3.16). Второй тип возмущений получен, прибавляя полюса к множеству $N(L)$ полюсов, представляющих гигантский излом. Реакция последних полюсов снова разделена в нечетные и четные функции, как может быть замечено из уравнений (3.21), (3.22). Вместе два типа возмущений рационализируют и объясняют все особенности собственных значений и собственных функций, полученных из стандартного линейного анализа стабильности.

## Глава 4

# Динамика и волновая деформация радиально распространяющихся фронтов выведенная из законов подобия, полученных для канальной геометрии.

## 4.1 Введение

Главная идея этой главы – это вывод о том что, чтобы получить законы подобия для неустойчивого распространения фронта в радиальной геометрии, полезно изучить сначала распространение в геометрии канала в присутствии шума, в которых динамика без шума обычно приводит к простым формам продвигающихся фронтов [1].

Для понимания радиальной геометрии требуется управление влиянием шума на неустойчивую динамику распространения. Особенно трудно достигнуть такого управления в радиальной геометрии из-за неопределенности различия между внешними шумами и зашумленными начальными условиями. Геометрия канала более проста, так как она устанавливает стационарное решение для роста в пределе без шума. Можно тогда изучить эффекты внешнего шума в такой геометрии без какой-либо двусмысленности. Если мы находим правила, чтобы перевести получающееся понимание эффектов шума для роста в канале к радиальной геометрии, то можно получить законы подобия в этой радиальной геометрии в удовлетворительной форме. Мы будем иллюстрировать детали такого перевода в контексте пламени в смеси газов, который существует как самоподдерживающийся фронт экзотермических химических реакций при газообразном сгорании. Но наше утверждение - то, что подобные идеи должны быть плодотворными также в других контекстах неустойчивого распространения фронта. Само собой, разумеется, есть и аспекты динамики фронта и статистики в радиальной геометрии, которые *не могут* быть объяснены из наблюдений за фронтом в геометрии канала; примеры таких аспектов обсуждены в конце этой главы.

Математически наш пример описан [11] уравнением движения для угловой зависимости модуля радиуса - вектора фронта пламени, $R(\theta, t)$:

$$\begin{aligned}
\frac{\partial R}{\partial t} &= \frac{U_b}{2R_0^{\,2}(t)}\left(\frac{\partial R}{\partial \theta}\right)^2 + \frac{D_M}{R_0^{\,2}(t)}\frac{\partial^2 R}{\partial \theta^2} \\
&\quad + \frac{\gamma U_b}{2R_0(t)}I(R) + U_b \, .
\end{aligned} \qquad (4.1)$$

Здесь $0 < \theta < 2\pi$ являются углом, а константы $U_b$, $D_M$ и $\gamma$ являются скоростью фронта для идеального цилиндрического фронта, коэффициентом диффузии Маркштейна и коэффициента теплового расширения соответственно. $R_0(t)$ средний радиус распространяющегося пламени:

$$R_0(t) = \frac{1}{2\pi}\int_0^{2\pi} R(\theta, t)d\theta \, . \qquad (4.2)$$

Оператор $I(R)$ лучше всего представлен в терминах его разложения в ряд Фурье. Его компонент Фурье $|k|R_k$, где $R_k$ компонент Фурье $R$. Моделирования этого уравнения, так же как эксперименты в параметрическом режиме, для которого это уравнение релевантно, указывают, что в течение больших времен $R_0$ растет во времени по степенному закону как

$$R_0(t) = (const + t)^\beta \, , \qquad (4.3)$$

с $\beta > 1$, и ширина поверхности раздела $W$ растет с $R_0$ как

$$W(t) \sim R_0(t)^\chi \, , \qquad (4.4)$$

с $\chi < 1$.

## 4.2 Геометрия развивающегося фронта пламени: анализ на основе разложения на полюса

Исследование растущих фронтов в нелинейной физике [1] предлагает очаровательные примеры непосредственной генерации фрактальной геометрии [2,3]. Продвигающиеся фронты редко остаются плоскими. Обычно они формируют или рекурсивные объекты с искаженным и разветвленным проявлением, как, например, картины Лапласовского роста или ограниченная диффузией агрегация (DLA) [31]. Или же они остаются графами, но "огрубляются" в смысле производства самоподобных фракталов, "ширина" которых расходится с линейным масштабом системы с неким характерным показателем степени.

Исследование роста поверхности раздела, где огрубление вызвано шумом окружающей среды, или с отожженным или с подавленным шумом, было предметом активного исследования в последние годы [32,33]. Эти исследования имели значительный успех, и есть существенное аналитическое понимание природы классов универсальности, которые могут ожидаться. Исследование огрубление фронта в системе, в которой плоская поверхность в принципе неустойчива, менее разработано. Одним интересным примером, который привлек внимание, является уравнение Курамото-Севашинского [9,34], которое, как известно, грубеет в 1+1 измерении, но, как утверждают, не грубеет в высших размерностях [42]. Другой выдающийся пример - Лапласовский рост [35]. Эта глава мотивирована новым примером динамики распространяющегося наружу пламени, чей фронт морщится и фрактализуется [11]. Мы увидим, что у этой задачи есть много особенностей, которые близко напоминают Лапласовский рост, включая существование единственного пальца при распространении в канале, расщепление такого пальца на ветки при цилиндрическом росте, направленном наружу, экстремальная чувствительность к шуму, и т.д. В случае фронтов пламени уравнение движения поддается аналитическим решениям, и в результате мы можем понять некоторые из этих проблем.

Физическая задача, которая мотивирует этот анализ, является задача о пламени в смеси газов, которое существует как самоподдерживающийся фронт экзотермических химических реакций при газообразном сгорании. Было известно в течение некоторого времени, что такое пламя - характерно неустойчивое [43]. Сообщалось, что такое пламя развивает характерные структуры, которыйе включают точки излома, и что при обычных экспериментальных условиях фронт пламени ускоряется во времени [10]. В недавней работе Филянд и др. [11] предложил уравнение движения, которое мотивировано физикой и, кажется, описывает многие существенные и наблюдаемые особенности. Уравнение написано в цилиндрической геометрии и для $R(\theta, t)$, который является модулем радиуса - вектора на фронте пламени:

$$\frac{\partial R}{\partial t} = \frac{U_b}{2R_0^2(t)}\left(\frac{\partial R}{\partial \theta}\right)^2 + \frac{D_M}{R_0^2(t)}\frac{\partial^2 R}{\partial \theta^2} \qquad (4.5)$$
$$+ \frac{\gamma U_b}{2R_0(t)}I(R) + U_b \ .$$

Здесь $0 < \theta < 2\pi$ являются углом, а константы $U_b$, $D_M$ и $\gamma$ являются скоростью фронта для идеального цилиндрического фронта, коэффициентом диффузии Маркштейна и коэффициентом теплового расширения, соответственно. $R_0(t)$ средний радиус распространяющегося пламени:

$$R_0(t) = \frac{1}{2\pi}\int_0^{2\pi} R(\theta, t)d\theta \ . \qquad (4.6)$$

Оператор $I(R)$ лучше всего представляется в терминах его Фурье разложения. Его компонента Фурье $|k|R_k$, где $R_k$ компонента Фурье $R$.

Численное моделирование того типа, о котором сообщали в ссылке [11], представлено на Рис. 10.1. Две самые характерные особенности этого моделирования – волновая деформация, появляющаяся в виде многих изломов на фронте, и его ускорение во времени. Можно наблюдать явление расщепление максимума на фронте, при котором новые точки излома добавляются к распространяющемуся фронту между существующими точками излома. И эксперименты и моделирования указывают, что в течение больших времен $R_0$ растет по степенному закону во времени.

$$R_0(t) = (const + t)^\beta \ , \qquad (4.7)$$

с $\beta > 1$, (порядка 1.5) и ширина поверхности раздела $W$ растет с $R_0$ как

$$W(t) \sim R_0(t)^\chi \ , \qquad (4.8)$$

с $\chi < 1$ (порядка 2/3). Понимание этих двух особенностей и вывод соотношения между степенями $\beta$ и $\chi$ являются главными целями этой главы.

Уравнение (4.6) может быть написано как однопараметрическое уравнение, перемасштабированием $R$ и $t$ согласно $r \equiv RU_b/D_M$, $\tau \equiv tU_b^2/D_M$. Вычисляя производную уравнения Eq.(4.6) относительно $\theta$ и подставляя безразмерные переменные, получаем:

$$\frac{\partial u}{\partial \tau} = \frac{u}{r_0^2}\frac{\partial u}{\partial \theta} + \frac{1}{r_0^2}\frac{\partial^2 u}{\partial \theta^2} + \frac{\gamma}{2r_0}I\{u\} \ . \qquad (4.9)$$

где $u \equiv \frac{\partial r}{\partial \theta}$. Чтобы сделать замкнутым это уравнение, мы нуждаемся во втором для $r_0(t)$, который получается усреднением (4.6) по углам и таким же перемасштабированием как сделаное выше. Результат:

$$\frac{dr_0}{d\tau} = \frac{1}{2r_0^2}\frac{1}{2\pi}\int_0^{2\pi} u^2 d\theta + 1 \ . \qquad (4.10)$$

Эти два уравнения - основание для дальнейшего анализа.

Следуя [12, 14–16, 38, 39] мы раскладываем теперь решение $u(\theta, \tau)$ в полюса, положения которых $z_j(\tau) \equiv x_j(\tau) + iy_j(\tau)$ в комплексной плоскости описываются временной зависимостью:

$$\begin{aligned}
u(\theta, \tau) &= \sum_{j=1}^{N} \cot\left[\frac{\theta - z_j(\tau)}{2}\right] + c.c. \qquad (4.11)\\
&= \sum_{j=1}^{N} \frac{2\sin[\theta - x_j(\tau)]}{\cosh[y_j(\tau)] - \cos[\theta - x_j(\tau)]} \ ,
\end{aligned}$$

$$r(\theta, \tau) = 2\sum_{j=1}^{N} \ln\left[\cosh(y_j(\tau)) - \cos(\theta - x_j(\tau))\right] + C(\tau) \ . \qquad (4.12)$$

В (4.12) $C(\tau)$ является функцией времени. Функция (4.12) является наложением квази-изломов (то есть изломов, которые округлены на их вершинах). Вещественная часть положения полюса (то есть $x_j$) описывает угловую координату максимума квази-излома, и мнимая часть положения полюса (то есть $y_j$) связана высота квази-излома. Когда $y_j$ уменьшается (увеличивается) высота излома увеличивается (уменьшается). Физическая мотивация для такого представления решений должна быть очевидна из Рис. 10.1.

Главное преимущество этого представления состоит в том, что распространение и волновая деформация фронта могут быть описаны теперь через динамику полюсов и $r_0(t)$. Подставляя (4.11) в (4.9), мы получаем следующие обыкновенные дифференциальные уравнения для положений полюсов:

$$-r_0^2\frac{dz_j}{d\tau} = \sum_{k=1, k\neq j}^{2N} \cot\left(\frac{z_j - z_k}{2}\right) + i\frac{\gamma r_0}{2}sign[Im(z_j)] \ . \qquad (4.13)$$

После подстановки (4.11) в (4.10) мы получаем, используя (4.13) обычное дифференциальное уравнение для $r_0$,

$$\frac{dr_0}{d\tau} = 2\sum_{k=1}^{N} \frac{dy_k}{d\tau} + 2\left(\frac{\gamma}{2}\frac{N}{r_0} - \frac{N^2}{r_0^2}\right) + 1 \ . \qquad (4.14)$$

В задаче роста, направленного наружу, вводятся важные поправки к результатам, полученным для канала. Число полюсов в устойчивой конфигурации пропорционально к радиусу $r_0$ вместо $L$, и он растет во времени. Система становится, поэтому, неустойчивой к добавлению новых полюсов. Если будет шум в системе, который может сгенерировать новые полюса, то они не будут уже отброшены в бесконечность вдоль $y$. Важно подчеркнуть, что любой бесконечно малый шум (или числовой или экспериментальный) достаточен, чтобы произвести новые полюса. Эти новые полюса не обязательно сливают свои $x$-положения с существующими точками излома. Даже притом, что есть притяжение вдоль действительной оси как и в случае канала, есть и растяжение расстояний между полюсами из-за радиального роста. Это может уравновесить притяжение. Наша первая новая идея - то, что эти две противодействующих тенденции определяют типичный масштаб, обозначенный как $\mathcal{L}$. Пусть у нас есть излом, который получен из $x$-слияния $N_c$ полюсов на линии $x = x_c$, и мы хотим знать, сольется ли $x$-соседний полюс с действительной координатой $x_1$ с этим большим изломом. Ответ зависит от расстояния $D = r_0|x_c - x_1|$. Существует длина $\mathcal{L}(N_c, r_0)$ такая что, если $D > \mathcal{L}(N_c, r_0)$, то одиночный излом никогда не сольется с большим изломом. В противоположном пределе одиночный излом сдвинет большой излом до их $x$-слияния положений, и большой излом будет иметь $N_c + 1$ полюсов.

Этот факт вытекает непосредственно из уравнений движения для $N_c$ полюсов в позиции $x$ и единственного полюса в точке $x_1$. Отметим, что из уравнения 4.7 (которое еще не объяснено) следует, что асимптотически $r_0(\tau) = (a+\tau)^\beta$ где $r_0(0) = a^\beta$. Затем начнем с 4.13 и напишем уравнение для углового расстояния $x = x_1 - x_c$. Из него следует для любой конфигурации $y_j$ вдоль мнимой оси

$$\frac{dx}{d\tau} \leq -\frac{2N_c \sin x [1 - \cos x]^{-1}}{(a+\tau)^{2\beta}} = -\frac{2N_c \cot(\frac{x}{2})}{(a+\tau)^{2\beta}} \ . \tag{4.15}$$

Для малых x мы получаем

$$\frac{dx}{d\tau} \leq -\frac{4N_c}{x(a+\tau)^{2\beta}} \ . \tag{4.16}$$

Решение этого уравнения:

$$x(0)^2 - x(\tau)^2 \geq \frac{8N_c}{2\beta - 1}(a^{1-2\beta} - (a+\tau)^{1-2\beta}) \ . \tag{4.17}$$

Чтобы найти $\mathcal{L}$, мы устанавливаем $x(\infty) > 0$, из которых мы находим это угловое расстояние останется конечным пока

$$x(0)^2 > \frac{8N_c}{2\beta - 1}a^{1-2\beta} \ . \tag{4.18}$$

Так как $r_0 \sim a^\beta$ мы находим пороговый угол $x^*$

$$x^* \sim \sqrt{N_c} r_0^{\frac{(1-2\beta)}{2\beta}} \ , \tag{4.19}$$

выше которого нет никакого слияния между гигантским изломом и изолированным полюсом. Чтобы найти фактическое расстояние $\mathcal{L}(N_c, r_0)$ мы умножаем угловое расстояние $r_0$ и находим

$$\mathcal{L}(N_c, r_0) \equiv r_0 x^* \sim \sqrt{N_c} r_0^{\frac{1}{2\beta}} \ . \tag{4.20}$$

Чтобы понять геометрическое значение этого результата, мы вспоминаем особенности решения для излома TFH. Имея типичную длину $L$ число полюсов в изломе линейно по $L$. Аналогично, если у нас будет в этой задаче два излома на расстоянии $2\mathcal{L}$ друг от друга, то число $N_c$ в каждом из них будет иметь порядок $\mathcal{L}$. Из (4.20) тогда следует, что

$$\mathcal{L} \sim r_0^{\frac{1}{\beta}} . \qquad (4.21)$$

Для $\beta > 1$ окружность растет быстрее, чем $\mathcal{L}$. Поэтому в некоторые временные моменты появляются полюса между двумя большими изломами и не притягиваются к кому-либо из них, а образуют новые изломы. Мы покажем позже, что самые неустойчивые положения к появлению новых изломов - точно середина между существующими изломами. Это - механизм для появления дополнительных изломов по аналогии с «пальцем», разделяющимся при Лапласовском росте.

Мы можем теперь оценить ширину фронта пламени как высоту наибольших изломов. Так как эта высота пропорциональна $\mathcal{L}$ (смотри свойство (v) решения TFH), уравнение (4.21) и уравнение (4.8) ведет к отношению между показателями степеней

$$\chi = 1/\beta . \qquad (4.22)$$

Этот закон подобия, как ожидается, будет выполняться полностью и для $\beta = 1$, для которого не ускоряется фронт пламени, и размер изломов становится пропорциональным $r_0$.

В каналах есть естественная характерная длина – это ширина $\tilde{L}$ канала. Использование результатов, полученных в канале, для радиальной геометрии будет основано на идентификации в этой радиальной геометрии характерного масштаба с временной зависимостью $\mathcal{L}(t)$, который играет роль $\tilde{L}$ в канальной геометрии. Чтобы сделать это, мы должны сначала рассмотреть кратко главные релевантные результаты для распространения пламени в канале в присутствии шума. В геометрии канала уравнение движения написано в терминах положения $h(x,t)$ фронта пламени выше $x$-оси. После соответствующих перемаштабирований [12] получается:

$$\frac{\partial h(x,t)}{\partial t} = \frac{1}{2}\left[\frac{\partial h(x,t)}{\partial x}\right]^2 + \nu \frac{\partial^2 h(x,t)}{\partial x^2} + I\{h(x,t)\} + 1. \qquad (4.23)$$

Удобно повторно перемасштабировать область определения к $0 < \theta < 2\pi$, и изменить переменные на $u(\theta, t) \equiv \partial h(\theta, t)/\partial \theta$. В терминах этой функции мы находим

$$\frac{\partial u(\theta,t)}{\partial t} = \frac{u(\theta,t)}{L^2}\frac{\partial u(\theta,t)}{\partial \theta} + \frac{\nu}{L^2}\frac{\partial^2 u(\theta,t)}{\partial \theta^2} + \frac{1}{L}I\{u(\theta,t)\} \qquad (4.24)$$

где $L = \tilde{L}/2\pi$. В условиях без шума это уравнение допускает точные решения, которые представлены в терминах $N$ полюсов, положения которых $z_j(t) \equiv x_j(t) + iy_j(t)$ в комплексной плоскости определяются следующей временной зависимостью:

$$u(\theta,t) = \nu \sum_{j=1}^{N} \cot\left[\frac{\theta - z_j(t)}{2}\right] + c.c. , \qquad (4.25)$$

Стационарное состояние для распространения в канале является единственным и линейно устойчивым; оно состоит из $N(L)$ полюсов, которые выровнены на одной линии, параллельной мнимой оси. Геометрическое проявление фронта пламени - гигантский излом, аналогичная единственному пальцу в случае Лапласовского роста в канале. Высота излома пропорциональна $L$, и скорость распространения постоянна. Число полюсов в гигантском изломе ли-нейно по $L$ согласно уравнению (1.10. Введение аддитивного случайного шума в динамику изменяет картину качественно. Удобно добавить шум к уравнению движения в Фурье представлении, добавляя белый шум $\eta_k$ для каждой $k$ моды. Корреляционная функция для шума удовлетворяет отношению $<\eta_k(t)\eta_{k'}(t')> = \delta_{k,k'}\delta(t-t')f/L$. Шум при нашем моделировании взят из плоского однородного распределения в интервале $[-\sqrt{2f/L}, \sqrt{2f/L}]$. Это гарантирует, что, когда размер системы изменяется, типичный шум на единицу длины фронта пламени остается постоянным. Это показано в главе 2 и [17], что для умеренного, но фиксированного уровня шума средняя скорость $v$ фронта увеличивается как функция $L$ по степенному закону. При нашем моделировании мы нашли

$$v \sim L^\mu, \quad \mu \approx 0.35 \pm 0.03 . \qquad (4.26)$$

Для фиксированного размера системы $L$ у скорости есть также зависимость в виде степенного закона от уровня шума, но с намного меньшим показателем степени: $v \sim f^\xi \quad \xi \approx 0.02$. Эти результаты были поняты теоретически, анализируя создание новых полюсов шумом, которые взаимодействуют с полюсами, определяющими гигантский излом (в главе 2 и [17]).

Затем мы проливаем свет на явление расщепления максимума. Здесь этот эффект проявляется как добавление новых изломов примерно посередине между уже существующими. Мы упоминали неустойчивость к добавлению новых полюсов. Мы покажем теперь, что максимум между изломами является самым нестабильным по отношению к появлению новых полюсов. Это можно показать и в канале и в радиальной геометрии. Например, рассмотрим TFH-гигантское решение для излома, в котором все полюса сосредоточены (без потери общности) на $x = 0$ линии. Прибавим новый полюс с комплексной позицией $(x_a, y_a)$ к существующим $N(L)$ полюсам, и изучим его судьбу. Можно показать, что в пределе $y_a \to \infty$ (в этом пределе соответствующее ему возмущение обращается в ноль) уравнение движения

$$\frac{dy_a}{dt} = \frac{2\pi\nu}{L^2}(2N(L)+1) - 1 \qquad y_a \to \infty . \qquad (4.27)$$

Так как $N(L)$ удовлетворяет (1.10), это уравнение может быть переписано как

$$\frac{dy_a}{dt} = \frac{4\pi\nu}{L^2}(1-\alpha) \qquad y_a \to \infty , \qquad (4.28)$$

где $\alpha = (L/(2\pi\nu) + 1)/2 - N(L)$. Очевидно, что $\alpha \leq 1$ и это точно 1 только, когда $L$ удовлетворяет $L = (2n+1)2\pi\nu$. Затем можно показать что для $y_a$ намного большего, чем $y_{N(L)}$, но не бесконечного верно следующее:

$$\frac{dy_a}{dt} > \lim_{y_a \to \infty} \frac{dy_a}{dt} \quad x_a = 0 \qquad (4.29)$$

$$\frac{dy_a}{dt} < \lim_{y_a \to \infty} \frac{dy_a}{dt} \quad x_a = \pi \qquad (4.30)$$

Мы узнаем из этих результатов, что существуют значения $L$, для которых полюс, который добавлен в бесконечности, будет иметь нулевое отталкивание ($dy_a/dt = 0$). Подобное понимание может быть получено и из стандартного анализа стабильности, не используя разложение на полюса. Возмущая решение в виде TFH-излома, мы находим линейные уравнения, собственные значения которых $\lambda_i$ мо-гут быть получены стандартными числовыми методами: (i) все $Re(\lambda_i)$ неположительные. (ii) при изолированных, значениях $L$, для которых $L = (2n+1)2\pi\nu$ $Re(\lambda_1)$ и $Re(\lambda_2)$ становятся нулем (отметим, что из-за логарифмического масштаба ноль не очевиден), (iii) существует общая тенденция для всех $Re(\lambda_i)$ приближаться к нулю по абсолютной величине как $\frac{1}{L^2}$, когда $L$ увеличивается. Это указывает на растущую чувствительность к шуму, когда размер системы увеличивается. (iv) существует мода Голдстоуна $\lambda_0 = 0$ из-за трансляционной инвариантности.

Результат этого обсуждения - то, что конечные возмущения (то есть полюса в при конечном $y_a$) будут расти, если $x$ положение полюса будет достаточно близко к максимуму. Положение $x = \pi$ (максимум пальца) является самым неустойчивым. В геометрии канала это означает, что шум приводит к появлению новых изломов в максимуме пальцев, но из-за притяжения к гигантскому излому, они движутся к $x = 0$ и исчезают в гигантском изломе. Фактически, при численном моделировании виден поезд из небольших изломов, которые перемещаются к гигантскому излому. Анализ показывает, что в то же самое время самый далекий полюс в $y_{N(L)}$ выбрасывается на бесконечность. Также и в цилиндрической геометрии большинство позиций неустойчивых к появлению новых изломов расположены между двумя существующими изломами независимо от того, является ли система предельной(общее количество полюсов соответствует радиусу), или неустойчивой (общее количество полюсов является слишком небольшим для данного радиуса). Ведет ли добавление нового полюса к расщеплению максимума, зависит от их $x$ положения. Так, когда расстояние между существующими изломами больше чем $\mathcal{L}$, новые полюса, которые произведены шумом, остаются около максимума между этими двумя изломами и вызывает расщепление максимума.

Используемая картина остается справедливой пока полюса, которые создаются шумом, не разрушают идентичность гигантского излома. Действительно, численное моделирование показывает, что в присутствии умеренного шума дополнительные полюса появляются как маленькие изломы, которые постоянно бегут к гигантскому излому. Наша цель тут, не предсказать числовые значения показателей степени в *канале* (это было сделано в [17] и главе 2), а использовать их, чтобы предсказать показатели степени, характеризующие ускорение и геометрию фронта пламени в *радиальной* геометрия.

На первый взгляд кажется, что в радиальной геометрии картина роста качественно отлична. Фактически, тщательное наблюдение за картиной роста (см. Рис. 10.1) показывает, что большую часть времени там существуют некие большие изломы, которые притягивают другие меньшие изломы, но что время от времени "новые"большие изломы формируются и начинают действовать как локальные поглотители небольших изломов, которые появляются беспорядочно. Понимание этого явления дает подсказку, как применить результаты, полученные для канала, для радиального роста.

Уравнения (4.9), (4.10) снова допускают точные решения в терминах полюсов, формы уравнения (4.13). Легко записать уравнения движения полюсов и проверить, что полюса притягиваются вдоль действительного направления. Это означает физически, что они притягиваются вдоль угловой координаты. но они отталкиваются вдоль мнимого направления, которое ассоциируется с радиальной координатой. Если бы не растяжение, которое вызвано увеличением радиуса (и с ним периметра), все полюса слились бы в один гигантский излом. Таким образом, у нас есть соревнование между притяжением полюсов и растяжением. Так как притяжение уменьшается с увеличением расстояния

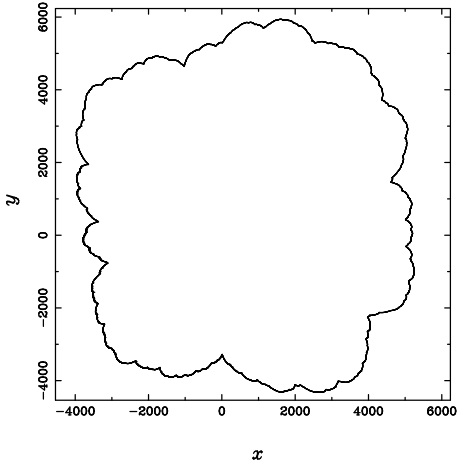

Рис. 4.1: Моделирования фронта пламени распространяющегося наружу. Отметим, что имеется широкое распределение размеров изломов.

между полюсами в угловых направлениях, всегда есть начальный критический масштаб длины, выше которого полюса не могут слиться вдоль действительной координаты, когда время прогрессирует.

Предположим теперь, что шум прибавляет новые полюса к системе. Полюса не обязательно сливают свои вещественные положения с существующими изломами. Пусть имеется большой излом, образованным из слияния действительных координат $N_c$ полюсов в точке $x_c$. Мы хотим знать, сольётся ли соседний полюс с действительной координатой $x_1$ с этим большим изломом. Ответ будет зависеть, конечно, от расстояния $D \equiv r_0|x_c - x_1|$. Прямое вычисление [19], используя уравнение движений для полюсов, показывает, что существует критическая длина $\mathcal{L}(r_0)$. Таким образом, если $D > \mathcal{L}(r_0)$ единственный полюс никогда не сливается с гигантским изломом. Результат вычисления - это

$$\mathcal{L} \sim r_0^{1/\beta} . \tag{4.31}$$

Заметим, что невозможность даже для одного полюса притянуться к большому излому означает, что раскол максимума произошел. Это - точный аналог раскола пальца в Лапласовском росте.

Теперь пришло время связать канал и радиальную геометрию. Мы идентифицируем типичный масштаб в радиальной геометрии как $\mathcal{L} \sim W \sim r_0^\chi$. С одной стороны, это приводит к соотношению между показателями степеней $\beta = 1/\chi$. С другой стороны, мы используем результат, уже установленный для канала, (4.26). Затем совмещаем его с этой идентификацией масштаба. В итоге мы находим $\dot{r}_0 = r_0^{\chi\mu}$. Сравнивая с (4.7) мы находим:

$$\beta = \frac{1}{(1 - \chi\mu)} . \tag{4.32}$$

Этот результат подводит нас к тому, чтобы ожидать два динамических режима для нашей задачи. Начинаясь с гладких начальных условий, в относительно короткие времена показатель степени грубости остается близким к

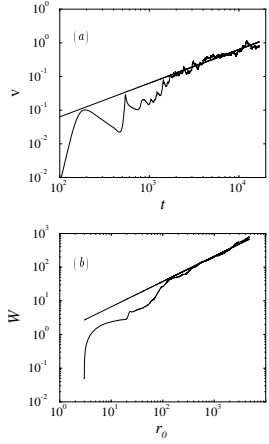

Рис. 4.2: Панель а: логарифмический график скорости как функции времени для радиальной системы. Параметры моделирования: $f = 10^{-8}$, $\gamma = 0.8$, $\nu = 1$. Панель b: Логарифмический график ширины фронта пламени как функция среднего радиуса.

единице. Это, главным образом, потому, что масштаб $\mathcal{L}$ еще не релевантен, и большинство полюсов, которые сгенерированы шумом, сливаются в несколько больших изломов. В более поздние времена грубеющий показатель степени устанавливается в его асимптотическом значении. Также и все асимптотические отношения подобия, используемые выше, становятся справедливыми. Мы, таким образом, ожидаем, что $\beta$ уменьшится с $1/(1 - \mu)$ до асимптотического значения, определяемого $\chi = 1/\beta$ в (4.32):

$$\beta = 1 + \mu \approx 1.35 \pm 0.03 \ . \tag{4.33}$$

Ожидаемая величина $\chi$ является, таким образом, $\chi = 0.74 \pm 0.03$. Мы проверяли эти предсказания при численном моделировании. Мы интегрировали уравнение (4.24), и на Рис. 10.2 мы показываем результаты для скорости роста как функцию времени. После ограниченной области экспоненциального роста мы наблюдаем непрерывное уменьшение во времени показателя степни. В начальной области мы получаем $\beta = 1.65 \pm 0.1$. В заключительной декаде временного диапазона мы находим $\beta = 1.35 \pm 0.1$. Мы считаем это хорошим совпадением с (4.33). Второе важное испытание предоставлено измерением ширины системы как функции радиуса. Результат можно видеть на Рис. 10.2b. Снова мы наблюдаем переход, связанный с начальной динамикой; В последней временной декаде показатель степени устанавливается в $\chi = 0.75 \pm .1$. Мы заключаем, что во времена достаточно большие, чтобы наблюдать асимптотику, наши предсказания проверены.

Наконец, мы подчеркиваем некоторые различия между радиальной и канальной геометриями. Фронты в канале показывают главным образом один гигантский излом, который только незначительно нарушен небольшими изломами, которые создаются шумом. В радиальной геометрии, как можно заключить из обсуждения выше, могут существовать в любое время изломы всех размеров, от наименьшего до наибольшего. Это широкое распределение изломов (и

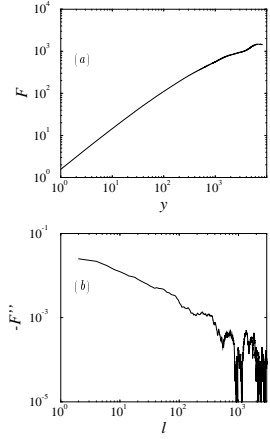

Рис. 4.3: Панель a: логарифмический график корреляционной функции F(y) поверхности раздела как функция расстояния y между точками. Панель b: Вторая производная корреляционной функции -F"(l) поверхности раздела как функция расстояния l между точками.

масштабов) должно влиять на корреляционную функцию способом, который отличается качественно от корреляционных функций, вычисленных в конфигурациях канала. Чтобы ясно понять этот момент, мы показываем на Рис. 10.3a структурную функцию

$$F(y) \equiv \sqrt{\langle |R(x+y) - R(x)|^2 \rangle} \qquad (4.34)$$

вычисленную для типичного радиального фронта, с $x = R\theta$. Чтобы подчеркнуть масштабированную область, мы показываем вторую производную этой функции в Рис. 10.3b. Нижний конец диаграммы может быть изображен хорошо степенным законом $y^{-\alpha}$ с $\alpha \approx 0.6$. Это указывает что $F(y) \approx Ay + By^{2-\alpha}$. В геометрии канала мы получаем полностью иную структурную функцию, которая не показывает такое степенное поведение как функция, вычисленная выше. Способ понять это поведение в радиальной геометрии состоит в том, чтобы рассмотреть распределение изломов, которые остаются отличными друг от друга, но чьи масштабы распределены согласно некоторому распределению $P'(\ell)$. Позвольте нам определить функцию распределения $P(\ell) \sim \ell P'(\ell)$. Эта функция дает нам вероятность того, что точка на цепи со средним радиусом лежит на основе излома с размером $\ell$. Каждый из этих изломов дает вклад в корреляционную функцию формы $f(y, \ell) \approx \ell g(y/\ell)$, где $g(x)$ является степенной функцией, $g(x) \approx x$ for $x < 1$ и $g(x) \approx$ постоянная для $x > 1$. Полная корреляция может быть оценена (когда полюса различны) как

$$F(y) \approx \sum_{\ell} P(\ell) \ell g(y/\ell) \,. \qquad (4.35)$$

Первая производная дает нам $\sum_{\ell} P(\ell) g'(y/\ell)$, и используя факт, что $g'$ стремится к нулю для $x > 1$, мы оцениваем $F'(y) = \sum_{\ell=y}^{W} P(\ell)$. Вторая производная дает $F''(y) \approx -P(y)$. Таким образом, структурная функция определена распре-делением масштабов изломов, и если это распределение описывается степенным законом, то это должно быть замечено во второй производной $F(y)$ как демонстрируется на Рис. 10.3. Заключение из этого анализа - то, что радиальный случай дает степенную функцию, которая характеризует распределением изломов, $P(\ell) \approx \ell^{-\alpha}$.

## 4.3 Выводы

Главная цель этой главы состояла в том, чтобы найти показатели степени задачи и найти связи между ними. Используя главный результат случая канала (зависимость скорости от размера канала) мы можем найти ускорение фронта пламени в радиальной геометрии $(R_0(t) = (const+t)^\beta)$

$$\beta = 1 + \mu \ . \tag{4.36}$$

где $\mu$ показатель степени для зависимости скорости от размера канала, найденном прежде, $\beta$ показатель степени ускорения, и $R_0$ средний радиус фронта пламени.

Зависимость ширины фронта пламени в радиальном случае $W(t)$ от среднего радиуса $(R_0(t) = (const+t)^\beta)$

$$W(t) \sim R_0(t)^\chi \ . \tag{4.37}$$

$$\chi = 1/\beta \ . \tag{4.38}$$

В итоге мы продемонстрировали, что возможно использовать информацию о динамике канала в присутствии шума, чтобы предсказать нетривиальные особенности радиальной эволюции, такие как ускорение и показатели степени для грубости поверхности. Стоило бы исследовать подобные идеи в контексте картин Лапласовского роста.

# Глава 5

# Лапласовский Рост

## 5.1 Введение.

Проблема образования сложных структур - одна из наиболее быстро развивающихся ветвей нелинейной науки сегодня [1]. Специальный интерес представляет исследование динамики фронта между двумя фазами (поверхность раздела), которая возникает во множестве неравновесных физических систем. Если, как это обычно случается, движение поверхности раздела медленное по сравнению с процессами, которые имеют место в большой части обеих фаз (таких как теплообмен, диффузия, и т.д.), скалярное поле, управляющее эволюцией поверхности раздела, является гармонической функцией. Это естественно тогда, назвать этот процесс . В зависимости от системы это гармоническое скалярное поле - температура (в замораживании жидкости или задача Стефана), концентрация (в затвердевании из пересыщенного раствора), электростатический потенциал (при электролитическом осаждении), давление (в потоках через пористые среды), вероятность (в диффузно-ограниченной агрегации), и т.д.

Математическая задача Лапласовского роста без поверхностного натяжения устанавливает семейство точных (аналитических) решений в терминах логарифмических полюсов в комплексной плоскости. Мы показываем, что у этого семейства решений есть замечательное свойство: общие начальные условия в геометрии канала, которые начинаются с произвольно большого числа особенностей, показывают обратный каскад в единственный палец.

В отсутствии поверхностного натяжения, эффект которого должен стабилизировать возмущения короткой длины волны поверхности раздела, двумерная задача Лапласовского роста описана следующим образом

$$(\partial_x^2 + \partial_y^2)u = 0 \ . \tag{5.1}$$

$$u\mid_{\Gamma(t)} = 0 \ , \partial_n u \mid_{\Sigma} = 1 \ . \tag{5.2}$$

$$v_n = \partial_n u \mid_{\Gamma(t)} \ . \tag{5.3}$$

Здесь $u(x, y; t)$ - упомянутое выше скалярное поле, $\Gamma(t)$ - движущаяся поверхность раздела, $\Sigma$ - фиксированная внешняя граница, $\partial_n$ - компонент градиента, нормального к границе (то есть производная по нормали), и $v_n$ - нормальная компонента скорости фронта.

Мы рассматриваем бесконечно длинную поверхность раздела, полученную периодическим продолжением поверхности раздела в канале с периодическими граничными условиями. Тогда мы представляем конформное отображение с временной зависимостью $f$ от более низкой половины "математической" плоскости, $\xi \equiv \zeta + i\eta$, к области физической плоскости, $z \equiv x + iy$, где уравнение Лапласа 5.1 определено как $\xi \xrightarrow{f} z$. Мы также требуем что $f(t,\xi) \approx \xi$ for $\xi \longrightarrow \zeta - i\infty$. Таким образом функция $z = f(t,\zeta)$ описывает движущуюся поверхность раздела. Из уравнений (5.1), (5.2), (5.3) для функции $f(t,\xi)$ мы получаем

$$Im(\frac{\partial f(\xi,t)}{\partial \xi}\overline{\frac{\partial f(\xi,t)}{\partial t}}) = 1 \mid_{\xi=\zeta-i0}, f_\zeta \mid_{\zeta-i\infty}= 1 . \tag{5.4}$$

Теперь мы расширим эти результаты, полученные для периодических граничных условий, к более физическим граничным условиям "без потока" (никакого потока через боковые границы канала). Это требует, чтобы движущаяся поверхность раздела ортогонально пересекла стены канала. Однако, в отличие от случая периодических граничных условий, у точек конца на этих двух границах не обязательно есть та же самая горизонтальная координата. Это - также периодическая задача, где период равняется дважды ширине канала. Анализ - такой же, как прежде, но теперь только половину решения нужно рассмотреть как физический канал, тогда как вторая половина является нефизическим зеркальным отображением.

Позвольте нам искать решение уравнения (5.4) в следующей форме

$$f(\xi,t) = \lambda \xi - i\tau(t) - i\sum_{l=1}^{N} \alpha_l \log(e^{i\xi} - e^{i\xi_l(t)}),$$
$$\sum_{l=1}^{N} \alpha_l = 1 - \lambda, -1 < \lambda < 1 , \tag{5.5}$$

где $\tau(t)$ некоторая действительная функция времени, $\lambda$ действительная постоянная, $\alpha_l$ комплексная постоянная, $\xi_l = \zeta_l + i\eta_l$ обозначает, что положение полюса с номером $l$ и $N$ - число полюсов.

Для граничного условия "без потоков" мы должны прибавить условие, что для каждого полюса $\xi_l = \zeta_l + i\eta_l$ с $\alpha_l$ существует полюс $\xi_l = -\zeta_l + i\eta_l$ с $\overline{\alpha_l}$. Так для функции $F(i\xi,t) = if(\xi,t)$

$$\overline{F(i\xi,t)} = F(\overline{i\xi},t) \tag{5.6}$$

Мы хотим доказать, что конечное состояние будет только одним пальцем.

## 5.2 Асимптотитечское поведение полюсов в математической плоскости.

Главная цель этой главы состоит в том, чтобы исследовать асимптотическое поведение полюсов в математической плоскости. Мы хотим продемонстрировать, что с течением времени $t \mapsto \infty$, все полюса идут в единственную точку (или две точки для граничных условий без потоков). Уравнение для поверхности раздела

$$f(\xi,t) = \lambda\xi - i\tau(t) - i\sum_{l=1}^{N}\alpha_l \log(e^{i\xi} - e^{i\xi_l(t)}),$$

$$\sum_{l=1}^{N}\alpha_l = 1 - \lambda, -1 < \lambda < 1 \ . \tag{5.7}$$

Подставляя уравнение (5.7) в

$$Im(\frac{\partial f(\xi,t)}{\partial \xi}\overline{\frac{\partial f(\xi,t)}{\partial t}}) = 1 \mid_{\xi=\zeta-i0} \ , \tag{5.8}$$

мы можем найти уравнения движения полюса:

$$const = \tau(t) + (1 - \sum_{k=1}^{N}\overline{\alpha_k})\log\frac{1}{a_l} + \sum_{k=1}^{N}\overline{\alpha_k}\log(\frac{1}{a_l} - \overline{a_k}) \tag{5.9}$$

и

$$\tau = t - \frac{1}{2}\sum_{k=1}^{N}\sum_{l=1}^{N}\overline{\alpha_k}\alpha_l \log(1 - \overline{a_k}a_l) + C_0 \ , \tag{5.10}$$

где $a_l = e^{i\xi_l}$.

Из уравнения (5.9) мы можем найти

$$C_1 = (1-\lambda)\tau - \sum_{l=1}^{N}\alpha_l \log a_l + \sum_{k=1}^{N}\sum_{l=1}^{N}\overline{\alpha_k}\alpha_l \log(1 - \overline{a_k}a_l) \ . \tag{5.11}$$

Из уравнений (5.10) и (5.11) мы можем получить

$$Im(\sum_{l=1}^{N}\alpha_l \log a_l) = constant \tag{5.12}$$

И

$$t = (\frac{1+\lambda}{2})\tau + \frac{1}{2}Re(\sum_{l=1}^{N}\alpha_l \log a_l) + C_1/2 \ , \tag{5.13}$$

где $\alpha_l$ постоянная, $\xi_l(t)$ положение полюсов, $a_l = e^{i\xi_l(t)}$, и $\frac{\lambda+1}{2}$ является частью канала, занятого движущейся жидкостью. Мы увидим, что для $\tau \mapsto \infty$ мы получаем один палец с широким $\frac{\lambda+1}{2}$.

В Приложении A мы докажем из уравнения (5.10), что $\tau \mapsto \infty$ если $t \mapsto \infty$ и если за конечное время не возникает сингулярности.

Уравнения движения полюса являются следующими из уравнения (5.9)

$$const = \tau + i\overline{\xi_k} + \sum_{l}\alpha_l \log(1 - e^{i(\xi_l - \overline{\xi_k})}), \tag{5.14}$$

$$const = \zeta_k + \sum_{l}(\alpha_l'' \log \mid 1 - e^{i(\xi_l - \overline{\xi_k})} \mid + \alpha_l' \arg(1 - e^{i(\xi_l - \overline{\xi_k})})), \tag{5.15}$$

$$const = \tau + \eta_k + \sum_{l}(\alpha_l' \log \mid 1 - e^{i(\xi_l - \overline{\xi_k})} \mid - \alpha_l'' \arg(1 - e^{i(\xi_l - \overline{\xi_k})})), \tag{5.16}$$

$$\xi_l = \zeta_l + i\eta_l, \eta_l > 0 \ . \tag{5.17}$$

$$\alpha_l = \alpha_l' + i\alpha_l'' \tag{5.18}$$

Позвольте нам проверить

$$\arg(1 - e^{i(\xi_l - \overline{\xi_k})}) = \arg([1 - e^{i(\zeta_l - \zeta_k)}e^{-(\eta_l + \eta_k)}]) = \arg[1 - a_{lk}e^{i\varphi_{lk}}] \tag{5.19}$$

$$\varphi_{lk} = \zeta_l - \zeta_k, a_{lk} = e^{-(\eta_l + \eta_k)} \tag{5.20}$$

$\arg[1 - a_{lk}e^{i\varphi_{lk}}]$ является однозначной функцией от $\varphi_{lk}$, то есть

$$-\frac{\pi}{2} \leq \arg[1 - a_{lk}e^{i\varphi_{lk}}] \leq \frac{\pi}{2} \ . \tag{5.21}$$

Из уравнения (5.16) единственный способ компенсировать расходимость члена $\tau$ состоит в том, что $\eta_k \mapsto 0$ для $\tau \mapsto \infty, 1 \leq k \leq N$.

Мы хотим исследовать асимптотическое поведение полюсов $\tau \mapsto \infty$. Чтобы устранить расходящийся член $\log |1 - e^{i(\xi_k - \overline{\xi_k})}|$, мы умножаем уравнение (5.16) by $\alpha_k''$ и уравнение (5.15) на $\alpha_k'$ и берем разность

$$const = \alpha_k' \zeta_k - \alpha_k'' \tau + \sum_{l \neq k}((\alpha_l'' \alpha_k' - \alpha_k'' \alpha_l') \log |1 - e^{i(\xi_l - \overline{\xi_k})}| +$$
$$(\alpha_l' \alpha_k' + \alpha_l'' \alpha_k'') \arg(1 - e^{i(\xi_l - \overline{\xi_k})})). \tag{5.22}$$

У нас есть расходящиеся члены $\alpha_k'' \tau$ в этом уравнении. Мы можем предположить, что для $t \mapsto \infty$, $N'$ группы полюсов существуют, чтобы устранить расходящиеся члены ($\varphi_{lk} \mapsto 0$ для всех элементов группы). $N_l$ - число полюсов в каждой группе, $1 < l < N'$. Для каждой группы суммированием уравнений (5.22) по всем групповым полюсам мы получаем

$$const = \alpha_k^{gr'} \zeta_k^{gr} - \alpha_k^{gr''} \tau + \sum_{l \neq k}((\alpha_l^{gr''} \alpha_k^{gr'} - \alpha_k^{gr''} \alpha_l^{gr'}) \log |1 - e^{i(\xi_l^{gr} - \overline{\xi_k^{gr}})}| +$$
$$(\alpha_l^{gr'} \alpha_k^{gr'} + \alpha_l^{gr''} \alpha_k^{gr''}) \arg(1 - e^{i(\xi_l^{gr} - \overline{\xi_k^{gr}})})). \tag{5.23}$$

$$\alpha_l^{gr''} = \sum_k^{N_l} \alpha_k'' \ , \tag{5.24}$$

$$\alpha_l^{gr'} = \sum_k^{N_l} \alpha_k' \ . \tag{5.25}$$

Нет никакого слияния этих групп для большого $\tau$. Мы исследуем движение полюсов при этом предположении.

$$|\zeta_l^{gr} - \zeta_k^{gr}| \gg \eta_l^{gr} + \eta_k^{gr}, 1 \leq l, k \leq N \ . \tag{5.26}$$

Для $l \neq k$, $\eta_k^{gr} \mapsto 0$, $\varphi_{lk}^{gr} = \zeta_l^{gr} - \zeta_k^{gr}$ мы получаем

$$\log\mid 1-e^{i(\xi_l^{gr}-\overline{\xi_k^{gr}})}\mid\approx\log\mid 1-e^{i(\zeta_l^{gr}-\zeta_k^{gr})}\mid=\log 2+\frac{1}{2}\log\sin^2\frac{\varphi_{lk}^{gr}}{2} \qquad (5.27)$$

и

$$\arg(1-e^{i(\xi_l^{gr}-\overline{\xi_k^{gr}})})\approx\arg(1-e^{i(\zeta_l^{gr}-\zeta_k^{gr})})=\frac{\varphi_{lk}^{gr}}{2}+\pi n-\frac{\pi}{2}\ . \qquad (5.28)$$

Мы выбираем $n$ в уравнении (5.28) так, чтобы уравнение (5.21) было правильно.

Подставляя эти результаты в уравнения (5.23) мы получаем

$$C_k=\alpha^{gr\prime}\zeta_k^{gr}-\alpha_k^{gr\prime\prime}\tau+\sum_{l\neq k}[(\alpha_l^{gr\prime\prime}\alpha_k^{gr\prime}-\alpha_k^{gr\prime\prime}\alpha_l^{gr\prime})\log\mid\sin\frac{\varphi_{lk}^{gr}}{2}\mid$$
$$+(\alpha_l^{gr\prime}\alpha_k^{gr\prime}+\alpha_l^{gr\prime\prime}\alpha_k^{gr\prime\prime})\frac{\varphi_{lk}^{gr}}{2}] \qquad (5.29)$$

## 5.3 Теорема о слиянии полюсов

Из уравнений (5.29) мы можем заключить

(i) Суммированием уравнений (5.29) (или точно из уравнения (5.12)) мы получаем

$$\sum_k\alpha_k^{gr\prime}\zeta_k^{gr}=const\ . \qquad (5.30)$$

(ii) Для $\mid\varphi_{lk}^{gr}\mid\mapsto 0,2\pi$, мы получаем $\log\mid\sin\frac{\varphi_{lk}^{gr}}{2}\mid\mapsto\infty$, подразумевая, что полюса не могут пересечь друг друга;

(iii) Из (ii) мы выводим $0<\mid\varphi_{lk}^{gr}\mid<2\pi$

(iv) Из (i) и (iii), $\zeta_k^{gr}\mapsto\infty$ невозможно;

(v) В уравнении (5.29) мы должны компенсировать второй расходящийся член. Из (iv) и (iii) мы можем сделать это только если $\alpha_l^{gr\prime\prime}=\sum_k^{N_l}\alpha_k^{\prime\prime}=0$ for all $l$.

Поэтому из eq. (5.29) мы получаем

$$\sum_k^{N_l}\alpha_k^{\prime\prime}=0\ , \qquad (5.31)$$

$$\dot{\varphi}_{lk}^{gr}=0\ , \qquad (5.32)$$

$$\varphi_{lk}^{gr}\neq 0\ , \qquad (5.33)$$

$$\dot{\zeta}_k^{gr}=0\ . \qquad (5.34)$$

Для асимптотического движения полюсов в группе $N_m$ мы получаем из уравнений (5.31), (5.32), (5.33), (5.34) беря ведущий член в уравнениях (5.13), (5.14)

$$\tau=\frac{2}{\lambda+1}t\ , \qquad (5.35)$$

$$0 = \dot{\tau} + \sum_{l}^{N_m} \alpha_l \frac{\dot{\eta}_k + \dot{\eta}_l + i(\dot{\zeta}_k - \dot{\zeta}_l)}{\eta_k + \eta_l + i(\zeta_k - \zeta_l)} \ . \tag{5.36}$$

Решением этих уравнений является

$$\eta_k = \eta_k^0 e^{-\frac{1}{\alpha_m^{gr\prime}}\frac{2}{1+\lambda}t} \ , \tag{5.37}$$

$$\varphi_{lk} = \varphi_{lk}{}^0 e^{-\frac{1}{\alpha_m^{gr\prime}}\frac{2}{1+\lambda}t} \ , \tag{5.38}$$

$$\dot{\zeta}_k = 0 \ . \tag{5.39}$$

Поэтому мы можем заключить, что для уничтожения расходящегося члена нам нужно

$$\alpha_l^{gr\prime\prime} = \sum_{k}^{N_l} \alpha_k'' = 0, \tag{5.40}$$

$$\alpha_l^{gr\prime}(1 + \lambda) > 0 \tag{5.41}$$

for all $l$.

## 5.4 Выводы

С периодическим граничным условием, уравнение (5.40), правильно для всех полюсов, таким образом, мы получаем $N' = 1$, $m = 1$ и $N_m = N$. Поэтому единственное решение:

$$\eta_k = \eta_k^0 e^{-\frac{2}{(1-\lambda^2)}t} \ , \tag{5.42}$$

$$\varphi_{lk} = \varphi_{lk}{}^0 e^{-\frac{2}{(1-\lambda^2)}t} \ , \tag{5.43}$$

$$\dot{\zeta}_k = 0 \ . \tag{5.44}$$

$$1 - \lambda^2 > 0 \tag{5.45}$$

С граничным условием без потоков у нас есть пары полюсов, для которых условия из уравнения (5.40) правильны, таким образом, все эти пары должны слиться. Из-за симметрии задачи эти полюса могут слиться только на границах канала $\zeta = 0, \pm\pi$. Поэтому мы получаем две группы полюсов на границах $N' = 2$, $m = 1, 2$, $N_1 + N_2 = N$, $\alpha_1^{gr\prime} + \alpha_2^{gr\prime} = 1 - \lambda$. Следовательно, мы получаем решение (на двух границах):

$$\eta_k^{(1)} = \eta_k^{(1),0} e^{-\frac{1}{\alpha_1^{gr\prime}}\frac{2}{1+\lambda}t} \ , \tag{5.46}$$

$$\varphi_{lk}^{(1)} = \varphi_{lk}{}^{(1),0} e^{-\frac{1}{\alpha_1^{gr\prime}}\frac{2}{1+\lambda}t} \ , \tag{5.47}$$

$$\dot{\zeta}_k^{(1)} = 0 \ ; \tag{5.48}$$

$$\eta_k^{(2)} = \eta_k^{(2),0} e^{-\frac{1}{\alpha_2^{gr\prime}}\frac{2}{1+\lambda}t} \;, \qquad (5.49)$$

$$\varphi_{lk}^{(2)} = \varphi_{lk}{}^{(2),0} e^{-\frac{1}{\alpha_2^{gr\prime}}\frac{2}{1+\lambda}t} \;, \qquad (5.50)$$

$$\zeta_k^{(2)} = \pm \pi \;; \qquad (5.51)$$

$$\alpha_1^{gr\prime}(1+\lambda) > 0, \qquad (5.52)$$

$$\alpha_2^{gr\prime}(1+\lambda) > 0. \qquad (5.53)$$

## 5.5 Приложение А

Мы должны доказать, что $\tau \mapsto \infty$, если $t \mapsto \infty$ и если не возникает сингулярности за конечное время. Это очевидно, если второй член в следующей формуле для $\tau$ больше чем ноль:

$$\tau = t + [-\frac{1}{2}\sum_{k=1}^{N}\sum_{l=1}^{N}\overline{\alpha_k}\alpha_l \log(1 - \overline{a_k}a_l)] + C_0 \;, \qquad (5.54)$$

где $\mid a_l \mid < 1$ for all $l$. Позвольте нам доказать это

$$\begin{aligned}
-\frac{1}{2}\sum_{k=1}^{N}\sum_{l=1}^{N}\overline{\alpha_k}\alpha_l \log(1 - \overline{a_k}a_l) &= -\frac{1}{2}\sum_{k=1}^{N}\sum_{l=1}^{N}\overline{\alpha_k}\alpha_l \sum_{n=1}^{\infty}(-\frac{(\overline{a_k}a_l)^n}{n}) \\
\frac{1}{2}\sum_{n=1}^{\infty}\frac{1}{n}(\sum_{k=1}^{N}\overline{\alpha_k}(\overline{a_k})^n)(\sum_{l=1}^{N}\alpha_l(a_l)^n) \\
&= \frac{1}{2}\sum_{n=1}^{\infty}\frac{1}{n}\overline{(\sum_{l=1}^{N}\alpha_l(a_l)^n)}(\sum_{l=1}^{N}\alpha_l(a_l)^n) > 0
\end{aligned} \qquad (5.55)$$

# Глава 6
# Резюме

Задача распространения пламени изучена как пример неустойчивых фронтов, которые деформируются во многих масштабах. Аналитический инструмент разложения по полюсам в комплексной плоскости используется, чтобы проанализировать взаимодействие неустойчивого процесса роста со случайными начальными условиями и возмущениями. Мы утверждаем, что эффект случайного шума огромный и что этим никогда нельзя пренебрегать в достаточно больших системах. Мы представляем моделирование, которые приводят к законам подобия для скорости и ускорения фронта как функция размера системы и уровня шума, и аналитические аргументы, которые объясняют эти результаты в терминах динамики полюсов, сгенерированных шумом.

Мы рассматриваем распространение фронта пламени в конфигурации канала. Решение для стационарного состояния в этой задаче является пространственно-зависимым, и поэтому линейный анализ стабильности описан частичным интегро-дифференциальным уравнением с пространственно-зависимым коэффициентом. Соответственно он включает сложные собственные функции. Мы показываем, что этот анализ может быть выполнен со всеми необходимым деталям, используя динамическую систему конечного порядка, в терминах динамики особенностей в комплексной плоскости, приводящую к детальному пониманию физики собственных функций и собственных значений.

Огрубление расширяющихся фронтов пламени появлением изломо-подобных особенностей - очаровательный пример взаимодействия между неустойчивостью, шумом и нелинейная динамикой, которая напоминает о само-фрактолизации в структурах Лапласовкого роста. Нелинейное интегро-дифференциальное уравнение, которое описывает динамику расширяющихся фронтов пламени, поддается аналитическому исследованию, используя разложение на полюса. Этот сильный метод позволяет развить удовлетворительное понимание качественного и небольшого количества количественных аспектов комплексной геометрии, которая развивается в расширяющихся фронтах пламени.

Распространение пламени используется как формирующий прототип и пример расширяющихся фронтов, которые деформируются беспредельно в радиальных конфигурациях, но достигают простой формы в геометрии канала. Мы показываем, что соответствующие законы подобия, которые управляют радиальным ростом, могут быть выведены, как только более простая геометрия канала понята подробно. В радиальных конфигурациях (на контрасте, чтобы провести канал через конфигурации) эффект внешнего шума крайне важен для ускорения и деформации фронтов. Однако, как только взаимосвязи между размером системы, скоростью распространения и шумовым уровнем понты в геометрии канала, законы подобия для радиального роста следуют из них.

Математическая задача Лапласового роста без поверхностного растяжения показывает семейство точных (аналитических) решений в терминах логарифмических полюсов в комплексной плоскости. Мы показываем, что у этого семейства решений есть замечательное свойство: произвольные начальные условия в геометрии канала, которые начинаются произвольно со многих особенностей, показывают обратный каскад в единственный палец.

# Литература